\documentclass[a4paper,fleqn,usenatbib]{mnras}

\usepackage[T1]{fontenc}
\usepackage{ae,aecompl}

\usepackage{graphicx}	
\usepackage{amsmath}	
\usepackage{amssymb}	
\usepackage{soul}	
\usepackage{color,xcolor}	

\title[Correlations between $L_\gamma$ and $\sigma_m$, $L_{inj}'$]{Correlations between $\gamma$-ray luminosity and magnetization of the jet as well as relativistic electron injection power:
cases for Mrk 421, 3C 454.3 and 3C 279}
\author[Hu, Yan, and Hu]{
Wen Hu,$^{1}$
Dahai Yan,$^{2}$\thanks{E-mail: yandahai@ynao.ac.cn}
and Qianglin Hu$^{1}$
\\
$^{1}$Department of Physics, Jinggangshan University, Jiangxi Province, Ji'an 343009, People's Republic of China\\
$^{2}$Key Laboratory for the Structure and Evolution of Celestial Objects, Yunnan Observatories, Chinese Academy of Sciences,\\
Kunming 650011, People's Republic of China\\
}

\date{Accepted 2021 February 12; Received 2021 February 11; in original form 2020 November 19}

\begin{document}
\label{firstpage}
\pagerange{\pageref{firstpage}--\pageref{lastpage}}
\maketitle

\begin{abstract}
By fitting high-quality and simultaneous multi-wavelength (MWL) spectral energy distributions (SEDs) at multiple epochs with a one-zone leptonic jet model,
we study jet properties of the three famous blazars Mrk 421, 3C 454.3 and 3C 279.
In the jet model, the emitting electron energy distributions (EEDs) are calculated by solving the kinetic equation of electron injection, escape, adiabatic and radiative energy losses.
To explore multi-dimensional parameter space systematically, we employ a Markov chain Monte Carlo (MCMC) fitting technique.
The properties of emission regions we derived here are consistent with those in previous studies, e.g., the particle-dominated and low-magnetization jet.
The new finding is that there is a tight correlation between $\gamma$-ray luminosity and electron injection power
and an anti-correlation between $\gamma$-ray luminosity and jet magnetization parameter.
The results suggest that same energy-dissipative mechanism (like a shock) could be operating in the jets of different types of blazars, and the origin of $\gamma$-ray flares is associated with the particle acceleration process.
\end{abstract}



\section{Introduction}\label{sec:intro}
Flat spectrum radio quasars (FSRQs) and BL Lacertae objects (BL Lacs) constitute a subset of active galactic nuclei (AGNs),
which are called blazars.
A blazar is characterized by flux variability at all wavelengths, 
high polarization at optical and radio frequencies \citep[e.g.,][]{Darcangelo2009,Marscher2010,Peceur2020},
apparently superluminal jet components \citep[e.g.,][]{Jorstad2004,Jorstad2005,Homan2009}, and non-thermal emission from a relativistic jet pointed close
to the observer's line of sight \citep{Urry1995}

Compared with BL Lacs with featureless optical spectra, 
FSRQs display prominent emission lines with the equivalent width $\geq 5 \rm \mathring{A}$
and the prominent ultraviolet excess which is attributed to thermal emission of accretion disk \citep{Urry1995}.
The multi-wavelength (MWL) SED of blazar is characterized by two prominent peaks.
The first peak, which is attributed to synchrotron emission of high-energy electrons in the jet, 
is generally located at infrared/optical bands.
The second peak, which could be generated by inverse Compton (IC) scattering off low energy photons by high-energy electrons,
is generally located at $\gamma$-ray energies.
The low photon fields for the IC scattering can include the synchrotron emission from the high-energy electrons themselves \citep[Synchrotron Self Compton: SSC; ][]{Bloom1996,Finke2008}
and external photon fields surrounding the jet (external Compton scattering: EC).
Depending on location of the emission region, the external photon field responsible for the EC emission could be either from dusty torus \citep[DT;][]{Blazejowski2000,Dermer2014,Yan2015,Hu2017,Wu2018},
and/or from broad-line region \citep[BLR;][]{Sikora1994,Yan2012,Bottcher2013,Hu2015}.

Electron acceleration in blazar jet is still an open question.
In general, it can be achieved by the energy dissipated in shocks and/or magnetic reconnections.
If the jet is dominated by the kinetic energy flux at the dissipation distance,
then shocks are natural candidates for powering the jet emission and accelerating electrons to
ultra-relativistic energy \citep{Marscher1985,Kirk2000,Spada2001,Sironi2009,Summerlin2012}. 
The shock scenario is supported by multiple observation of $\gamma-$ray outbursts in coincidence with the emergence of a jet perturbation in or close to radio core of blazars \citep[e.g.,][]{Marscher2008,Jorstad2013,Abeysekara2018}.
While, if the jet remains Poynting flux dominated at the energy dissipation distance,
shocks are generally expected to be weak, and magnetic reconnection is a more plausible candidate for acceleration of electrons
\citep[e.g.,][]{Cerutti2012,Sironi2013,Sironi2014,Guo2014,Guo2015}.
The ratio between the Poynting flux and kinetic energy flux,
namely magnetization parameter $\sigma_m\equiv B'^2/4\pi\rho' c^2$,
is a crucial quantity to discriminate the two scenarios \footnote{Here, $B'$ and $\rho'$ are the magnetic field strength and the rest-mass density in the rest frame of the jet, respectively.}.

Recently, particle-in-cell (PIC) simulations of relativistic magnetic reconnection have demonstrated that electrons can be efficiently accelerated
to form a non-thermal distribution for $\sigma_m\gtrsim1$ \citep[see][and references therein]{Petropoulou2019}.
Magnetic reconnection scenario has been proposed  to account for many of the extreme spectral and temporal properties of blazars \citep{Giannios2009a,Giannios2013,Petropoulou2016,Christie2019}.
The magnetic reconnection scenario predicts a near equipartition between the energies carried by magnetic field and emitting electrons \citep{Sironi2015,Petropoulou2019}.

Modeling the broad-band SED of a blazar is frequently used to probe 
blazar jet physics \citep[e.g.,][]{Ghisellini2010,Ghisellini2015,Yan2013,Yan2014,Inoue2016,Chen2018}.
Since blazar emission often shows strong variabilities at all wavelength,
the evolution of well-sampled simultaneous SEDs at different epochs are important.
In the work, we investigate the jet physics in three typical blazars (3C 454.3, Mrk 421 and 3C 279)
through modeling their quasi-simultaneous SEDs in multiple activity states.

The outline of this paper is as follows.
Section \ref{model} describes the jet model and the fitting method.
In Section \ref{results}, we report our results.
Finally in Section \ref{sec:summ}, we summarize our discussions and conclusions.
Throughout the paper, we used a flat cosmology model with the following parameters:
$\rm H_0=70 ~km~s^{-1}~Mpc^{-1}$, $\rm \Omega_M=0.3$, and $\Omega_\Lambda = 0.7$ \citep{Hinshaw2013,Bennett2013}.

\section{Model}\label{model}

We adopt a one-zone homogeneous leptonic jet model.
It is assumed that emissions are produced in a spherical blob of radius $R^\prime$
which moves relativistically with bulk Lorentz factor $\Gamma$ in the observer's frame
at an angle $\theta_{\rm obs}\sim\Gamma^{-1}$ with respect to our line of sight.
The emission region is filled with a uniform and tangled magnetic field of strength $B'$.
The radius of the emission region is estimated from the minimum variability timescale $t_{\rm var}$, through $R'=c\delta_D t_{\rm var}/(1+z)$,
where $z$ is the redshift of the source and Doppler factor $\delta_D$ is approximated as $\Gamma$.
Throughout this paper, all primed quantities refer to the comoving frame of the emission region and unprimed quantities denote the observer's frame.

It is assumed that a population of ultra-relativistic non-thermal electrons is continuously
injected into the blob with a rate of $Q_e'(\gamma')$ in units of $\rm s^{-1}$.
The injected electrons lose energy through radiation and adiabatic processes.
The kinetic equation that governs the evolution of electrons can be described by the time-dependent Fokker-Planck equation \citep{Coppi1990,Chiaberge1999}.

With respect to our previous work, we further refine the model,
including a physically realistic, stratified BLR model proposed by \cite{Finke2016}.
A modification is including adiabatic expansion, which may be important in modeling the SEDs of blazars \citep[e.g.,][]{Lewis2016,Lewis2018}.
The radiation coolings from both EC-BLR and EC-DT processes are considered to calculate the radiating electron distribution.

We consider a single power-law (PL) of electron injection, which is given by
\begin{equation}
{\gamma'}^2Q_e^\prime(\gamma^\prime)=Q_0\gamma^{\prime2-n},~\gamma_{\rm min}'\le\gamma'\le\gamma_{\rm max}',
\end{equation}
with
\begin{equation}
Q_0' =\left\{
             \begin{tabular}{l}
             $\frac{L_{\rm inj}'}{m_ec^2}\frac{2-n}{{\gamma_{\rm max}'}^{2-n}-{\gamma_{\rm min}'}^{2-n}};~n\neq2$ \\
             $\frac{L_{\rm inj}'}{m_ec^2\ln\left(\gamma_{\rm max}'/\gamma_{\rm min}'\right)};~n=2$ \\
             \end{tabular}
            \right. ,
\end{equation}
where $\gamma_{\rm min}'$ and $\gamma_{\rm max}'$ are respectively the low- and high-energy cutoffs,
$L_{\rm inj}'$ is injection power of electrons in units of ergs/s, $n$ is the power-law index ,
$m_e$ is the rest mass of electron, and $c$ is the speed of light.

\subsection{Emitting Electron Energy Distribution}

The kinetic equation governing the evolution of the electron energy distribution, $N_e'(\gamma')$, is given by
\begin{equation}\label{kineticEQ}
 \begin{aligned}
\frac{\partial{N_e^\prime(\gamma')}}{\partial{t'}}&=
-\frac{\partial}{\partial{\gamma'}}\left[\dot\gamma'N_e^\prime(\gamma')\right]-\frac{N_e^\prime(\gamma')}{t'_{\rm esc}}+Q_e'(\gamma') ,
\end{aligned}
\end{equation}
where $t'_{\rm esc}$ is the escape timescale of electrons, and $\dot\gamma'$ is the energy-loss rate of the electrons.

In a blazar jet,  the electrons lose energy through synchrotron and IC scattering on internal and external photon fields.
The external radiation fields surrounding the jet include BLR  and DT photon fields.
Therefore, the electron radiative cooling rate can be written as
$\dot\gamma'_{\rm rad}= \dot\gamma_{\rm syn}'+\dot\gamma_{\rm ssc}'+\dot\gamma_{\rm BLR}'+\dot\gamma_{\rm DT}'$,
where $\dot\gamma'_{\rm syn}, ~\dot\gamma'_{\rm ssc}, ~\dot\gamma'_{\rm BLR}$ and $\dot\gamma'_{\rm DT}$
are the cooling rate due to the synchrotron, SSC, EC-BLR and EC-DT radiation, respectively.

The synchrotron energy-loss rate is given by
\begin{equation}
-\dot\gamma'_{\rm syn}=\frac{4\sigma_T}{3m_ec}{U_B'\gamma'}^2,
\end{equation}
where $U_B'={B'}^2/8\pi$ is the magnetic field energy density.

The SSC energy-loss rate is given by
\begin{equation}
-\dot\gamma'_{\rm ssc}=\frac{4\sigma_T}{3m_ec}{\gamma'}^2\int_0^\infty d\epsilon' u_{syn}'(\epsilon')f_{kn}(\epsilon',\gamma'),
\end{equation}
where $u_{\rm syn}'(\epsilon')$ is the spectral energy density of the synchrotron radiation,
and \begin{equation}
f_{kn}(\epsilon',\gamma')=\frac{9}{16}\int_{\gamma'_{\rm l}}^{\gamma'}d\gamma''F_{c}(x,q)\frac{\gamma'-\gamma''}{{\epsilon'}^2{\gamma'}^4},
\end{equation}
where the lower limit for the integration is $\gamma'_{\rm l}\simeq\gamma'+\epsilon'-\frac{4{\gamma'}^2\epsilon'}{1+4\gamma'\epsilon'}$, and
the kernal function is
\begin{equation}\label{fc1}
 \begin{aligned}
F_{c}(x,q) = & \Big[2q\ln{q}+q+1-2q^2 + \\
&\frac{(xq)^2}{2(1+xq)}(1-q)\Big] \times H[q;\frac{1}{4\gamma'^2},1],
\end{aligned}
\end{equation}
\citep{Jones1968,Blumenthal1970}.
Here, $H(x; a, b)$ is the Heaviside function defined as H = 1 if $a \le x \le b$ and H = 0 otherwise.
In the equation,  $x=4\epsilon'\gamma'$, $q=\frac{\epsilon_\gamma'/\gamma'}{x(1-\epsilon_\gamma'/\gamma')}$,
and the scattered photon energy is $\epsilon_\gamma'=\gamma'+\epsilon'-\gamma''$.

The EC-DT energy-loss rate is given by
\begin{equation}
-\dot\gamma'_{\rm ec}=\frac{4\sigma_T}{3m_ec}\gamma^2\int_0^\infty d\epsilon u_{ext}(\epsilon)f_{kn}(\epsilon,\gamma)\ ,
\end{equation}
where the quantities $\gamma=\delta_D\gamma'$ and $\epsilon$ refer to the stationary frame with respect to the black hole,
and $u_{\rm ext}(\epsilon)$ is the spectral energy density of external photon field.
Here, the DT radiation is assumed to be described as a dilute blackbody spectrum with a peak frequency of $\nu_{DT} = 3\times10^{13}$ Hz in the lab frame \citep[e.g.,][]{Tavecchio2008,Ghisellini2009}, and the spectral energy density is given by
\begin{equation}
u_{ext}(\epsilon)=\frac{15U_0}{(\pi \Theta)^4}\frac{\epsilon^3}{\exp\left(\epsilon/\Theta\right)-1}\ ,
\end{equation}
where $\rm \Theta=h\nu_{DT}/2.82m_ec^2$ and $\rm U_0\simeq2.1\times10^{-4}~erg/cm^3$ are the dimensionless temperature and energy density of the DT radiation field, respectively.

 The EC-BLR loss rate is given by
  \begin{equation}
\dot\gamma'_{\rm BLR}=-\frac{4\sigma_T}{3m_ec}{\gamma}^2\sum_{i=1}^{n=26}\oint d\Omega u_{BLR}(R_{loc}, \epsilon_i,\Omega)f_{kn}(\epsilon_i,\gamma),
\end{equation}
where $R_{loc}$ is the location of the emission region, $\epsilon_i$ is the dimensionless energy of the lines, n denotes the number of different lines in the BLR model, and
\begin{equation}
f_{kn}(\epsilon,\gamma)=\frac{9}{32}\int_{\gamma_{\rm l}}^{\gamma}d\gamma''F_{c}(x,q)\frac{\gamma-\gamma''}{{\epsilon}^2{\gamma}^4},
\end{equation}
where  $\gamma_{\rm l}\simeq\gamma+\epsilon-\frac{2{\gamma}\overline\epsilon}{1+2\overline\epsilon}$, and
\begin{equation}
\label{fc}
 \begin{aligned}
F_{c}(x,q) = &\Big[y+y^{-1}-\frac{2\epsilon_\gamma}{\gamma\overline\epsilon y}+\left(\frac{\epsilon_\gamma}{\gamma\overline\epsilon y}\right)^2\Big]\\
&\times H[\epsilon_\gamma;\frac{\overline\epsilon}{2\gamma},\frac{2\gamma\overline\epsilon}{1+2\overline\epsilon}]\ ,\
\end{aligned}
\end{equation}
where  $y=1-(\epsilon_\gamma/\gamma)$, $\overline\epsilon=\gamma\epsilon(1-\mu\mu_{obs})$
and $\epsilon_\gamma=\gamma+\epsilon-\gamma''$ \citep{Dermer1993,Dermer2009}.
In the calculation, we adopt a stratified BLR model, with 26 different lines emitting at different radii, in a spherical shell configuration \citep{Finke2016}.
In the model, the location and luminosities of the various lines are estimated by using empirical relations derived from reverberation mapping,
when the disk luminosity $L_d$ has been specified.

In addition to the radiative losses,  electrons also loss energy through adiabatic expansion of the outflowing plasma blob.
The adiabatic losse is evaluated through
\begin{equation}
-\dot\gamma'_{\rm adi}=\frac{3c}{R'\delta_D}\gamma'\ ,
\end{equation}
assuming a conical jet with opening angle  $\theta_{op}\sim1/\Gamma$ \citep{Bottcher2013}.

Therefore, the total cooling rate for electrons is $\dot\gamma'=\dot\gamma'_{\rm adi}+\dot\gamma'_{\rm rad}$.

We parameterize the escape timescale in terms of the light crossing timescale as $t_{\rm esc}'=\eta R'/c$ with $\eta>1$.
Here, we adopted a typical value of $\eta = 10$ \citep{Bottcher2002,Hu2020}.

With the above information, equation~(\ref{kineticEQ}) is numerically solved by using the full implicit scheme described by \cite{Graff2008} to
calculate the steady-state EED.
 Subsequently, the observed SEDs of synchrotron and inverse-Compton (IC) emissions are calculated by using the formulas in \citet{Finke2008} and \citet{Dermer2009}.
 Here, the synchrotron self-absorption (SSA) process is considered \citep{Rybicki1979,crusius1986}.
We also consider the contribution from an accretion disk following \cite{Dermer2002},
which is assumed to be a \cite{Shakura1973} disk.

 In summary, the model is characterized by eight parameters, i.e.,  $B', n, \rm t_{var}, \delta_D, L'_{inj}, \gamma_{min}', \gamma_{max}'$ and $R_{\rm loc}$.

 \subsection{Fitting Methodology}
By applying the model to the observed SED, the free parameters and their uncertainties are estimated by performing the MCMC fitting method,
which is a powerful tool to explore the multi-dimensional parameter space in blazar science \citep{Yan2013,Yan2015}.
The details on MCMC technique can be found in \cite{Lewis2002,Yuan2011,Liu2012}.

 \begin{table*}
\caption{Mean values and 1$\sigma$ errors of the parameters for 3C 454.3, Mrk 421 and 3C 279.\label{tab1}}
\begin{tabular}{lccccccc}
\hline
state	           & $B^\prime$ (G)	  & $\delta_{\rm D}$  	&$\log_{10} L'_{\rm inj}$  & $\log_{10} \gamma_{\rm min}'$ & $n$ & $\log_{10} R_{\rm loc}$   & $\chi_{r}^2$  \\
\hline
3C 454.3\\
Low     &$ 0.96 \pm 0.22	$&$ 30.90 \pm 2.25  $&$ 42.55 \pm 0.09  $&$ 2.80 \pm 0.09  $&$ 3.08 \pm 0.35  $&$ --	    $& 1.32\\
Low$^\dag$     &$ 0.70 \pm 0.12	$&$ 22.23 \pm 1.69  $&$ 43.14 \pm 0.09  $&$ 2.89 \pm 0.12  $&$ 2.90 \pm 0.37  $&$ --	    $& 1.44\\
06/11   &$ 0.98 \pm 0.21	$&$ 37.39 \pm 2.68  $&$ 42.81 \pm 0.07  $&$ 2.63 \pm 0.06  $&$ 3.41 \pm 0.32  $&$ --	    $&  1.03\\
27/11   &$ 0.60 \pm 0.11	$&$ 34.19 \pm 2.70  $&$ 43.35 \pm 0.12  $&$ 2.93 \pm 0.12  $&$ 2.63 \pm 0.35  $&$ --  	    $&  1.02\\
01/12   &$ 0.78 \pm 0.10	$&$ 40.44 \pm 2.12  $&$ 43.20 \pm 0.06  $&$ 2.81 \pm 0.07  $&$ 3.20 \pm 0.23  $&$ --  	    $&  1.06\\
02/12   &$ 1.04 \pm 0.16	$&$ 54.85 \pm 4.52  $&$ 43.13 \pm 0.09  $&$ 2.30 \pm 0.14  $&$ 3.06 \pm 0.16  $&$ 18.00 \pm 0.06 $&  1.37\\
03/12   &$ 0.74 \pm 0.07	$&$ 42.57 \pm 2.20  $&$ 43.28 \pm 0.06  $&$ 2.73 \pm 0.08  $&$ 3.01 \pm 0.10  $&$ 18.14 \pm 0.14 $&   1.37\\
\\
Mrk 421 \\
Quiet   &$ 0.06 \pm 0.01	$&$ 59.86 \pm 2.54 $&$ 39.66 \pm 0.01  $&$ 2.88\pm0.05 $&$ 2.29\pm 0.04 $&$ 5.65 \pm 0.02 $& 2.61\\
Quiet$^\dag$	&$ 0.03 \pm 0.003	$&$ 23.54 \pm 0.96 $&$ 40.78 \pm 0.03  $&$ 3.20\pm0.05 $&$ 2.04\pm 0.04 $&$ 6.02 \pm 0.02 $& 3.12\\
55266   &$ 0.03 \pm 0.01	$&$ 73.46 \pm 6.04 $&$ 39.60 \pm 0.02  $&$ 2.78\pm0.08 $&$ 2.11\pm 0.05 $&$ 6.05 \pm 0.02 $& 2.47\\ 
55270   &$ 0.07 \pm 0.02	$&$ 54.25 \pm 4.98 $&$ 39.76 \pm 0.04  $&$ 2.82\pm0.08 $&$ 2.19\pm 0.07 $&$ 5.75 \pm 0.04 $& 0.96\\ 
55277   &$ 0.13 \pm 0.06	$&$ 48.15 \pm 9.78 $&$ 39.68 \pm 0.10  $&$ 2.61\pm0.15 $&$ 2.05\pm 0.14 $&$ 5.63 \pm 0.07 $& 1.15\\ 
\\
3C 279\\
Period A   &$ 1.28 \pm 0.09	$&$ 37.73 \pm 1.12 $&$ 41.85 \pm 0.03  $&$ 2.62 \pm 0.02  $&$ 3.68 \pm 0.08 $&$ --	    $& 1.54\\
Period A$^\dag$  &$ 0.98 \pm 0.09	$&$ 23.13 \pm 0.79 $&$ 42.80 \pm 0.04  $&$ 2.63 \pm 0.04  $&$ 3.62 \pm 0.09 $&$ 17.47\pm0.10$& 0.80\\
Period C   &$ 1.13 \pm 0.09	$&$ 39.40 \pm 1.12 $&$ 42.08 \pm 0.03  $&$ 2.69 \pm 0.02  $&$ 3.58 \pm 0.08 $&$ --	    $& 1.70\\
Period D   &$ 0.52 \pm 0.06	$&$ 42.86 \pm 1.59 $&$ 42.55 \pm 0.06  $&$ 2.98 \pm 0.06  $&$ 3.10 \pm 0.10 $&$ --	    $& 1.54\\
\hline
\end{tabular}
\begin{flushleft}
Notes: \dag  denotes the results obtained with $t_{\rm var}=1$ day. 
\end{flushleft}
\end{table*}

\section{Results}\label{results}

\begin{figure*}
  \centering
  \includegraphics[width=\textwidth]{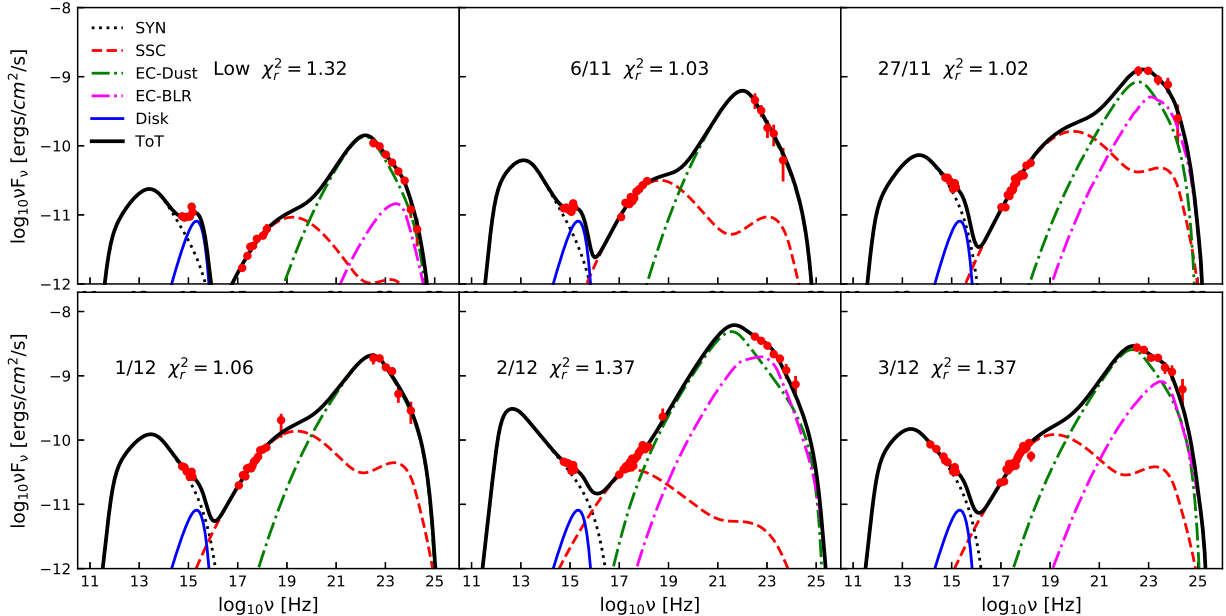}
  \caption{Comparisons of the best-fitting SEDs with observed data of 3C 454.3.
The different components are labelled in the legend.
   }\label{figure1}
\end{figure*}

We apply the model described in Section \ref{model} to the well-sampled SEDs of three famous blazars 3C 454.3, Mrk 421 and 3C 279.
The results of the fits to the observed SEDs are shown in the Figures \ref{figure1}-\ref{figure3}.
The corner plots of the free parameters are displayed in the left panels of Figures \ref{distr1}-\ref{distr4} in Appendix \ref{appA}. 
The fitted parameter values are tabulated in Table \ref{tab1}.
 In the calculations, a relative systematic uncertainty of 5\% was added in quadrature to the statistical error of the IR-optical-UV and X-rays data \citep{Poole2008, Abdo2011a}.

\subsection{SEDs modelling}

\subsubsection{3C 454.3}

3C 454.3 is the brightest $\gamma$-ray FSRQ with a redshift $z=0.859$ \citep{Jackson1991}.
In recent years, the source attracts much attention, because of its remarkably high activity over the entire electromagnetic spectrum\citep[e.g.,][]{Villata2006,Raiteri2008,Vercellone2009,Vercellone2010,Vercellone2011,Abdo2011,Shah2017}
and its broken PL $\gamma$-ray spectrum \citep{Abdo2009}.
Here, we focus on a simultaneous MWL campaign organized during 2009 November and December.
The MWL SEDs are collected from \cite{Bonnoli2011}.
For investigating the change of the parameters in different active states,
we also model the SED at the lowest $\gamma$-ray state since the beginning of Fermi/LAT observations.

The bolometric luminosity of the accretion disc $\rm L_{d}$ is $3\times10^{46}$ erg/s \citep{Raiteri2007}, and
the mass of black hole $\rm M_{BH}$ is $5\times10^8M_\odot$ \citep{Bonnoli2011}.
To further reduce the number of model parameters, we take $\rm t_{\rm var}$ = 6 hours according to the analysis of $\gamma$-ray variability \citep{Tavecchio2010,Jorstad2013}.
Thus, there are six free parameters in the model, i.e., $\rm B', \delta_D, L_{inj}', \gamma_{min}', n$ and $R_{loc}$.
The best-fitting values of the parameters are summarized in Table \ref{tab1}.

From Figure \ref{figure1}, one can see that the model provides satisfactory fits to six SEDs.
In the states of 27/11, 2/12 and 3/12,
the $\gamma$-rays can be well interpreted as the superposition of EC-BLR and EC-DT radiations;
while in the other states, the $\gamma$-rays are attributed to the EC-DT emission.
 The X-ray emission in highest $\gamma$-ray flare is dominated by EC-DT component.
In the other five states, the X-ray emission is almost attributed to SSC component only.

\begin{figure*}
  \centering
  \includegraphics[width=\textwidth]{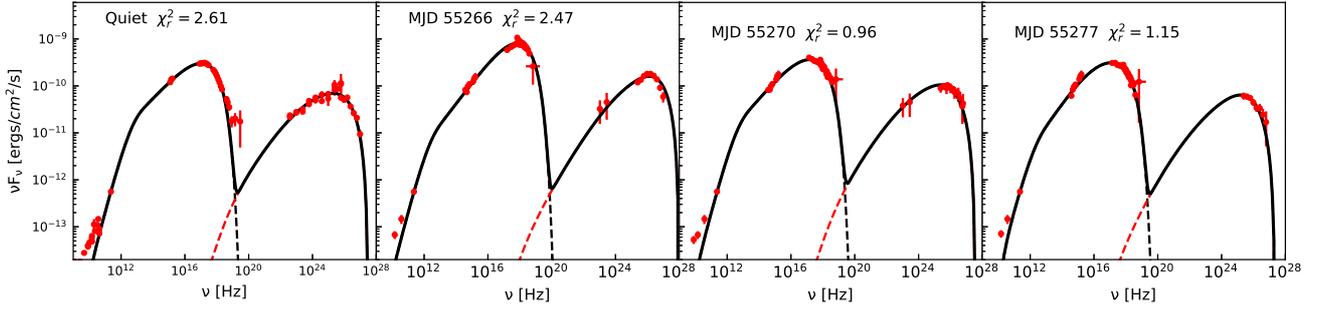}\\
  \caption{Comparisons of the best-fitting SEDs with observed data of Mrk 421.
The black and red dashed lines refer to the synchrotron and SSC emission, respectively. }\label{figure2}
\end{figure*}

\subsubsection{Mrk 421}

Mrk 421 ($z = 0.031$) is the first established extragalactic TeV gamma-ray blazar \citep{Punch1992},
and is categorized as a high-frequency-peaked BL Lac (HBL).
Here we focus on a MWL campaign conducted during a high activity in 2010 March.
The observational data are taken from \cite{Aleksic2015}.
For comparison, the data from 4.5 month long MWL campaign \citep{Abdo2011a} are also included in the work, 
which provides an unprecedented, complete look at the quiescent SED for the source.
In order to account for the rapid variability, the minimum variability timescale $t_{\rm var}$= 1 hour is taken in the fittings \citep[see][for discussion]{Aleksic2015}.

The best-fitting SEDs and observational data points are shown Figure \ref{figure2} and the obtained parameters are reported in Table \ref{tab1}.
EC processes are neglected in the source.
The radio data point from SMA reported in \cite{Abdo2011a} is used to constrain $\gamma_{\rm min}'$ of the PL electron injection.
From Figure \ref{figure2}, we can see that the fitting to each SED is successful. 

\begin{figure*}
  \centering
  \includegraphics[width=\textwidth]{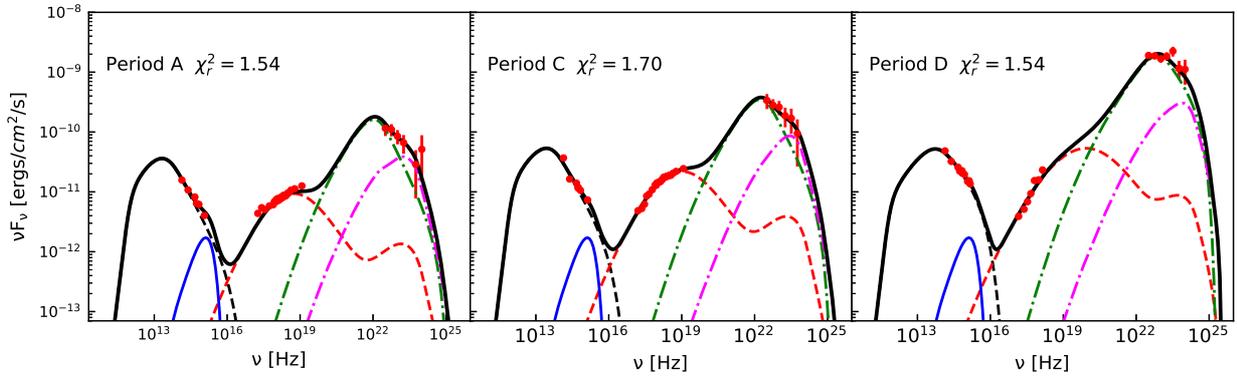}\\
  \caption{Comparisons of the best-fitting SEDs with observed data of 3C 279.
   }\label{figure3}
\end{figure*}

\subsubsection{3C 279}

For comparison, we also revisit the jet properties of the famous FSRQ 3C 279 ($z$=0.538),
which was studied with a similar model in our previous work \citep{Hu2020}.
Here, we focus on the results of a MWL observing campaign conducted during a phase of increased activity from 2013 December to 2014 April.
The MWL SEDs are taken from \cite{Hayashida2015}.
In the fitting, we adopt $\rm t_{var}=2$ hours, which was obtained in the period D in \cite{Hayashida2015}.
$\rm L_d=2\times10^{45}$ ergs/s \citep{Pian1999} and $\rm M_{BH}=5\times10^8M_\odot$ \citep{Gu2001,Woo2002,Nilsson2009} are adopted.

Figure \ref{figure3} shows the SED fitting results, with parameters summarized in Table \ref{tab1}.
The SEDs can be fitted well by the model.
During the period A, we note that both the EC-DT and EC-BLR components are required to reproduce the $\gamma$-ray emission,
while during period D, the $\gamma$-ray emission is dominated by the EC-DT component.
During the period C, an EC-BLR component is needed to account for the $\gamma$-ray emission.
For all three states, the X-ray spectrum is attributed to SSC emission.

 \begin{table*}
\caption{Mean values and 1$\sigma$ errors of the derived parameters for 3C 454.3, Mrk 421 and 3C 279.\label{tab2}}
\begin{tabular}{lcccccc}
\hline
state	           & $\log_{10}P_B$ 	  & $\log_{10}P_e$  	&$\log_{10}P_p$  & $\log_{10}P_r$  & $U_B'/U_e'(10^{-1})$   & $\sigma_m(10^{-2})$  \\
\hline
3C 454.3\\
Low 	&$ 44.56 \pm 0.27  $&$ 45.24 \pm 0.06	$&$ 46.58 \pm 0.06   $&$ 44.88 \pm 0.03     $&$ 2.85\pm2.84     $&$ 2.22\pm1.43	$\\
Low$^\dag$	&$ 44.92 \pm 0.22  $&$ 45.35 \pm 0.02	$&$ 46.75 \pm 0.07   $&$ 45.19 \pm 0.03     $&$ 4.27\pm2.46     $&$ 3.18\pm1.34	$\\
06/11   &$ 44.90 \pm 0.26  $&$ 45.68 \pm 0.05	$&$ 47.21 \pm 0.06   $&$ 45.31 \pm 0.03     $&$ 2.13 \pm 1.55   $&$ 1.14 \pm 0.60$\\
27/11   &$ 44.33 \pm 0.26  $&$ 45.90 \pm 0.04	$&$ 47.28 \pm 0.07   $&$ 45.80 \pm 0.06     $&$ 0.35 \pm 0.32   $&$ 0.25 \pm 0.15$\\
01/12   &$ 44.86 \pm 0.18  $&$ 45.94 \pm 0.03	$&$ 47.48 \pm 0.05   $&$ 45.79 \pm 0.03     $&$ 0.92 \pm 0.49   $&$ 0.51 \pm 0.18$\\
02/12   &$ 45.63 \pm 0.25  $&$ 46.11 \pm 0.07	$&$ 48.10 \pm 0.13   $&$ 45.98 \pm 0.03	    $&$ 4.00 \pm 2.52   $&$ 0.76 \pm 0.38$\\
03/12   &$ 44.91 \pm 0.16  $&$ 46.01 \pm 0.02	$&$ 47.64 \pm 0.06   $&$ 45.92 \pm 0.02     $&$ 0.86 \pm 0.33   $&$ 0.38 \pm 0.09$\\
\\
Mrk 421\\
Quiet	&$ 42.23\pm0.03     $&$ 43.88\pm0.04	 $&$ 43.90\pm0.04	$&$ 41.59\pm0.04    $&$ 0.23\pm0.04     $&$ 4.33\pm0.65    $\\
Quiet$^\dag$	&$ 42.83\pm0.03     $&$ 43.89\pm0.04	 $&$ 43.68\pm0.04	$&$ 42.40\pm0.03    $&$ 0.88\pm0.13     $&$ 28.69\pm3.75   $\\
55266   &$ 42.07\pm0.08     $&$ 44.01\pm0.09	 $&$ 43.93\pm0.06	$&$ 41.78\pm0.08    $&$ 0.12\pm0.05     $&$ 2.90\pm0.90    $\\
55270   &$ 42.23\pm0.08     $&$ 43.84\pm0.09	 $&$ 43.88\pm0.06	$&$ 41.79\pm0.08    $&$ 0.26\pm0.10     $&$ 4.64\pm1.44	   $\\
55277   &$ 42.46\pm0.14     $&$ 43.56\pm0.17	 $&$ 43.75\pm0.09	$&$ 41.80\pm0.17    $&$ 0.98\pm0.53     $&$ 11.62\pm4.79   $\\
\\
3C 279\\
Period A&$ 44.39\pm0.06     $&$ 44.98\pm0.02	 $&$ 46.31\pm0.03	$&$ 44.31\pm0.03    $&$ 2.63\pm0.39     $&$ 2.39\pm0.26	   $\\
Period A$^\dag$&$ 45.47\pm0.11     $&$ 45.12\pm0.02	 $&$ 46.77\pm0.05	$&$ 44.87\pm0.02    $&$ 23.07\pm6.19    $&$ 10.09\pm1.82   $\\
Period C&$ 44.35\pm0.05     $&$ 45.16\pm0.02	 $&$ 46.50\pm0.03	$&$ 44.59\pm0.03    $&$ 1.56\pm0.23     $&$ 1.43\pm0.17	   $\\
Period D&$ 43.82\pm0.14     $&$ 45.47\pm0.04	 $&$ 46.72\pm0.05	$&$ 45.18\pm0.04    $&$ 0.24\pm0.11     $&$ 0.26\pm0.07	   $\\
\hline
\end{tabular}
\begin{flushleft}
Notes: Column 2-5 are the jet power carried by magnetic field, electrons, cold proton and radiation, respectively.
All the values are in logarithmic space. Column 6 is equipartition parameter.
Column 7 is the magnetization parameter. 
\end{flushleft}
\end{table*}

\subsection{Physical properties of the jets}

Evolution of model parameters could allow us to get a deep insight into the cause of such a activity that may be associated
with changes in the physical conditions of jet, e.g., the injection rate, Doppler factor, magnetic field strength
\citep[e.g.,][]{Bottcher2002,Graff2008,Hu2015,Hu2020},
 and/or change in the acceleration process \citep{Yan2013}.

 \subsubsection{Locations of the radiation region}

Our results show that the locations of emission regions $R_{\rm loc}$ can be only constrained well
for the data of 3C 454.3 on 02/12 and 03/12 (see Figures \ref{distr1}, \ref{distr2} and \ref{distr4} in Appendix \ref{appA}).
It can be found from Table \ref{tab1} that the marginalized 95\% CIs of $R_{\rm loc}$ in logarithmic space
are [17.88-18.12] cm and [17.95-18.52] cm, respectively.
It agrees with the result derived by \cite{Nalewajko2014} who used an independent method to constrain $\gamma$-ray emission site.
For the other states of 3C 454.3 and 3C 279, a meaningful constraint on $R_{\rm loc}$ can not be obtained,
and the 95\% lower limits are reported in Table \ref{tab1}.
Note that an upper limit, $R_{loc}\leq R_{DT}=2.5\times10^{18}L_{d,45}^{1/2}$, can be imposed by the adopted DT geometry.
Using the vales of $L_d$ obtained from observations,  we have $R_{DT}\simeq1.36\times10^{19}$ cm for 3C 454.3.

 \subsubsection{Magnetic field strength and Doppler factor}

The magnetic field strength $B'$ and the Doppler factor $\delta_{\rm D}$ are well constrained
 (see Figure \ref{distr1}-\ref{distr4} in Appendix \ref{appA}).

 The value of $B'$ is $\sim1$ G in all states for 3C 454.3 and 3C 279,
 and it varies in the range of [0.03, 0.1] G for Mrk 421.
 This is in agreement with that deduced from the modeling of SEDs presented
 in previous studies \citep[e.g.,][]{Yan2013,Bottcher2019,Hu2020}.
 On the other hand, the values of $B'$ are more or less similar to that in the
 pc-scale jet as estimated from the core-shift measurement \citep{Pushkarev2012,Kutkin2014,Mohan2015}.
This implies that the $\gamma$-ray emission regions may be located at pc-scale.

$\delta_D$ is found to be larger than 30.
The high $\delta_D$-values are roughly consistent with that estimated from the radio variability time-scales  \citep{Hovatta2009}. 
Moreover, it is also supported by the studies of the kinematics of the jet of 3C 279  \citep{Lister1997,Jorstad2004} and 3C 454.3 \citep{Lister2009,Jorstad2010,Jorstad2013}.
For Mrk 421, the high $\delta_D$-values are consistent with the results reported in \cite{Hervet2019}.
Note that $\delta_D$ for Mrk 421 is generally larger than that for 3C 454.3 and 3C 279.

\begin{figure}
 \includegraphics[width=0.45\textwidth]{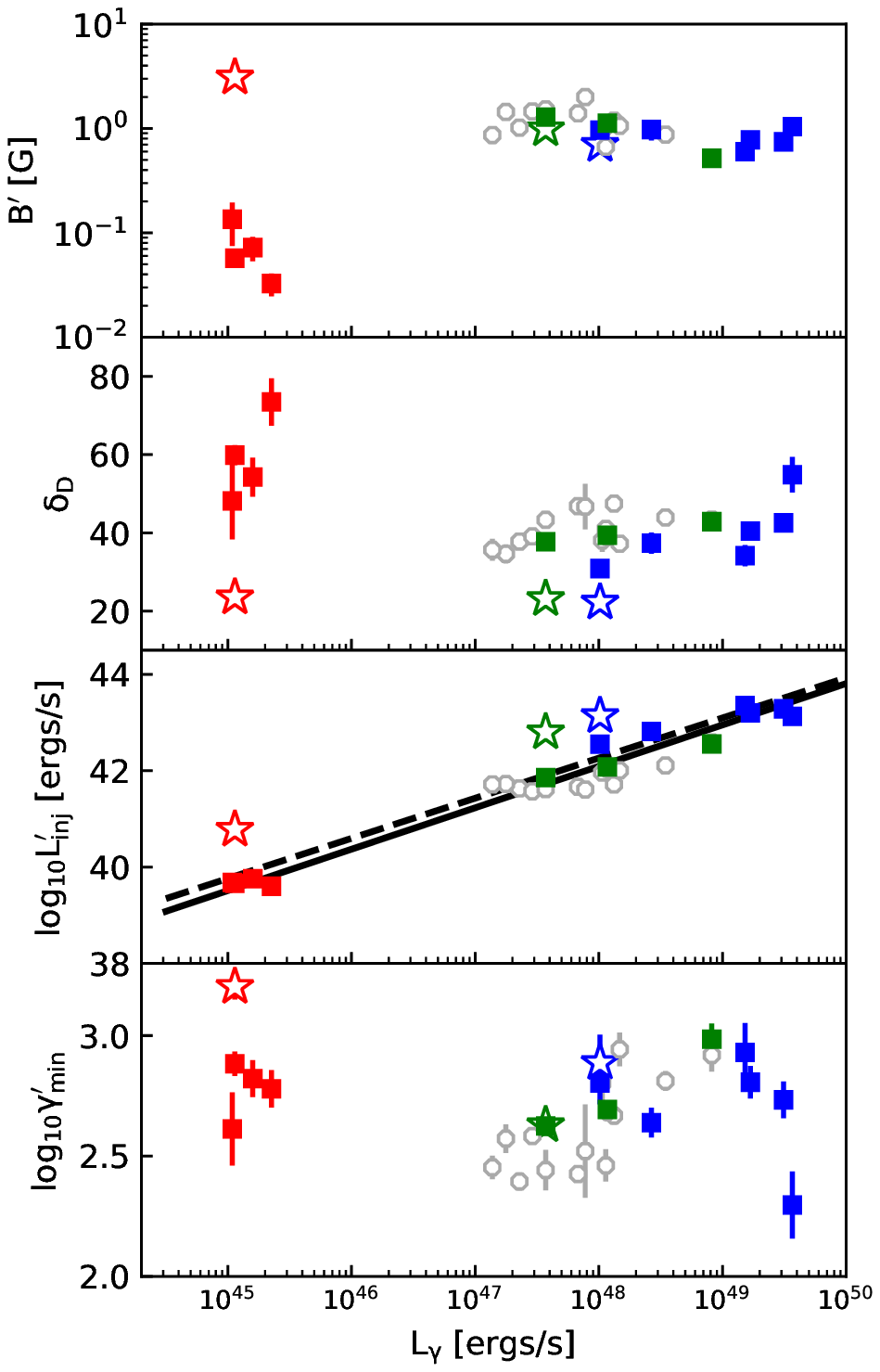}
\caption{Evolutions of the model parameters as a function of the observed $\gamma$-rays luminosity.
The red, blue and green squares are the results of Mrk 421, 3C 454.3 and 3C 279, respectively. 
The open stars denote the results obtained with $t_{\rm var}=1$ day for the quiescent states of the three sources.
For 3C 279, the model parameters reported in Hu et al. (2020) are also shown in the each panel, and are denoted by the gray open circles.
In the $L_\gamma-L_{\rm inj}'$ plot, the solid and dashed lines denote the best linear fits to the results with a short and a long variability timescale for the quiescent states, respectively. 
 \label{figure4}}
\end{figure}

 \subsubsection{Injection electron spectrum}

 We find that the parameters of the injection electron spectrum, i.e., $L_{\rm inj}', \gamma_{\rm min}'$ and $n$, 
 are constrained very well for all SEDs (see Figure \ref{distr1}-\ref{distr4} in Appendix \ref{appA}).
From Table \ref{tab1}, it can be seen that the power-law indexes of injection electron spectrum are restricted to relatively small ranges.
It is $n\simeq2.0-2.3$ for Mrk 421, $n\simeq3.0-3.3$ for 3C 454.3, and $3.1-3.7$ for 3C 279.

For the FSRQs, $L_{\rm inj}'$ varies from $\sim7\times10^{41}$ to $2\times10^{43}$ ergs/s, and it is $\sim5\times10^{39}$ ergs/s for Mrk 421.
In $L_\gamma-L_{\rm inj}'$ plot, we find that the powers of injected electrons $L_{\rm inj}'$ increase with $L_\gamma$ in 3C 454.3 and 3C 279.
For the three sources, the best linear fit in log scale gives $L_{\rm inj}'\propto L_\gamma^{0.86\pm0.03}$ 
with a Pearson correlation coefficient of $r=0.99$ and a chance probability of $p=4.66\times10^{-11}$.

Our results show that $\gamma_{\rm min}'$ is in the range of $[340,740]$ for Mrk 421,
and it is in the range of $[400,10^3]$ for the two FSRQs.
Moreover, we note that $\gamma_{\rm min}'$ increases with $L_\gamma$ in 3C 279 \citep{Hu2020}, 
and there are no such a trend in 3C 454.3 and Mrk 421.
Notice that for the two FSRQs $\gamma_{\rm min}'$ is in the fast-cooling regime, while for Mrk 421 it is in the slow-cooling regime.

\subsubsection{Jet powers}\label{jetpower}

We evaluate the powers of the relativistic jet from our spectral fits.
To consider the uncertainties on the transport parameters, we obtain the values of the derived parameters by using the MCMC code adopted.
In the right panels of Figures \ref{distr1}-\ref{distr4} in Appendix \ref{appA}, we display the corner plots of the derived parameters,
and the mean values and $1\sigma$ uncertainties are listed in Table \ref{tab2}.

Assuming that there is one proton per radiating electron, the jet powers carried by the magnetic field $P_B$, relativistic electrons $P_e$ and protons $P_p$,
as well as radiation $P_r$, are evaluated through the method implemented by \cite{Celotti1993}.

The jet powers of FSRQs 3C 454.3 and 3C 279 are significantly larger than that of HBL Mrk 421.
Further, one can find $P_{p}\simeq P_e> P_B > P_r$ for Mrk 421, and $P_p>P_e> P_r \gtrsim P_B$ for 3C 454.3 and 3C 279.
It generally agrees with previous works for BL Lacs \citep[e.g.,][]{Zhang2012,Yan2014} and FSRQs \citep[e.g.,][]{Celotti2008,Ghisellini2010,Ghisellini2015}.

\subsubsection{Equipartition and magnetization}

Using our modeling results, we can obtain the equipartition parameter $U_B'/U_e'$,
where $U_B'$ and $U_e'$ are respectively the magnetic field and relativistic electrons energy densities in the rest frame of the jet.
In addition, we calculate magnetization parameter $\sigma_m\equiv(\upsilon_A/c)^2$
\footnote{Notice that the definition of $\sigma_m$ can be related to the non-relativistic magnetization defined by \cite{Baring2017}, through the relation $\sigma_m=\sigma\Gamma_{sh}/(m_p/m_e)$. On the other hand, the definition of $\sigma_m$ may be equivalent to the definition of $\sigma\equiv P_B/P_m$, where $P_m=P_e+P_p$ is the kinetic power of matter. For Mrk 421, one can obtain $\sigma_m\simeq P_B/P_p\simeq P_B/P_e$, since $P_e\sim P_p$. For 3C 454.3 and 3C 279,  one can obtain $\sigma_m\simeq 2P_B/P_p$, since $P_p\gg P_e$.} \citep{Cerutti2012,Sironi2013,Sironi2014},
where $\upsilon_A$ is equal to $B'/\sqrt{4\pi n_pm_p}$, with $n_p$ and $m_p$ denoting the thermal proton number density and rest mass, respectively.

\begin{figure}
 \includegraphics[width=0.45\textwidth]{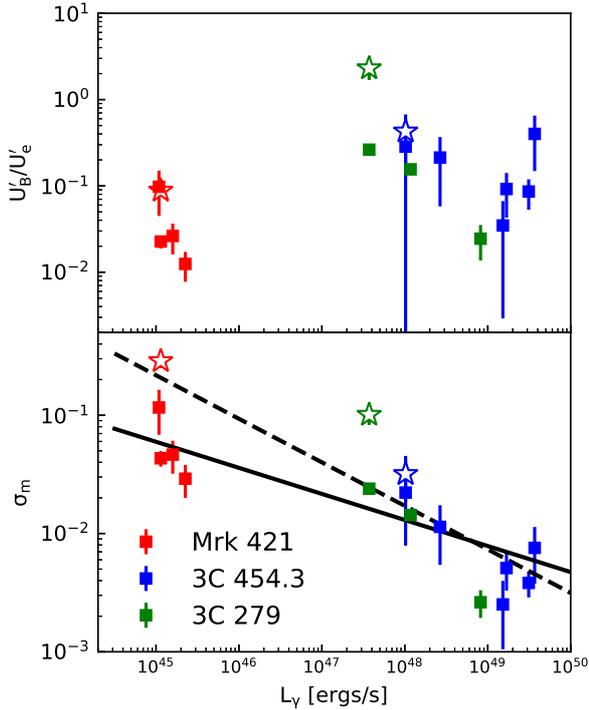}\\
   \caption{Evolutions of the equipartition (upper) and magnetization (lower) parameters as a function of the observed $\gamma$-rays luminosity.
The open stars denote the results derived with $t_{var}=1$ day for the quiescent states of the three sources.
 In the $L_\gamma-\sigma_m$ plot, the solid and dashed lines denote the best linear fits to the results with a short and a long variability timescale for the quiescent states, respectively.
   \label{figure5}}
\end{figure}

The variations of the equipartition ($U_B'/U_e'$) and magnetization ($\sigma_m$) with activities are shown in Figure \ref{figure5}
with the values summarized in Table \ref{tab2}.
$U_B'/U_e'$ ranges from 0.01 to 0.3, deviating from equipartition.
No difference is found between the values for HBL and FSRQ.
$U_B'/U_e'$ is not correlated with $L_{\gamma}$.

With a large sample of $\gamma$-ray blazars,
\citet{Chen2018} found that the equipartition parameter of blazars have a large width in its distribution,
and BL Lacs have much smaller $U_B'/U_e'$, comparing with FSRQs.

$\sigma_m$ is in the range of 0.002 and 0.1.
Mrk 421 has larger $\sigma_m$ than that of 3C 454.3 and 3C 279.
$\sigma_m$ varies from 0.03 to 0.2 for Mrk 421, which is in good agreement with the results of \cite{Lewis2016}
obtained by comparing the theoretical model with the
X-ray time lags during the 1998 April 21 flare.
In particular, using a relativistic oblique shock acceleration + radiation-transfer model,
\cite{Bottcher2019} successfully explained the SEDs and variabilities of 3C 279
during the flaring activity considered in the work.
They obtained the non-relativistic magnetization $\sigma=3.42$ in period A and 1.84 in period C.
Using the relation $\sigma_m=\sigma\Gamma_{\rm sh}/(m_p/m_e)$ and $\Gamma_{\rm sh}=\delta_{\rm D}$,
we obtain $\sigma_m=6.9\times10^{-2}$ and $3.9\times10^{-2}$, respectively.
They are also in good agreement with our results.

There is a good correlation between $\sigma_m$ and $L_\gamma$.
It yields $\sigma_m\propto L_\gamma^{-0.22\pm0.04}$ with a Pearson correlation coefficient of $r=-0.86$ and a chance of probability of $p=1.61\times10^{-4}$.

\section{Discussion and Conclusions}\label{sec:summ}

Using a single-zone leptonic model and the MCMC fitting technique,
we model the high-quality and simultaneous MWL SEDs at multiple epochs of three typical blazars
(3C 454.3, 3C 279 and Mrk 421).
For 3C 454.3, the SEDs in a low $\gamma$-ray state and five highest $\gamma$-ray flaring states \citep{Bonnoli2011} are considered.
For Mrk 421, we consider the SED in a $\gamma$-ray quiescent state \citep{Abdo2011a} and the
SEDs in three highest $\gamma$-ray flaring states reported in \citet{Aleksic2015}.
For 3C 279, we consider three SEDs in \cite{Hayashida2015}, which are the representatives for the 14 SEDs studied in \cite{Hu2020}.
The jet model used here is a refined one of \cite{Hu2020}.

 We find that the model can explain the SEDs well.
The properties of the emission regions obtained by using our model are consistent with previous studies.
Mrk 421 has smaller magnetic field strength and larger Doppler factor than 3C 454.3 and 3C 279.
The electron injection spectrum of 3C 279 and 3C 454.3 is steeper than that of Mrk 421.
Mrk 421 has smaller electron injection power and larger magnetization than 3C 454.3 and 3C 279.
The three jets are found to be low-magnetization and non-equipartition,
suggesting that electrons may be accelerated by shock in the jet.

The new findings are the tight correlations between $L_{\gamma}$ and $L'_{\rm inj}$, $\sigma_m$.
The correlation between $L_{\gamma}$ and $L'_{\rm inj}$ has been found in modeling the 14 SEDs of 3C 279 \citep{Hu2020}.
Here, after including the HBL Mrk 421 and 3C 454.3, this relation still holds.
 $L'_{\rm inj}$ can be associated with acceleration in the jet.
 The correlation between $L_{\gamma}$ and $L'_{\rm inj}$ suggests that
  particle acceleration process play an important role in driving $\gamma$-ray flares of blazars.

 As mentioned above, the low magnetization ($\sigma_m\le0.1$) indicates that shocks may be responsible for the acceleration of the electrons in the jets \citep[e.g.,][]{Sironi2015}.
 The anti-correlation between $L_{\gamma}$ and $\sigma_m$ is firstly found in modeling blazars SEDs.
 This suggests that the same energy-dissipative mechanism behind the jet emission could be at work in different types of blazars.

In the above modelings, we used the minimum variability timescale measured from the $\gamma$-ray flares for each source.
The same variability timescale is adopted for the SED in the quiescent state.
This implies that the gamma-rays in quiescent states and flaring states are produced in the same region. 
Alternatively, gamma-rays in different states may be produced in different regions \citep[e.g.,][]{2021MNRAS.500.5297A}.
In this case, a longer variability timescale should be used for the quiescent state.
In order to investigate this scenario,
we also performed the fitting to the SED at quiescent state for each source using $t_{\rm var}=1$ day. 
The modeling results are shown in Figure \ref{figure6}, and the corresponding corner plots of the input and output parameters are shown in Figure \ref{distr5}
with the parameters values reported in Tables \ref{tab1} and \ref{tab2}. 
We find that the increasing of $t_{\rm var}$ could primarily lead to decrease in $\delta_{\rm D}$, and increase in $L_{\rm inj}'$.
$B',~\gamma_{\rm min}',~n$ as well as $R_{\rm loc}$ are found to change slightly.
For comparison, the parameters values are plotted in the Figures \ref{figure4} and \ref{figure5} (open stars). 
Interestingly, it is found that the correlation between $L_{\rm inj}'$ and $L_\gamma$ is still significant, with $r=0.91$ and $p=1.57\times10^{-5}$.
In the $\sigma_{\rm m}-L_\gamma$ panel of Figure \ref{figure5}, the Pearson test also gives a significant correlation with $r=-0.87$ and $p=1.05\times10^{-4}$.
The results may indicate that the correlations are independent of the location of the gamma-ray emission region.

The variability may be caused by changes in physical condition of the jet, e.g., injection, magnetic field, Doppler factor \citep{Diltz2014}, 
with possible intervention of shock waves or turbulence \citep[e.g.,][]{Sikora2001,Marscher2014}.
In the frame of the one-zone leptonic model, our results indicate that the electron injection may be the main driver of the variability.
In other words, the acceleration of electrons causes the $\gamma$-ray variability.

In addition to the intrinsic origin of the variability,
it could also be attributed to a geometrical effect of changing in the viewing angle \citep{Raiteri2009,Raiteri2017,Villata2007,Liodakis2020b}.
It is interesting to make a distinction between two explanations of variability.
However, it is difficult to do such a distinction based on the current observations.

It should be pointed out that the collimation parameter of the emitting region $\Gamma\theta_{\rm obs}=1$ is fixed in the fitting.
This assumption is frequently adopted in blazar SED modeling \citep[e.g.,][]{Zhang2012,Bottcher2013,Yan2014,Yan2015,Nalewajko2014}.
 However, the VLBI observations indicate that $\Gamma\theta_{\rm obs}\ll1$ \citep[e.g.,][]{Jorstad2017} for some blazars.
The assumption should not affect significantly the spectral fits to the observed SEDs, 
and has little impact on our main conclusions.
This is mainly due to that the emitting EEDs and the resulting SEDs are directly related to $\delta_{\rm D}$, 
which can keep constant by decreasing $\Gamma$ and $\theta_{\rm obs}$
\footnote{In the limit $\Gamma\gg1$, $\theta_{\rm obs}\ll1$, the Doppler factor $\delta_{\rm D}$ can be related to $\Gamma$ 
through the relationship $\delta_{\rm D}=2\Gamma/(1+\Gamma^2\theta_{obs}^2)$ \citep{Dermer2014}.}.
However, we would like to stress that the decreasing of $\Gamma\theta_{\rm obs}$ will reduce the intrinsic jet power.
Therefore, it should be considered with caution when the jet powers from the spectral fitting are compared to 
the powers predicted by the Blanford-Payne \citep[BP;][]{Blandford1982} 
or Blanford-Znajek \citep[BZ;][]{Blandford1982} mechanisms.

From Table \ref{tab2}, one can find that the averaged jet powers estimated from the SED fittings 
are $4.14\times10^{47},~3.74\times10^{46}$  and $1.48\times10^{44}$ erg/s for 3C 454.3, 3C 279 and  Mrk 421, respectively.
On the other hand, the jet powers can also be estimated by modeling the observed SEDs with hadronic models \citep[e.g.,][]{Bottcher2013,Petropoulou2016b,Barkov2010}.
In particular, the hadronic models received a wide attention since the detection of coincident neutrinos and $\gamma$-rays from blazar TXS 0506+056 \citep{Aartsen2018a,Aartsen2018b}.
 The required power in relativistic protons $L_p$ is
of the order of magnitude of $10^{49}$ erg/s for 3C 454.3 and $10^{48}$ erg/s for 3C 279 \citep{Bottcher2013},
while for Mrk 421 $L_p$ ranges from $2.3\times10^{47}$ to $7.8\times10^{47}$ erg/s \citep{Petropoulou2016a}.
It can be seen that the jet powers estimated from hadronic models are much greater than that estimated from leptonic model.

We can compare the jet power against the power of the BZ process \citep{Blandford1977}, which is believed to be the plausible explanation for the jet launch.
Generally, the predicted jet power can be written as $P_{BZ}=\eta_jP_{acc}$, where $\eta_j$ is the jet production efficiency, 
and  the accretion power is $P_{acc}=\dot{M}c^2$ with $\dot{M}$ denoting the mass accretion rate. As it has been theoretically estimated and numerically confirmed,
$\eta_j$ in a magnetically choked accretion flow scenario can exceed unity \citep{Tchekhovskoy2010,McKinney2012}, and reaches $\sim1.9$ for maximal BH spins \citep{Sikora2013}. 
In FSRQs, the accretion powers can be calculated as $P_{acc}=L_d/\eta_d$, where $\eta_d$ is the accretion disk radiative efficiency.
With assumption of $\eta_d=0.1$ and $0.3$ \citep{Ghisellini2014}, we find that $P_{acc}\simeq(1-3)\times10^{47}$ erg/s for 3C 454.3, and $P_{acc}\simeq(0.67-2)\times10^{46}$ erg/s for 3C 279.
Compared with FSRQs, BL Lacs are believed to have low accretion rates \citep[e.g.,][]{Wang2002,Xu2009}.
For Mrk 421, we estimate the accretion power through the relation
$P_{acc}=\dot{m}L_{Edd}\simeq1.26\times10^{46}\dot{m}M_8$ ergs/s,
where $L_{Edd}$ is the Eddington luminosity, $\dot{m}$ is the mass accretion rate in units of $\dot{M}_{Edd}=L_{Edd}/c^2$, and $M_8$ is the BH mass in units of $10^8M_\odot$ ($M_\odot$ denotes the solar mass.).  
Assuming $\dot{m}\sim(3-10)\times10^{-3}$ \citep{Ghisellini2008,Meyer2011}, we obtain $P_{acc}\simeq(0.72-2.4)\times10^{44}$ erg/s for Mrk 421, 
when we take $\log M_{\rm BH}/M_\odot=8.28$ derived from the measurement of stellar velocity dispersion \citep{Woo2002}.
Therefore, it seems that the relativistic jets in the three sources may be governed by the BZ process.

Our results show that the energy density ratio of magnetic field and radiating electrons $U_B'/U_e'$ varies from $\sim0.01$ to $\sim0.5$.
This is consistent with results from modeling the SEDs of the large sample blazars \citep[e.g.,][]{Celotti2008,Ghisellini2014,Chen2018} 
or individual sources \citep[e.g.,][]{Yan2013,Dermer2014,Hu2015}.
Interestingly, we note that in the quiescent states of the three sources $U_B'/U_e'$ is closer to equipartition compared to the flaring states.
In fact, the results may be supported by the radio observations.
The studies of the core-shift-effect have shown that the distance of the core from the jet base $r_c$, 
the core size $W$ and the light-curve time lag $\Delta t$ all depend on the observation frequency $\nu$ as $r_c\propto W\propto\Delta t\propto\nu^{-1/k}$ with $k\simeq1$
\citep[e.g.,][]{Sokolovsky2011,Pushkarev2012,Zamaninasab2014,Mohan2015,Agarwal2017}. 
This is in agreement with prediction of a synchrotron self-absorbed conical jet model of \cite{Blandford1979} and \cite{Konigl1981} in the case of equipartition.
\cite{Plavin2019} argued that any fixed frequency dependence such as $r_c\propto\nu^{-1}$ is disrupted during flares.
The authors concluded that the observed flux density variability and the variations of the core position in a flaring jet 
are mainly caused by significant increase in emitting electron density and slight decrease in the magnetic field. 
This indicates that the electron energy density dominates over magnetic field energy density during a flaring activity. 
Moreover, the extreme brightness temperatures $\ge10^{13}$ K observed by RadioAstron also support that equipartition may be violated during flares \citep[e.g.,][]{Gomez2016,Bruni2017,Pilipenko2018,Kutkin2018}.

\section*{Acknowledgements}
We thank the reviewer for constructive suggestions and comments. We acknowledge the National Natural Science Foundation of China (NSFC-11803081, NSFC-12065011,NSFC-U1831124) and
the joint foundation of Department of Science and Technology of Yunnan Province and Yunnan University [2018FY001(-003)].
The work of D. H. Yan is also supported by the CAS Youth Innovation Promotion Association and Basic research Program of Yunnan Province (202001AW070013).

\section*{data availability}
No new data were generated or analysed in support of this research.



\begin{thebibliography}{99}
\bibitem[\protect\citeauthoryear{Abdo et al.}{2009}]{Abdo2009}Abdo A. A. et al., 2009, ApJ, 699, 817
\bibitem[\protect\citeauthoryear{Abdo et al.}{2011}]{Abdo2011}Abdo A. A. et al., 2011, ApJ, 733, L26
\bibitem[\protect\citeauthoryear{Abdo et al.}{2011}]{Abdo2011a} Abdo A. A., Ackermann M., Ajello M. et al.,  2011a, ApJ, 736, 131
\bibitem[\protect\citeauthoryear{Abeysekara et al.}{2018}]{Abeysekara2018} Abeysekara A. U., Benbow W., Bird R. et al., 2018, ApJ, 856, 95
\bibitem[\protect\citeauthoryear{Acharyya, Chadwick, \& Brown}{2021}]{2021MNRAS.500.5297A} Acharyya A., Chadwick P.~M., Brown A.~M., 2021, MNRAS, 500, 5297
\bibitem[\protect\citeauthoryear{Aleksi\'c et al.}{2015}]{Aleksic2015}Aleksi\'c J. et al., 2015, A\&A 578, A22   
\bibitem[\protect\citeauthoryear{Aartsen et al.}{2018a}]{Aartsen2018a}Aartsen M. G., Ackermann M., Adams, J. et al., 2018a, Sci, 361, eaat1378
\bibitem[\protect\citeauthoryear{Aartsen et al.}{2018b}]{Aartsen2018b}Aartsen M. G., Ackermann M., Adams, J. et al., 2018b, Sci, 361, 147
\bibitem[\protect\citeauthoryear{Agarwal et al.}{2017}]{Agarwal2017}Agarwal A., Mohan P., Gupta A. C. et al., 2017, MNRAS, 469, 813
\bibitem[\protect\citeauthoryear{Baring et al.}{2017}]{Baring2017}Baring M. G., B\"{o}ttcher M., Summerlin E. J., 2017, MNRAS, 464, 4875
\bibitem[\protect\citeauthoryear{Barkov, Aharonian \& Bosch-Ramon}{2010}]{Barkov2010}Barkov M. V., Aharonian F. A., Bosch-Ramon V., 2010, ApJ, 724, 1517
\bibitem[\protect\citeauthoryear{Bennett et al.}{2013}]{Bennett2013}Bennett C. L., Larson D., Weiland J. L. et al., 2013, ApJS, 208, 20
\bibitem[\protect\citeauthoryear{Blandford \& K\"{o}nigl}{1979}]{Blandford1979}Blandford R. D., K\''{o}nigl A., 1979, ApJ, 232, 34
\bibitem[\protect\citeauthoryear{Blandford \& Payne}{1982}]{Blandford1982}Blandford R. D., Payne D. G., 1982, MNRAS, 199, 883
\bibitem[\protect\citeauthoryear{Blandford \& Znajek}{1977}]{Blandford1977}Blandford R. D., Znajek R. L., 1977, MNRAS, 179, 433
\bibitem[\protect\citeauthoryear{Blazejowski et al.}{2000}]{Blazejowski2000}Blazejowski M., Sikora M. et al., 2000, ApJ, 545, 107
\bibitem[\protect\citeauthoryear{Bloom \& Marscher}{1996}]{Bloom1996}Bloom S. D., Marscher A. P., 1996, ApJ, 461, 657
\bibitem[\protect\citeauthoryear{Blumenthal \& Gould}{1970}]{Blumenthal1970} Blumenthal G. R., Gould R. J., 1970, RvMP, 42, 237
\bibitem[\protect\citeauthoryear{Bonnoli et al.}{2011}]{Bonnoli2011}Bonnoli G., Ghisellini G., Foschini L. et al., 2011, MNRAS, 411, 368
\bibitem[\protect\citeauthoryear{B\"{o}ttcher \& Chiang}{2002}]{Bottcher2002}B\"{o}ttcher M., Chiang J., 2002, ApJ, 581, 127
\bibitem[\protect\citeauthoryear{B\"{o}ttcher et al.}{2013}]{Bottcher2013}B\"{o}ttcher M., Reimer A., Sweeney K., Prakash A., 2013, ApJ, 768, 54
\bibitem[\protect\citeauthoryear{B\"{o}ttcher \& Baring}{2019}]{Bottcher2019}B\"{o}ttcher M., Baring M. G., 2019, ApJ, 887, 133
\bibitem[\protect\citeauthoryear{Bruni et al.}{2017}]{Bruni2017}Bruni G., G\'{o}mez J. L., Casadio C. et al., 2017, A\&A, 604A,111B
\bibitem[\protect\citeauthoryear{Celotti \& Fabian}{1993}]{Celotti1993}Celotti A., Fabian A. C., 1993, MNRAS, 264, 228
\bibitem[\protect\citeauthoryear{Celotti \& Ghisellini}{2008}]{Celotti2008}Celotti A., Ghisellini G., 2008, MNRAS, 385, 283
\bibitem[\protect\citeauthoryear{Cerutti et al.}{2012}]{Cerutti2012}Cerutti B., Uzdensky D. A., Begelman M.C., 2012, ApJ, 746,148
\bibitem[\protect\citeauthoryear{Chen}{2014}]{Chen2014} Chen L., 2014, ApJ, 788, 179
\bibitem[\protect\citeauthoryear{Chen}{2018}]{Chen2018} Chen L., 2018, ApJ, 235, 39
\bibitem[\protect\citeauthoryear{Chiaberge \& Ghisellini}{1999}]{Chiaberge1999}Chiaberge M., Ghisellini G., 1999, MNRAS, 306, 551
\bibitem[\protect\citeauthoryear{Christie et al.}{2019}]{Christie2019}Christie I. M., Petropoulou M., Sironi L., Giannios D., 2019, MNRAS, 482, 65
\bibitem[\protect\citeauthoryear{Coppi et al.}{1990}]{Coppi1990}Coppi P. S., Blandford R. D., 1990, MNRAS, 245, 453
\bibitem[\protect\citeauthoryear{Crusius \& Schlickeiser}{1986}]{crusius1986}Crusius A., Schlickeiser R., 1986, A\&A, 164, L16
\bibitem[\protect\citeauthoryear{D'arcangelo et al.}{2009}]{Darcangelo2009}D'arcangelo, F. D., Marscher, A. P., Jorstad, S. G., et al. 2009, ApJ, 697, 985
\bibitem[\protect\citeauthoryear{Dermer \& Schlickeiser}{2002}]{Dermer2002}Dermer C. D., Schlickeiser R., 2002, ApJ, 757, 667
\bibitem[\protect\citeauthoryear{Dermer \& Schlickeiser}{1993}]{Dermer1993}Dermer C. D., Schlickeiser R., 1993, ApJ, 416, 458D
\bibitem[\protect\citeauthoryear{Dermer et al.}{2009}]{Dermer2009}Dermer C. D., Fink J. D., Krug H., B\"{o}ttcher M., 2009, ApJ, 692, 32
\bibitem[\protect\citeauthoryear{Dermer et al.}{2014}]{Dermer2014}Dermer C. D., Cerruti M., Lott B. et al., 2014, ApJ, 782, 82
\bibitem[\protect\citeauthoryear{Dermer et al.}{2015}]{Dermer2015}Dermer C. D., Yan D. H., Zhang L. et al., 2015, ApJ, 809, 174
\bibitem[\protect\citeauthoryear{Diltz \& B\"{o}ttcher}{2014}]{Diltz2014}Diltz C.,  B\"{o}ttcher M., 2014, JHEAp, 1, 63D
\bibitem[\protect\citeauthoryear{Finke et al.}{2008}]{Finke2008}Finke J. D., Dermer C. D., B\"{o}ttcher M., 2008, ApJ, 686, 181
\bibitem[\protect\citeauthoryear{Finke}{2016}]{Finke2016}Finke J. D., 2016, ApJ, 830, 94
\bibitem[\protect\citeauthoryear{Ghisellini \& Tavecchio}{2008}]{Ghisellini2008}Ghisellini G., Tavecchio, F. 2008, MNRAS, 387, 1669
\bibitem[\protect\citeauthoryear{Ghisellini \& Tavecchio}{2009}]{Ghisellini2009}Ghisellini G., Tavecchio F., 2009, MNRAS, 397, 985
\bibitem[\protect\citeauthoryear{Ghisellini et al.}{2010}]{Ghisellini2010}Ghisellini G., Tavecchio F., Foschini L. et al., 2010, MNRAS, 402, 497
\bibitem[\protect\citeauthoryear{Ghisellini et al.}{2014}]{Ghisellini2014}Ghisellini, G., Tavecchio, F., Maraschi, L., et al. 2014, Natur, 515, 376
\bibitem[\protect\citeauthoryear{Ghisellini \& Tavecchio}{2015}]{Ghisellini2015}Ghisellini G., Tavecchio F., 2015, MNRAS, 448, 1060
\bibitem[\protect\citeauthoryear{Giannios et al.}{2009}]{Giannios2009a}Giannios D., Uzdensky D. A., Begelman M. C., 2009, MNRAS, 395, L29
\bibitem[\protect\citeauthoryear{Giannios}{2013}]{Giannios2013}Giannios D., 2013, MNRAS, 431, 355
\bibitem[\protect\citeauthoryear{G\'{o}mez et al.}{2016}]{Gomez2016}G\'{o}mez J. L., Lobanov, A. P., Bruni, G. et al., 2016, ApJ, 817, 96G
\bibitem[\protect\citeauthoryear{Graff et al.}{2008}]{Graff2008}Graff P. B., Georganopoulos M. et al.,  2008, ApJ, 689, 68G
\bibitem[\protect\citeauthoryear{Gu et al.}{2001}]{Gu2001}Gu M., Cao X., Jiang D. R., 2001, MNRAS, 327, 1111
\bibitem[\protect\citeauthoryear{Guo et al.}{2014}]{Guo2014}Guo F., Li H., Daughton W., Liu Y.-H., 2014, PhRvL, 113, 155005
\bibitem[\protect\citeauthoryear{Guo et al.}{2015}]{Guo2015}Guo F., Liu Y.-H., Daughton W., Li H., 2015, ApJ, 806, 167
\bibitem[\protect\citeauthoryear{Hayashida et al.}{2015}]{Hayashida2015}Hayashida M., Nalewajko K., Madejski G. M. et al., 2015, ApJ, 807, 79
\bibitem[\protect\citeauthoryear{Hervet et al.}{2019}]{Hervet2019}Hervet O., Williams D. A., Falcone A. D. et al., 2019, ApJ, 877, 26H
\bibitem[\protect\citeauthoryear{Hinshaw et al.}{2013}]{Hinshaw2013}Hinshaw G., Larson D., Komatsu E. et al., 2013, ApJS, 208, 19
\bibitem[\protect\citeauthoryear{Homan et al.}{2009}]{Homan2009}Homan D. C., Kadler M., Kellermann K. I. et al., 2009, ApJ, 706, 1253
\bibitem[\protect\citeauthoryear{Hovatta et al.}{2009}]{Hovatta2009} Hovatta T., Valtaoja E., Tornikoski M., L\''{a}hteenm\''{a}ki A., 2009, A\&A, 494, 527
\bibitem[\protect\citeauthoryear{Hu et al.}{2015}]{Hu2015}Hu W., Fan Z.-H., Dai B.-Z., 2015, RAA, 15, 1455
\bibitem[\protect\citeauthoryear{Hu et al.}{2017}]{Hu2017}Hu W., Dai B.-Z, Zeng W., Fan Z.-H, Zhang L., 2017a, NewA, 52, 82H
\bibitem[\protect\citeauthoryear{Hu et al.}{2020}]{Hu2020}Hu W., Yan D. H., Dai B.-Z, Zeng W., Hu Q. L., 2020, MNRAS, 493, 410
\bibitem[\protect\citeauthoryear{Inoue \& Tanaka}{2016}]{Inoue2016}Inoue Y., Tanaka Y. T., 2016, ApJ, 828, 13
\bibitem[\protect\citeauthoryear{Jackson \& Browne}{1991}]{Jackson1991}Jackson N., Browne W. A., 1991, MNRAS, 250, 414
\bibitem[\protect\citeauthoryear{Jim\'{e}nez-Fern\'{a}ndez \& van Eerten}{2021}]{Jimenez2021}Jim\'{e}nez-Fern\'{a}ndez B., van Eerten H. J., 2021, MNRAS, 500, 3613J
\bibitem[\protect\citeauthoryear{Jones}{1968}]{Jones1968}Jones F. C.,  1968, PhRv, 167, 1159J
\bibitem[\protect\citeauthoryear{Jorstad et al.}{2004}]{Jorstad2004} Jorstad S. G., Marscher A. P., Lister M. L. et al., 2004, AJ, 127, 3115
\bibitem[\protect\citeauthoryear{Jorstad et al.}{2005}]{Jorstad2005} Jorstad S. G., Marscher A. P., Lister M. L. et al., 2005, ApJ, 130, 1418
\bibitem[\protect\citeauthoryear{Jorstad et al.}{2010}]{Jorstad2010} Jorstad S. G., Marscher A. P., Larionov V. M. et al., 2010, ApJ, 715, 362
\bibitem[\protect\citeauthoryear{Jorstad et al.}{2013}]{Jorstad2013} Jorstad S. G., Marscher A. P., Smith P. S. et al., 2013, ApJ, 773, 147
\bibitem[\protect\citeauthoryear{Jorstad et al.}{2017}]{Jorstad2017} Jorstad S. G., Marscher A. P., Morozova D. A. et al., 2017, ApJ, 846, 98J
\bibitem[\protect\citeauthoryear{Kirk et al.}{2000}]{Kirk2000}Kirk J. G., Guthmann A. W., Gallant Y. A., Achterberg A., 2000, ApJ, 542, 235
\bibitem[\protect\citeauthoryear{Komissarov et al.}{2007}]{Komissarov2007}Komissarov S. S., Barkov M. V., Vlahakis N., K\''{o}nigl A., 2007, MNRAS, 380, 51
\bibitem[\protect\citeauthoryear{K\"onigl}{1981}]{Konigl1981}K\"onigl A., 1981, ApJ, 243, 700
\bibitem[\protect\citeauthoryear{Kutkin et al.}{2014}]{Kutkin2014}Kutkin A. M. et al., 2014, MNRAS, 437, 3396
\bibitem[\protect\citeauthoryear{Kutkin et al.}{2018}]{Kutkin2018}Kutkin A. M. et al., 2018, MNRAS, 475, 4994
\bibitem[\protect\citeauthoryear{Lewis \& Bridle}{2002}]{Lewis2002}Lewis A., Bridle S., 2002, Phys. Rev. D, 66, 103511
\bibitem[\protect\citeauthoryear{Lewis et al.}{2016}]{Lewis2016}Lewis T. R.,, Becker P. A., Finke J. D., 2016, ApJ, 824, 108
\bibitem[\protect\citeauthoryear{Lewis et al.}{2018}]{Lewis2018}Lewis T. R., Finke J. D.,  Becker P. A., 2018, ApJ, 853, 6
\bibitem[\protect\citeauthoryear{Liodakis et al.}{2020}]{Liodakis2020b}Liodakis I., Blinov D., Jorstad S. G. et al., 2020, ApJ, 902, 61L
\bibitem[\protect\citeauthoryear{Liodakis \& Petropoulou}{2020}]{Liodakis2020a}Liodakis I., Petropoulou M. et al., 2020, ApJL, 893, L20
\bibitem[\protect\citeauthoryear{Lister \& Marscher}{1997}]{Lister1997}Lister M. L., Marscher A. P., 1997, ApJ, 476, 572
\bibitem[\protect\citeauthoryear{Lister et al.}{2009}]{Lister2009}Lister M. L.,  Cohen M. H., Homan D. C. et al., 2009, ApJ, 138, 1874
\bibitem[\protect\citeauthoryear{Liu et al.}{2012}]{Liu2012}Liu J., Yuan Q., Bi X. J., Li H., Zhang X.M., 2012, Phys. Rev. D, 85, d3507
\bibitem[\protect\citeauthoryear{Marscher \& Gear}{1985}]{Marscher1985}Marscher A. P., Gear W. K., 1985, ApJ, 298, 114
\bibitem[\protect\citeauthoryear{Marscher et al.}{2008}]{Marscher2008}Marscher A. P., Jorstad S. G., D'Arcangelo F. D. et al., 2008, Nature, 452, 966
\bibitem[\protect\citeauthoryear{Marscher  et al.}{2010}]{Marscher2010}Marscher A. P., Jorstad S. G.,  Larionov V. M., et al., 2010, ApJL, 710, 126
\bibitem[\protect\citeauthoryear{Marscher}{2014}]{Marscher2014}Marscher, A. P. 2014, ApJ, 780, 87,
\bibitem[\protect\citeauthoryear{McKinney, Tchekhovskoy \& Blandford}{2012}]{McKinney2012}McKinney J. C., Tchekhovskoy A., Blandford R. D., 2012, MNRAS,423, 3083
\bibitem[\protect\citeauthoryear{Meyer et al.}{2011}]{Meyer2011}Meyer E. T., Fossati G., Georganopoulos M., Lister M. L., 2011, ApJ,740, 98
\bibitem[\protect\citeauthoryear{Mohan et al.}{2015}]{Mohan2015}Mohan P., Agarwal A., Mangalam A. et al., 2015, MNRAS, 452, 2004
\bibitem[\protect\citeauthoryear{Nalewajko et al.}{2014}]{Nalewajko2014}Nalewajko K., Begelman M. C., Sikora M., 2014, ApJ, 789, 161
\bibitem[\protect\citeauthoryear{Nilsson et al.}{2009}]{Nilsson2009}Nilsson K., Pursimo T., Villforth C. et al., 2009, A\&A, 505, 601
\bibitem[\protect\citeauthoryear{Peceur et al.}{2020}]{Peceur2020}Peceur N. M., Tayler A. R., Kraan-Korteweg R. C., 2020, MNRAS, 495, 2162
\bibitem[\protect\citeauthoryear{Petropoulou, Coenders \& Dimitrakoudis}{2016}]{Petropoulou2016a}Petropoulou M., Coenders S., Dimitrakoudis S., 2016, APh, 80, 115P
\bibitem[\protect\citeauthoryear{Petropoulou et al.}{2016}]{Petropoulou2016}Petropoulou M., Giannios D., Sironi L., 2016, MNRAS, 462, 3325
\bibitem[\protect\citeauthoryear{Petropoulou \& Dermer}{2016}]{Petropoulou2016b} Petropoulou M., Dermer C. D., 2016, ApJL, 825, L11
\bibitem[\protect\citeauthoryear{Petropoulou et al.}{2019}]{Petropoulou2019}Petropoulou M., Sironi L., Spitkovsky A., Giannios D., 2019, ApJ, 880, 37
\bibitem[\protect\citeauthoryear{Pian et al.}{1999}]{Pian1999}Pian E., Urry C. M., Maraschi L. et al., 1999, ApJ, 521, 112
\bibitem[\protect\citeauthoryear{Pilipenko et al.}{2018}]{Pilipenko2018}Pilipenko S. V. et al., 2018, MNRAS, 474, 3523
\bibitem[\protect\citeauthoryear{Plavin et al.}{2019}]{Plavin2019}Plavin A. V.,‹ Kovalev Y. Y., Pushkarev A. B., 2019, MNRAS, 485, 1822
\bibitem[\protect\citeauthoryear{Punch et al.}{1992}]{Punch1992} Punch M. et al., 1992, Nature, 358, 477
\bibitem[\protect\citeauthoryear{Poole et al.}{2008}]{Poole2008} Poole T. S., Breeveld A. A., Page M. J. et al., 2008, MNRAS, 383, 627P
\bibitem[\protect\citeauthoryear{Pushkarev et al.}{2012}]{Pushkarev2012}Pushkarev A. B., Hovatta T., Kovalev Y. Y., 2012, A\&A, 545, A113
\bibitem[\protect\citeauthoryear{Raiteri et al.}{2007}]{Raiteri2007}Raiteri C. M., Villata M., Larionov V. M. et al., 2007, A\&A, 473, 819
\bibitem[\protect\citeauthoryear{Raiteri et al.}{2008}]{Raiteri2008}Raiteri C. M. et al., 2008, A\&A, 485, L17  
\bibitem[\protect\citeauthoryear{Raiteri et al.}{2009}]{Raiteri2009}Raiteri C. M., Villata M., Capetti A. et al., 2009, A\&A, 507, 769
 \bibitem[\protect\citeauthoryear{Raiteri et al.}{2017}]{Raiteri2017}Raiteri C. M., Villata M., Acosta-Pulido J. A. et al., 2017, Natur, 552, 374R
\bibitem[\protect\citeauthoryear{Rybicki \& Lightman}{1979}]{Rybicki1979}Rybicki G. B., Lightman A. P., 1979, Radiative Processes in Astrophysics. Wiley, New York
\bibitem[\protect\citeauthoryear{Shah et al.}{2017}]{Shah2017}Shah Z., Sahayanathan S., Mankuzhiyil N. et al., 2017, MNRAS, 470, 3283S
\bibitem[\protect\citeauthoryear{Shakura \& Sunyaev}{1973}]{Shakura1973}Shakura N. I., Sunyaev R. A., 1973, A\&A, 24, 337
\bibitem[\protect\citeauthoryear{Sikora et al.}{1994}]{Sikora1994}Sikora M., Begelman M. C., Rees M. J., 1994, ApJ, 421, 153
\bibitem[\protect\citeauthoryear{Sikora et al.}{2001}]{Sikora2001}Sikora, M., B{\l}a\`zejowski, M. et al., 2001, ApJ, 554, 1
\bibitem[\protect\citeauthoryear{Sikora \& Begelman}{2013}]{Sikora2013}Sikora M., Begelman M. C., 2013, ApJL, 764, L24
\bibitem[\protect\citeauthoryear{Sironi \& Spitkovsky}{2009}]{Sironi2009}Sironi L., Spitkovsky A., 2009, ApJ, 698, 1523
\bibitem[\protect\citeauthoryear{Sironi et al.}{2013}]{Sironi2013}Sironi L., Spitkovsky A., Arons, J., 2013, ApJ, 771, 54
\bibitem[\protect\citeauthoryear{Sironi et al.}{2015}]{Sironi2015}Sironi L., Petropoulou M., Giannios D., 2015, MNRAS, 450, 183
\bibitem[\protect\citeauthoryear{Sironi \& Spitkovsky}{2014}]{Sironi2014}Sironi L.,  Spitkovsky A., 2014, ApJ, 783, L21
\bibitem[\protect\citeauthoryear{Sokolovsky et al.}{2011}]{Sokolovsky2011}Sokolovsky K. V., Kovalev Y. Y., Pushkarev A. B., Lobanov A. P., 2011, A\&A, 532, A38
\bibitem[\protect\citeauthoryear{Spada et al.}{2001}]{Spada2001}Spada M., Ghisellini G., Lazzati D., Celotti A., 2001, MNRAS, 325, 1559
\bibitem[\protect\citeauthoryear{Summerlin \& Baring}{2012}]{Summerlin2012}Summerlin E. J., Baring M. G., 2012, ApJ, 745, 63
\bibitem[\protect\citeauthoryear{Tavecchio \& Ghisellini}{2008}]{Tavecchio2008}Tavecchio F.,  Ghisellini G., 2008, MNRAS, 386, 945T
\bibitem[\protect\citeauthoryear{Tavecchio et al.}{2010}]{Tavecchio2010}Tavecchio F., Ghisellini G., Bonnoli, G. et al., 2010, MNRAS, 405, L94
\bibitem[\protect\citeauthoryear{Tchekhovskoy, McKinney \& Narayan}{2009}]{Tchekhovskoy2009}Tchekhovskoy A., McKinney J. C., Narayan R., 2009, ApJ, 699, 1789
\bibitem[\protect\citeauthoryear{Tchekhovskoy, Narayan \& McKinney}{2010}]{Tchekhovskoy2010}Tchekhovskoy A., Narayan R., McKinney J. C., 2010, ApJ, 711, 50
\bibitem[\protect\citeauthoryear{Urry \& Padovani}{1995}]{Urry1995}Urry C. M., Padovani, P., 1995, PASP, 107, 803
\bibitem[\protect\citeauthoryear{van den Berg et al.}{2019}]{Van2019}van den Berg J. P., B\"{o}ttcher M. et al., 2019, ApJ, 874, 47
\bibitem[\protect\citeauthoryear{Vercellone et al.}{2009}]{Vercellone2009}Vercellone S., Chen A. W., Vittorini V. et al., 2009, ApJ, 690,1018
\bibitem[\protect\citeauthoryear{Vercellone et al.}{2010}]{Vercellone2010}Vercellone S., DAmmando F., Vittorini V. et al., 2010, ApJ,712, 405
\bibitem[\protect\citeauthoryear{Vercellone et al.}{2011}]{Vercellone2011}Vercellone S. et al., 2011, ApJL, 736, L38
\bibitem[\protect\citeauthoryear{Villata et al.}{2006}]{Villata2006}Villata M. et al. 2006, A\&A, 453, 817
\bibitem[\protect\citeauthoryear{Villata et al.}{2007}]{Villata2007}Villata M., Raiteri C. M., Aller M. F. et al., 2007, A\&A, 464, L5
\bibitem[\protect\citeauthoryear{Vlahakis}{2015}]{Vlahakis2015}Vlahakis N., 2015, ASSL, 414, 177
\bibitem[\protect\citeauthoryear{Wang, Staubert \& Ho}{2002}]{Wang2002}Wang J.-M., Staubert R., Ho L. C., 2002, ApJ, 579, 554
\bibitem[\protect\citeauthoryear{Woo \& Urry}{2002}]{Woo2002}Woo J. H., Urry C. M., 2002, ApJ, 579, 530
\bibitem[\protect\citeauthoryear{Wu et al.}{2018}]{Wu2018}Wu Lin-hui, Wu Qingwen, Yan Da-hai, et al., 2018, ApJ, 852, 45
\bibitem[\protect\citeauthoryear{Xu, Cao \& Wu}{2009}]{Xu2009}Xu, Y.-D., Cao, X., Wu, Q., 2009, ApJL, 694, L107
\bibitem[\protect\citeauthoryear{Yan et al.}{2012a}]{Yan2012}Yan D. H., Zeng, H. D., Zhang, L., 2012, PASJ, 64, 80
\bibitem[\protect\citeauthoryear{Yan et al.}{2014}]{Yan2014}Yan D. H., Zeng H. D., Zhang L., 2014, MNRAS, 439, 2933
\bibitem[\protect\citeauthoryear{Yan et al.}{2013}]{Yan2013}Yan D. H., Zhang L., Yuan Q., Fan Z. H., Zeng H. D., 2013, ApJ, 765, 122
\bibitem[\protect\citeauthoryear{Yan et al.}{2015}]{Yan2015}Yan D. H., Zhang L., Zhang S. N., 2015, MNRAS, 454, 1310
\bibitem[\protect\citeauthoryear{Yuan et al.}{2011}]{Yuan2011}Yuan Q., Liu S., Fan Z., Bi X., Fryer C., 2011, ApJ, 735, 120   
\bibitem[\protect\citeauthoryear{Zamaninasab et al.}{2014}]{Zamaninasab2014}Zamaninasab M., Clausen-Brown E., Savolainen T. et al., 2014, Natur, 510,126Z
\bibitem[\protect\citeauthoryear{Zhang et al.}{2012}]{Zhang2012}Zhang J., Liang E.-W., Zhang S.-N., \& Bai J. M., 2012, ApJ, 752, 157
\end{thebibliography}



\appendix

\section{Corner plots of model parameters} \label{appA}

\begin{figure*}
 \includegraphics[width=0.49\textwidth]{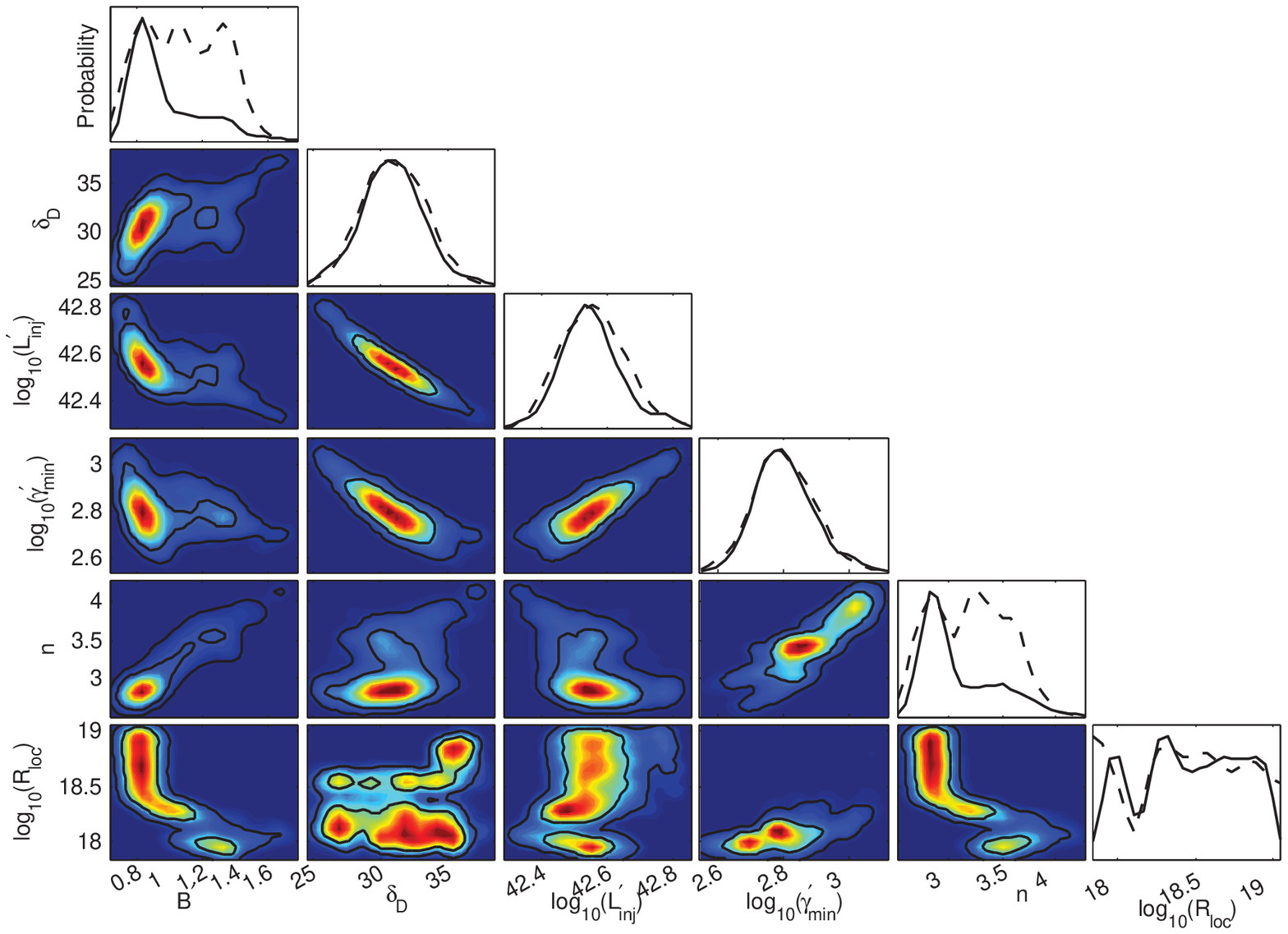} \includegraphics[width=0.49\textwidth]{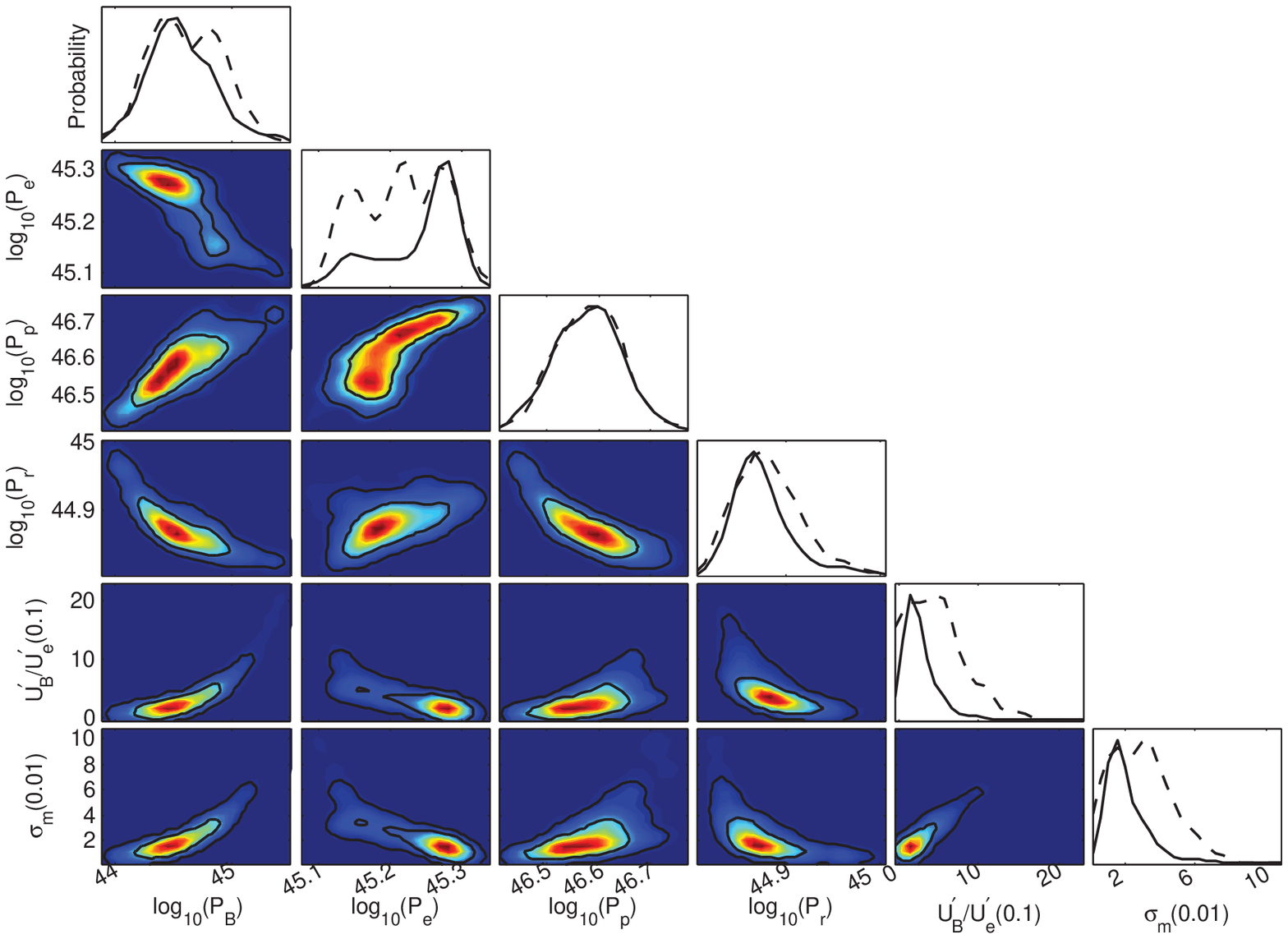}
 \includegraphics[width=0.49\textwidth]{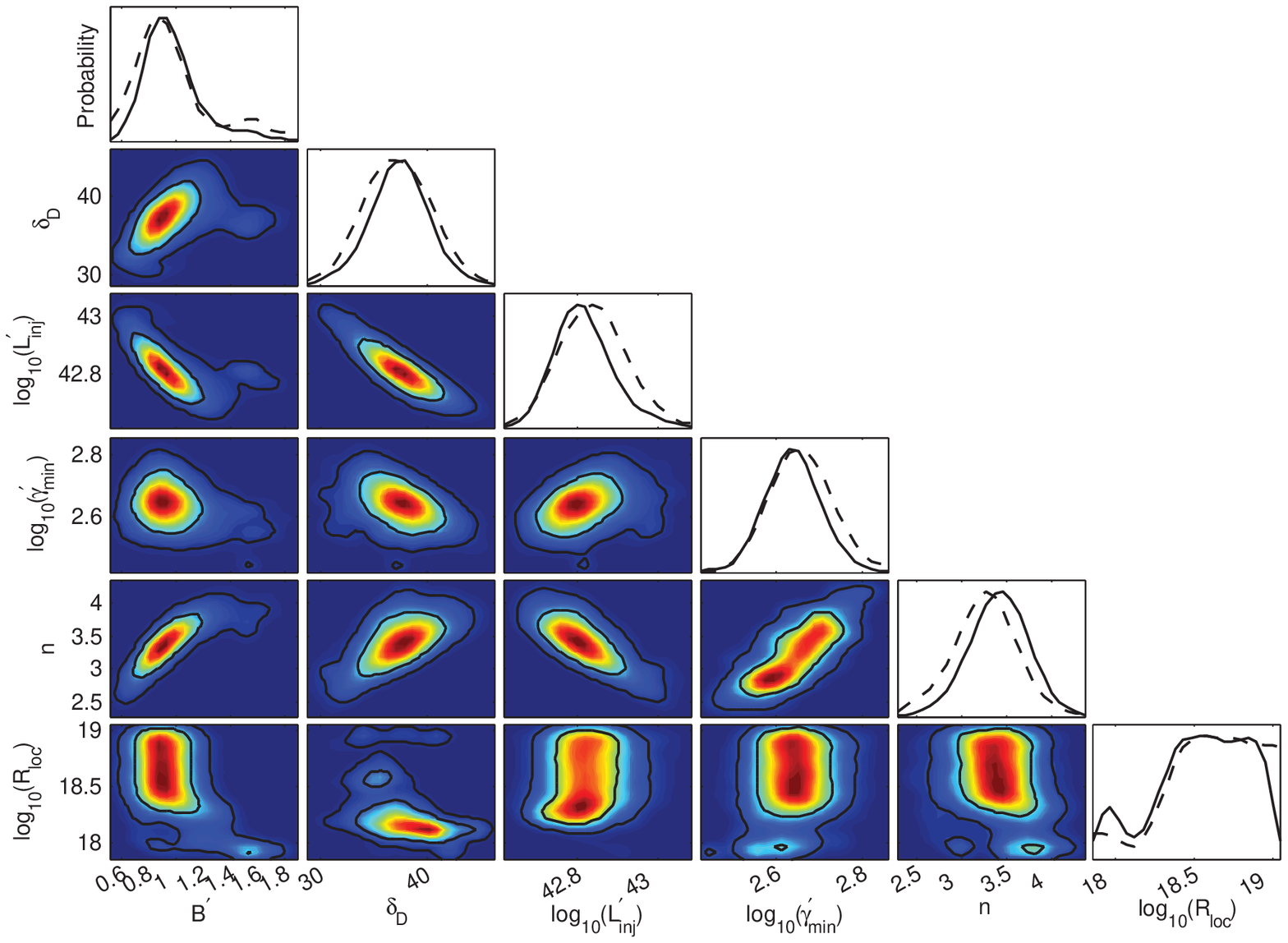} \includegraphics[width=0.49\textwidth]{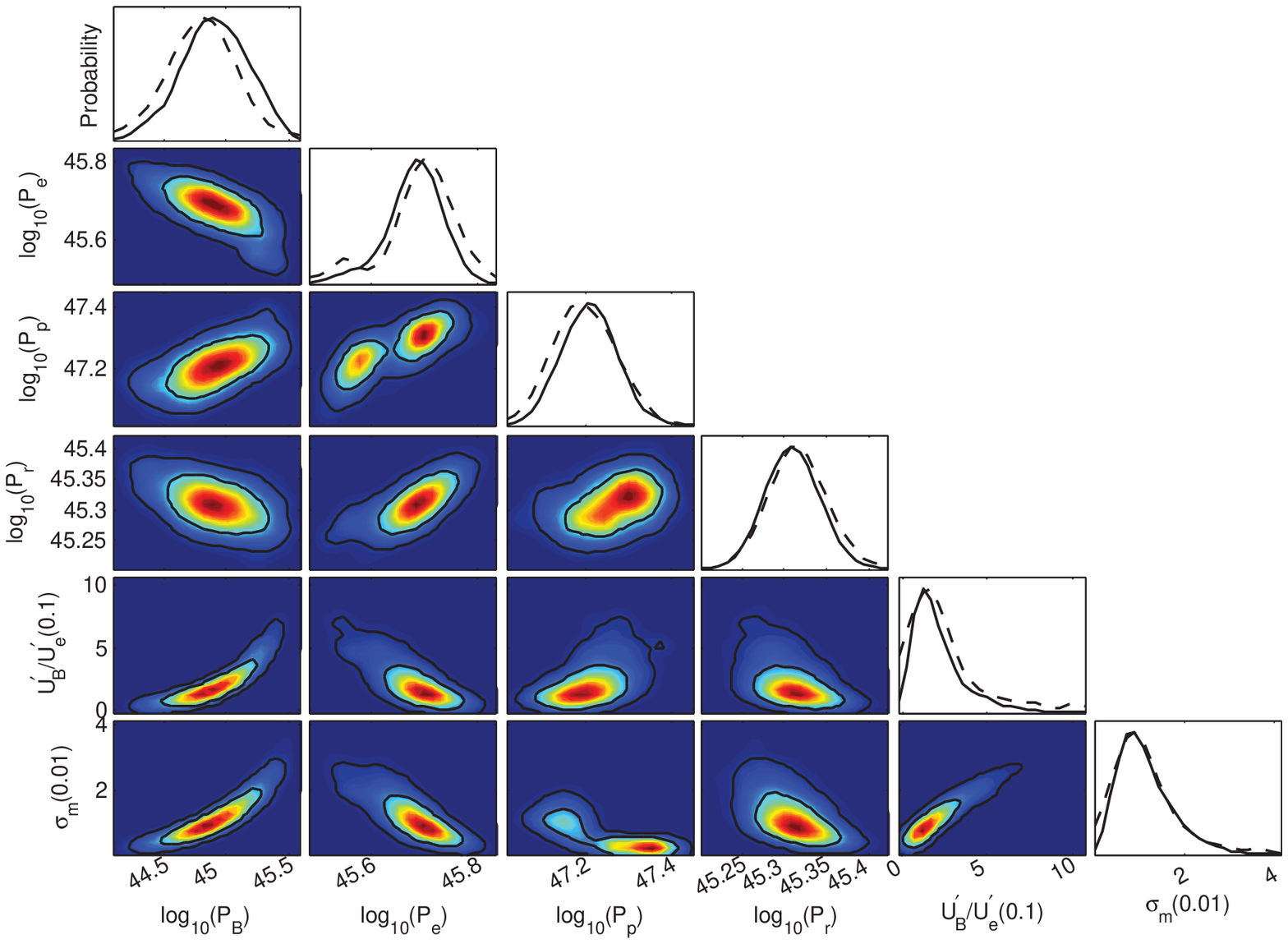}
 \includegraphics[width=0.49\textwidth]{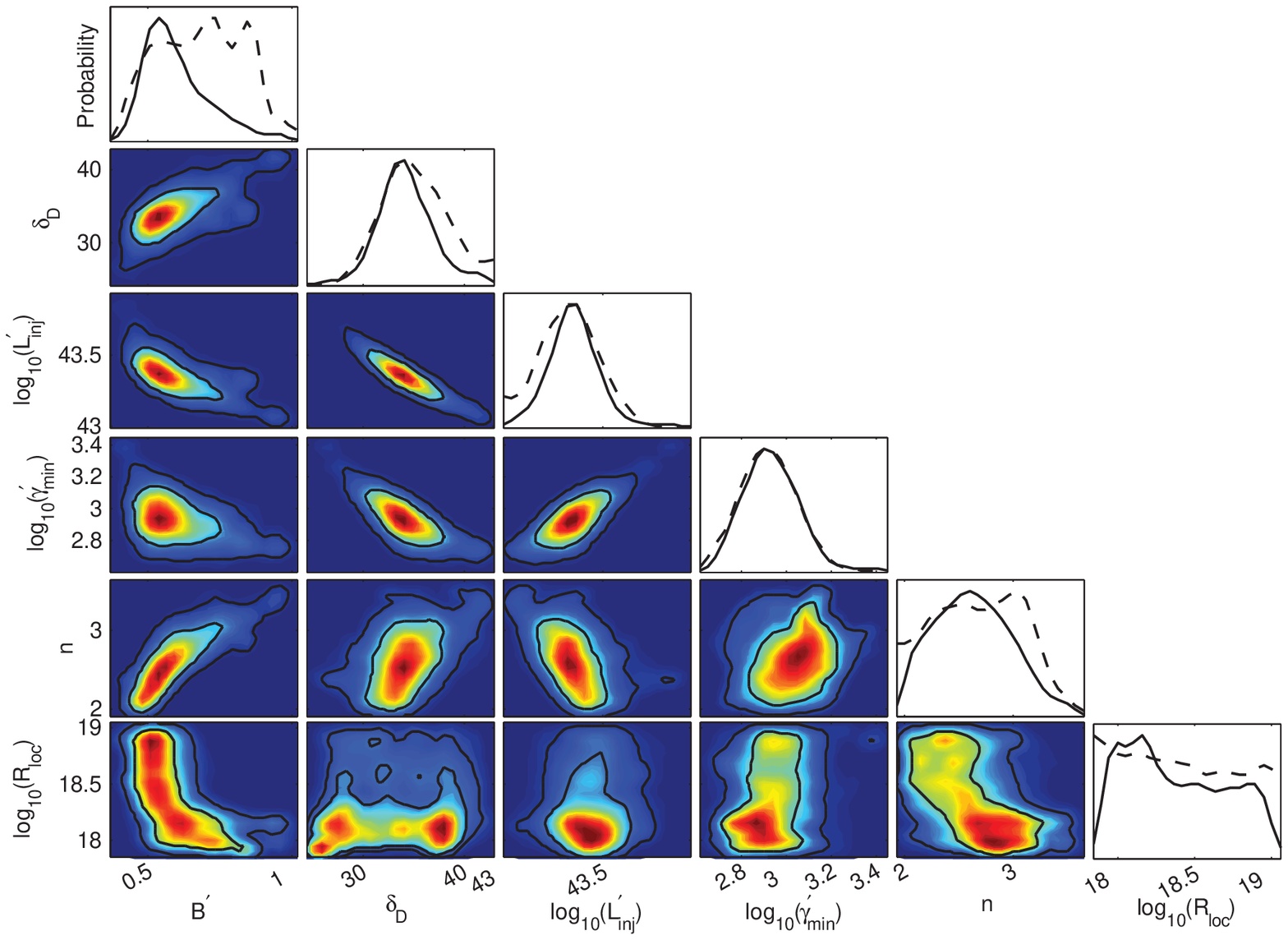} \includegraphics[width=0.49\textwidth]{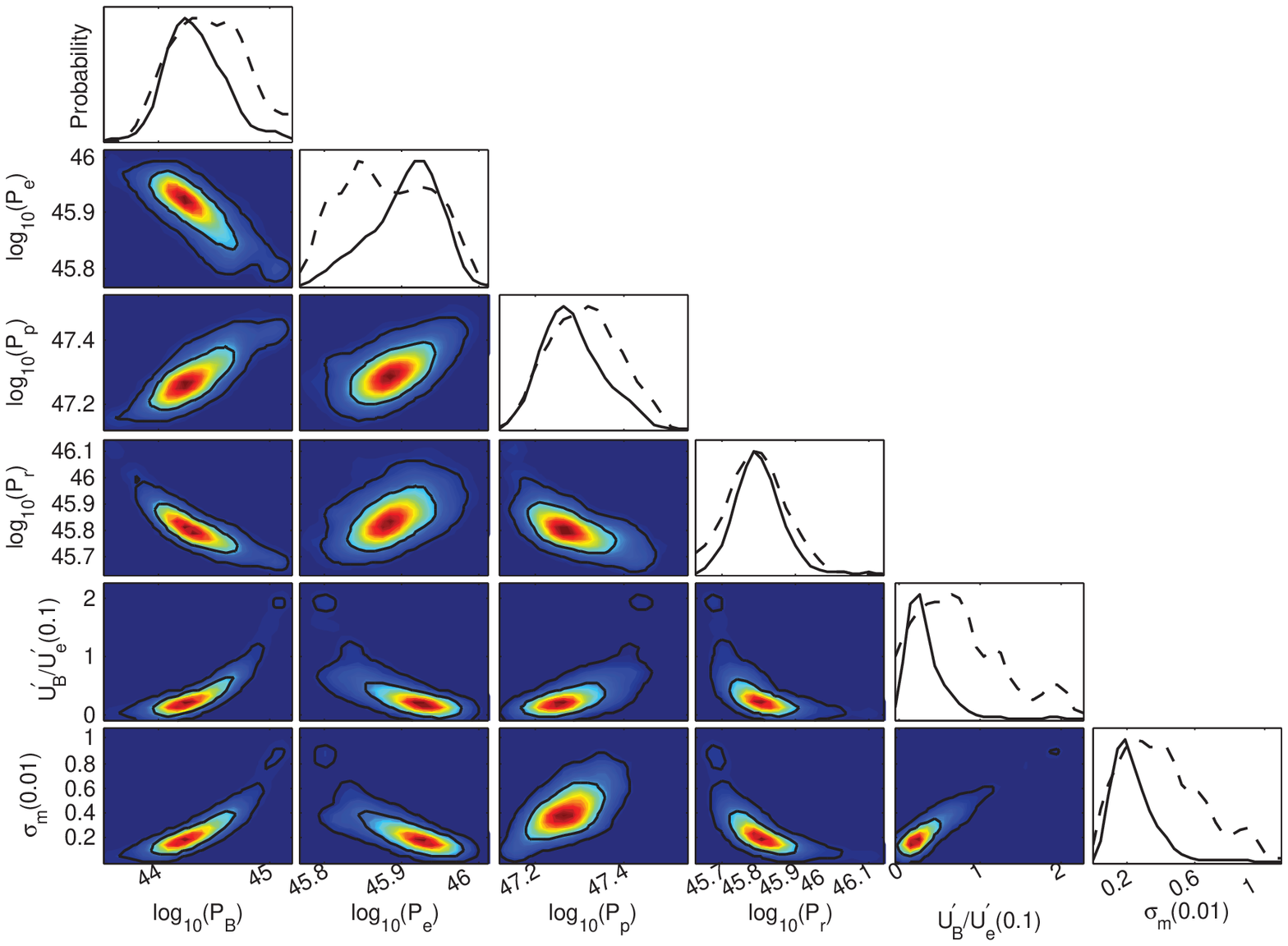}
\caption{Corner plots of the free model parameters (left) and derived parameters (right) for FSRQs 3C 454.3.
For the two-dimensional confidence contours of the parameters, the inner and outer contours denote the 68 and 95 per cent confidence intervals, respectively. For the one-dimensional probability distributions of the parameters, the dashed lines show the maximum likelihood distributions and solid lines show the marginalized probability distributions.
From top to bottom, the plots are the results obtained from fitting SEDs at Low $\gamma$-ray state, 11/6 and 27/11 reported in Bonnoli et al. (2011), respectively.
}\label{distr1}
\end{figure*}

\begin{figure*}
 \includegraphics[width=0.49\textwidth]{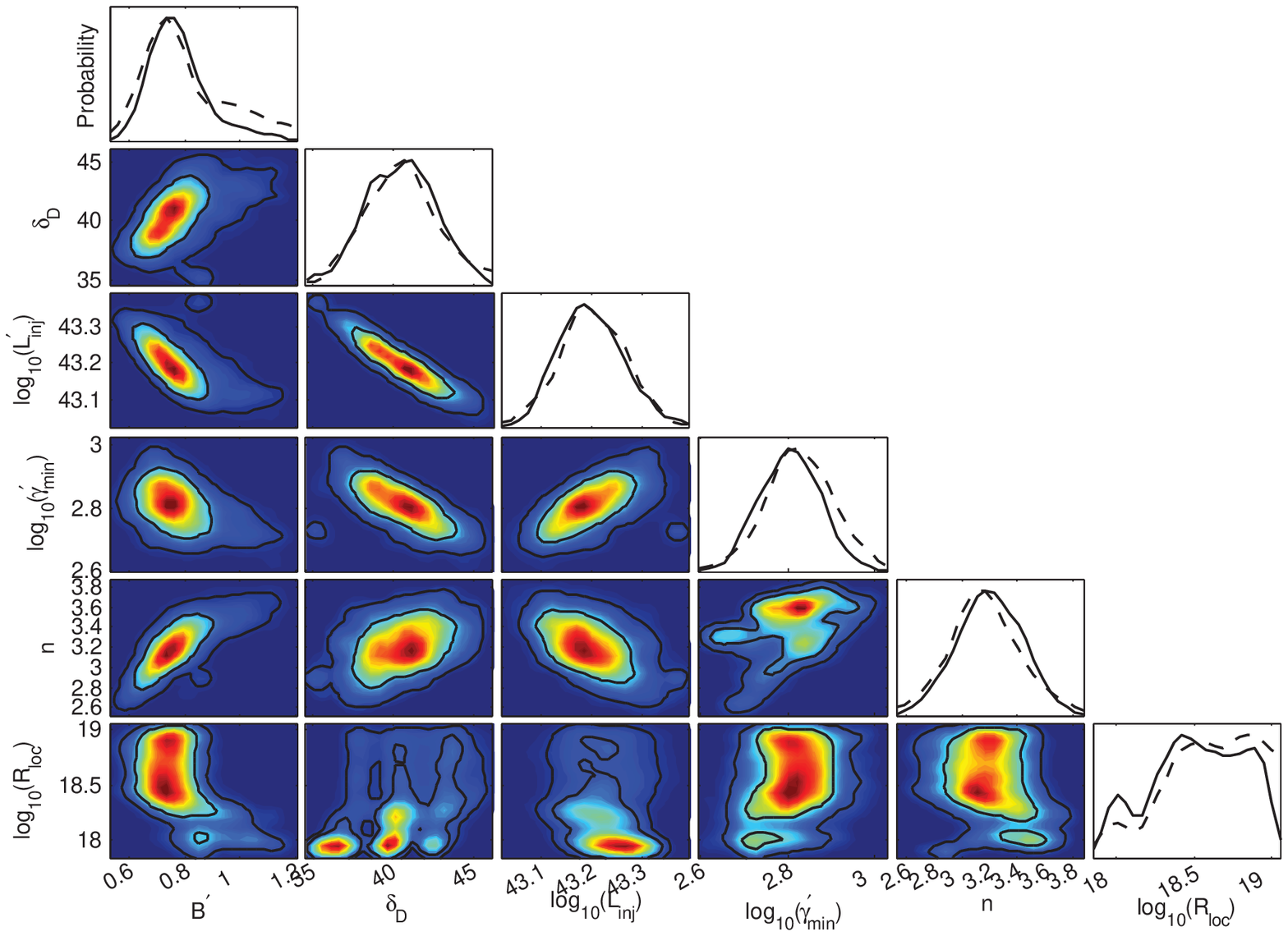} \includegraphics[width=0.49\textwidth]{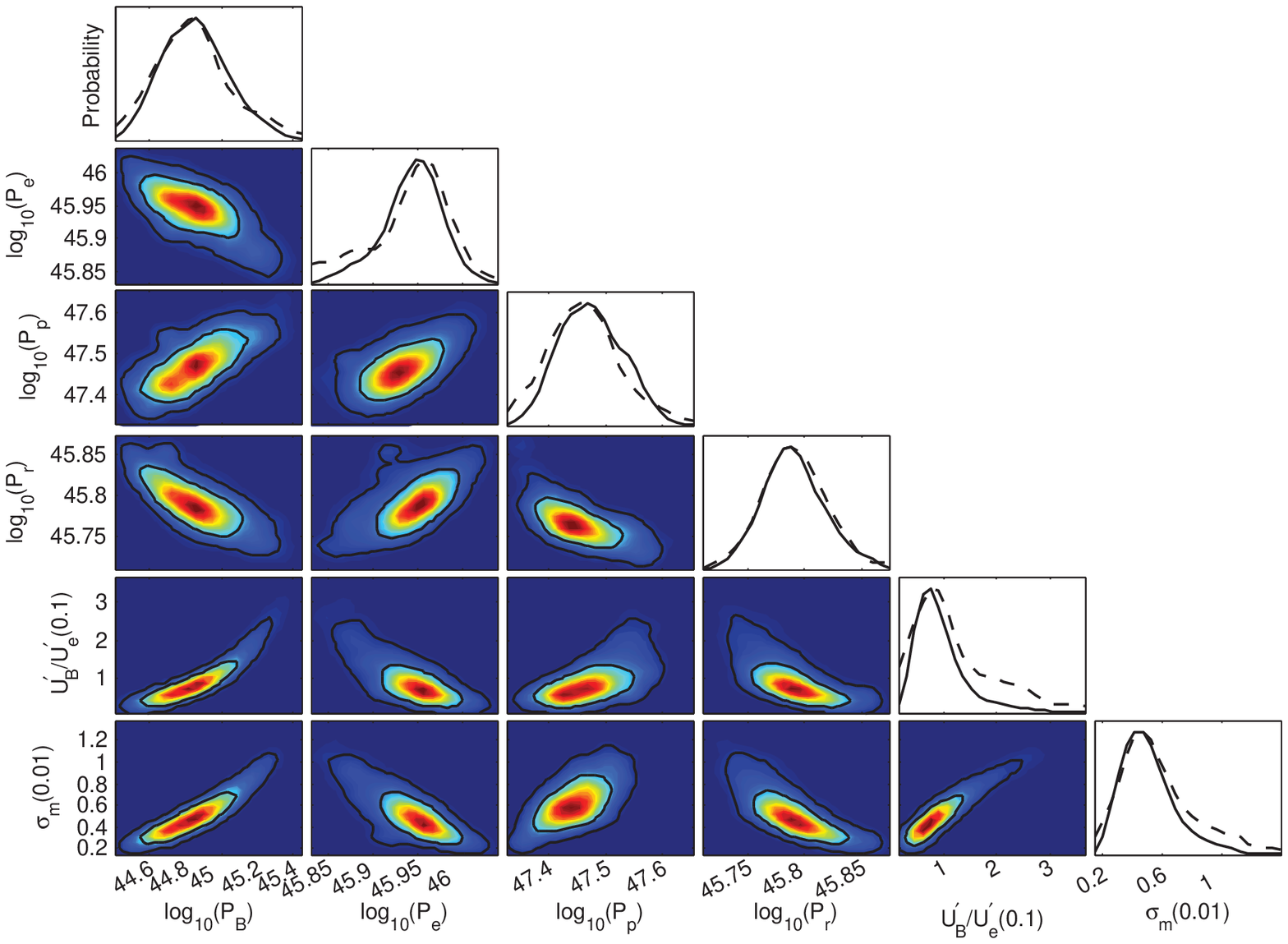}
 \includegraphics[width=0.49\textwidth]{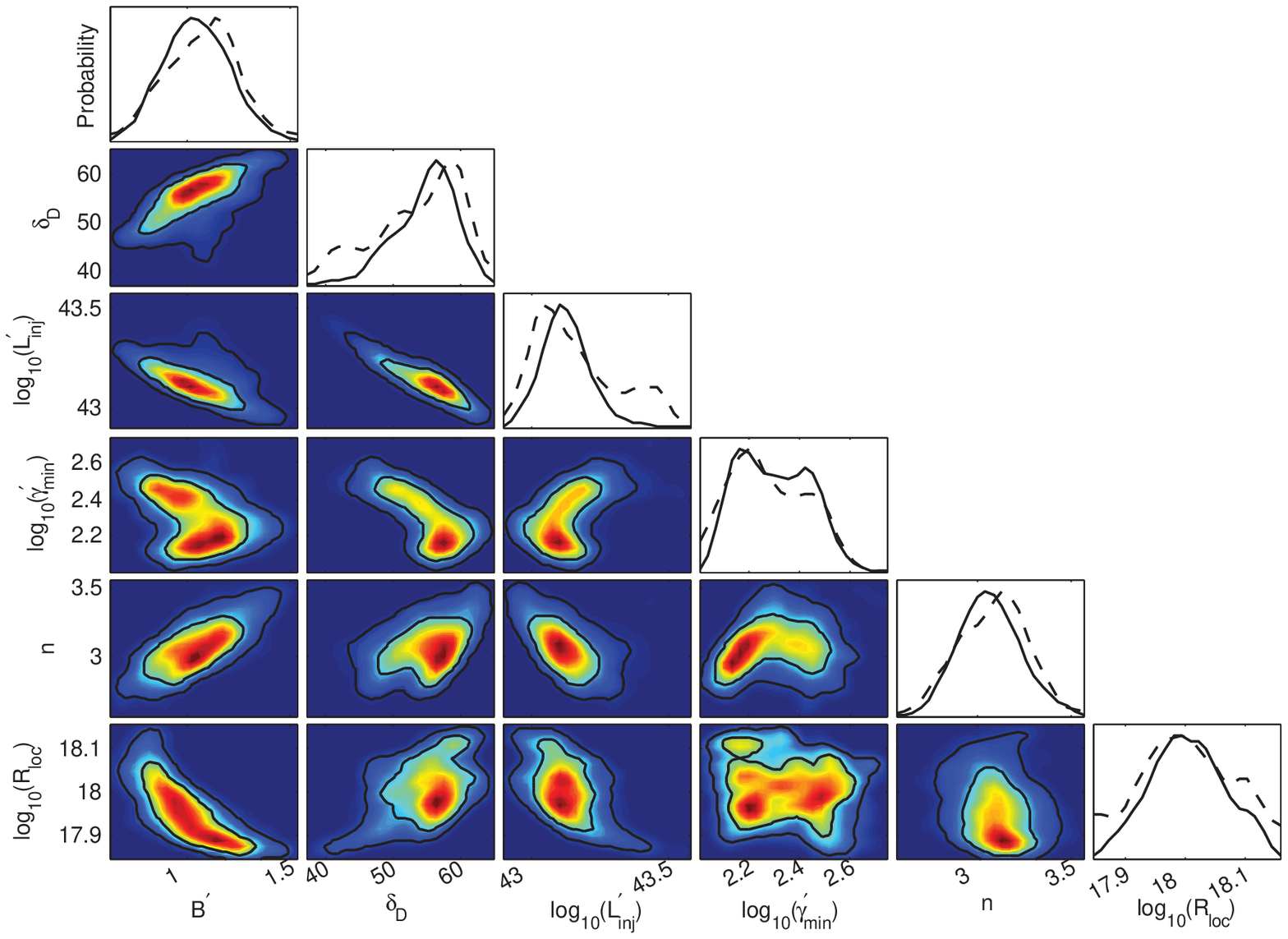} \includegraphics[width=0.49\textwidth]{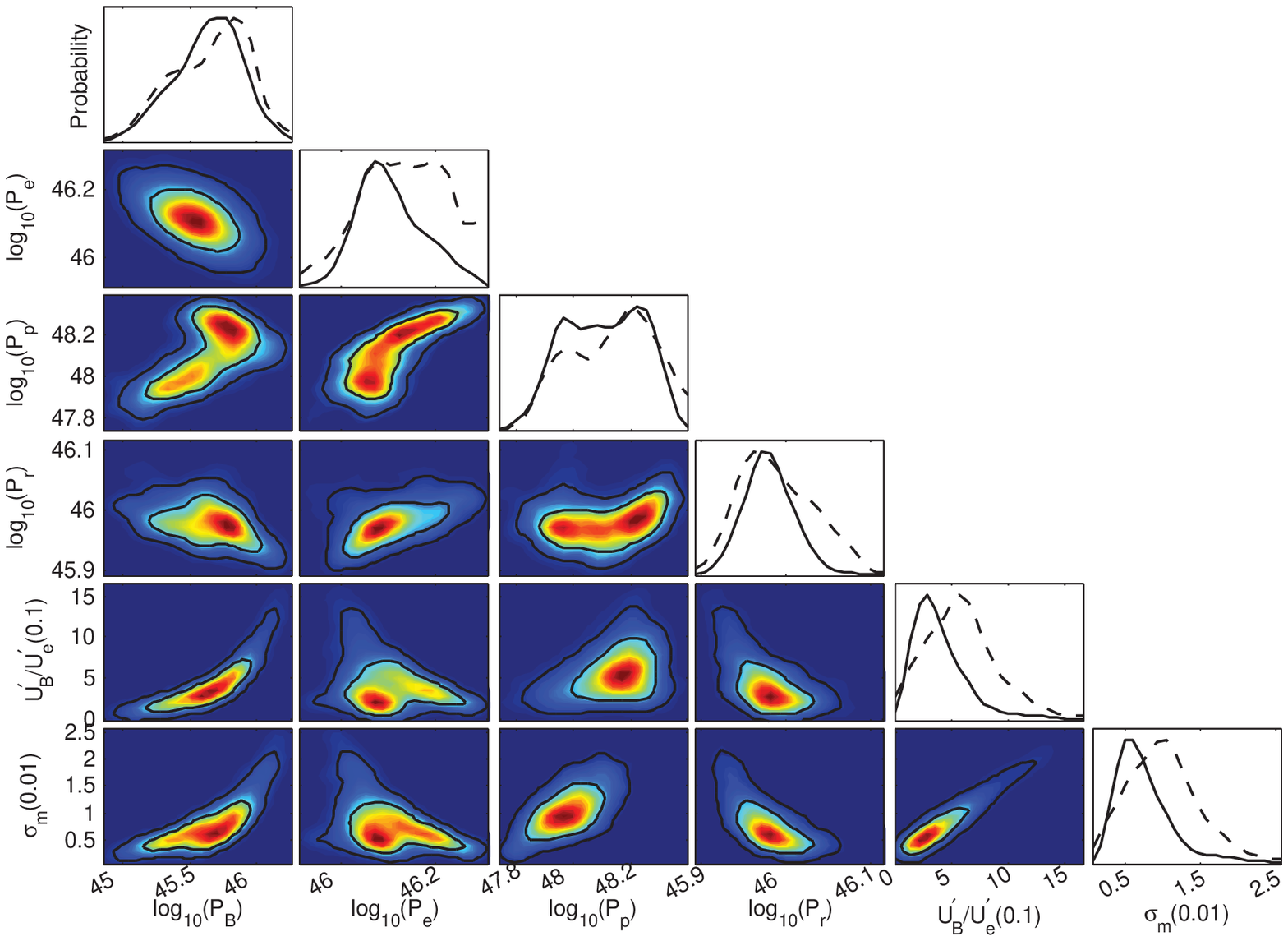}
 \includegraphics[width=0.49\textwidth]{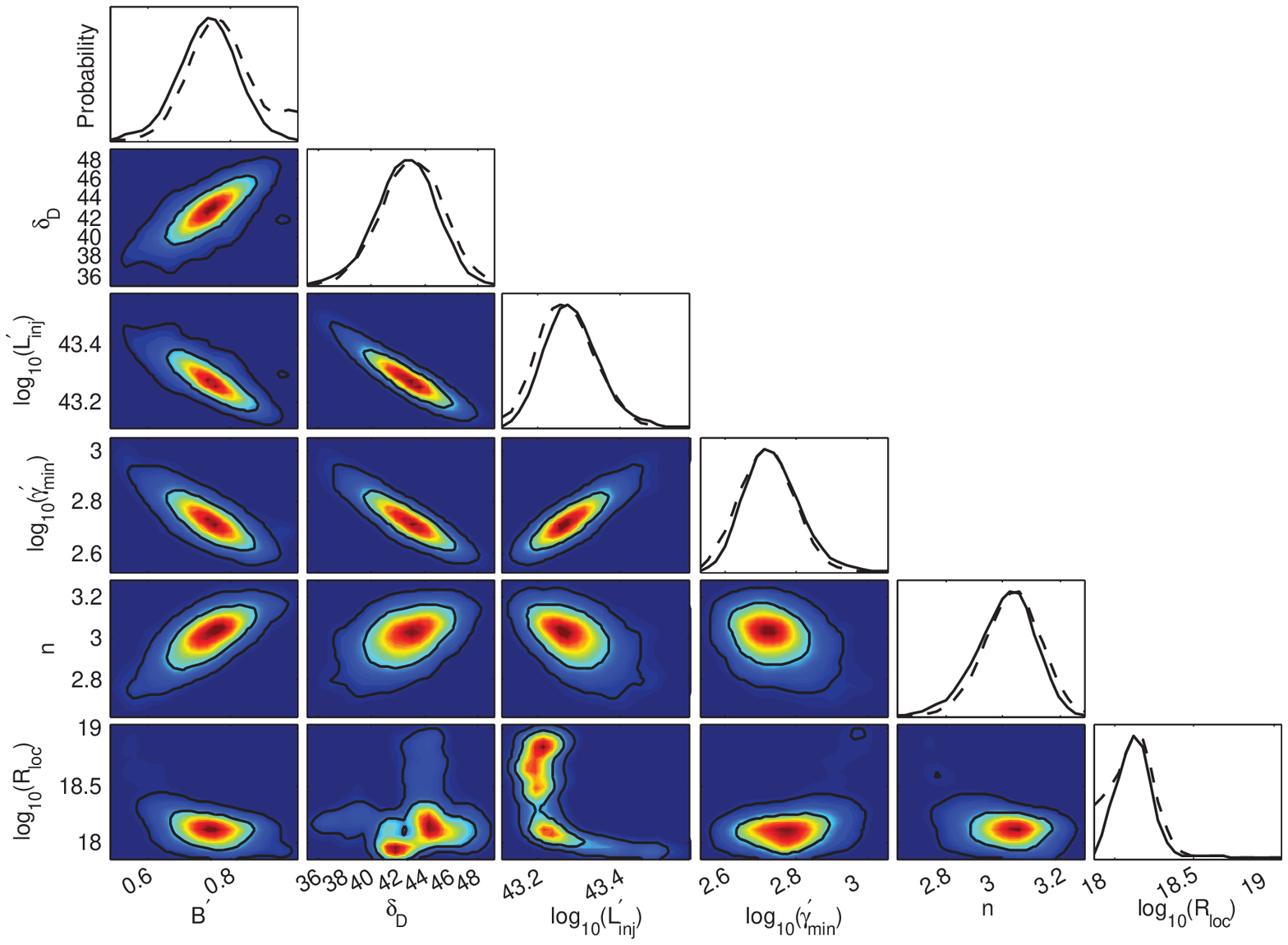} \includegraphics[width=0.49\textwidth]{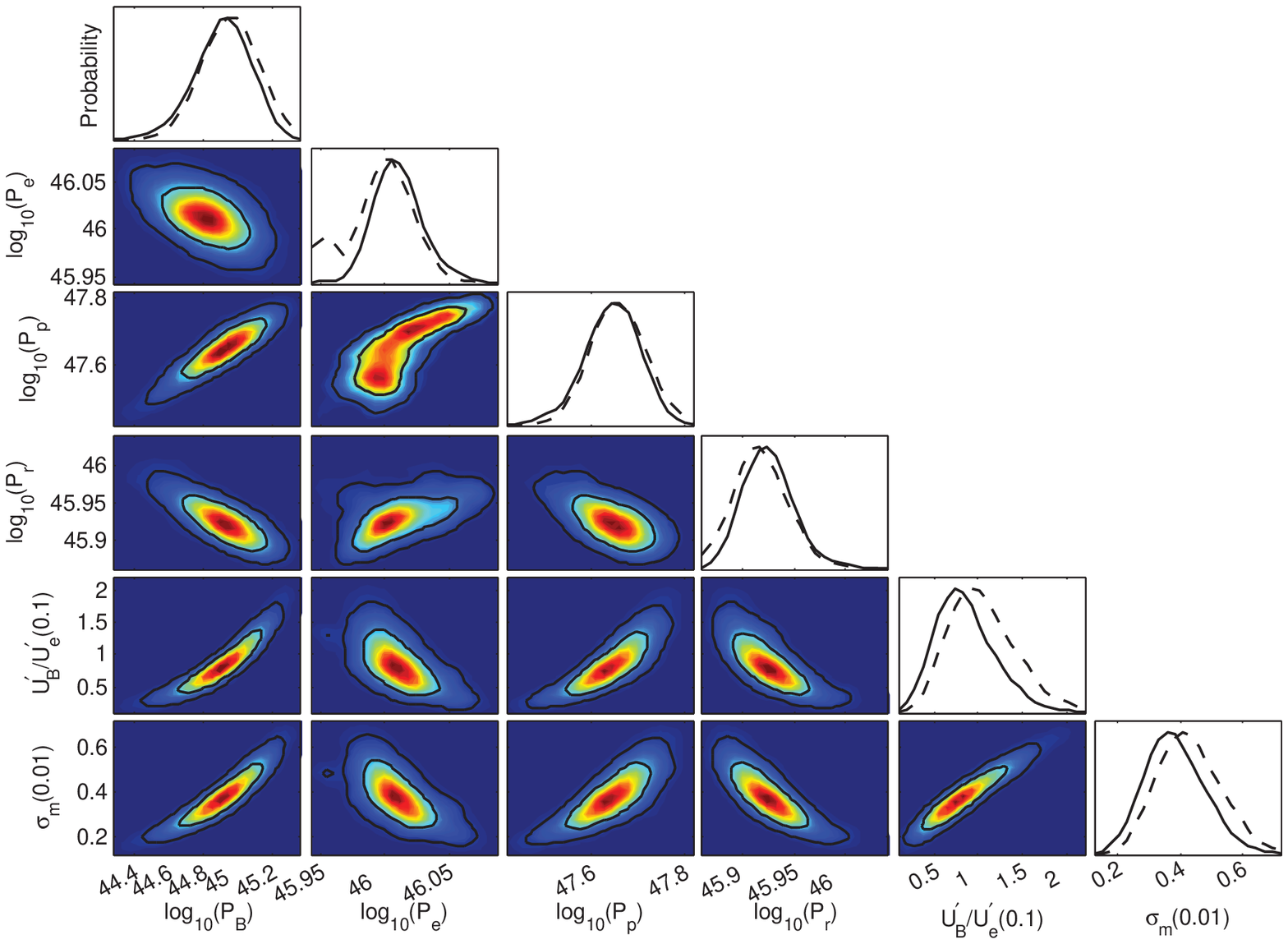}
\caption{Same as Figure~\ref{distr1}, but for the SEDs on 01/12 (top), 02/12 (middle) and 03/12 (bottom).
 \label{distr2}}
\end{figure*}

\begin{figure*}
 \includegraphics[width=0.43\textwidth]{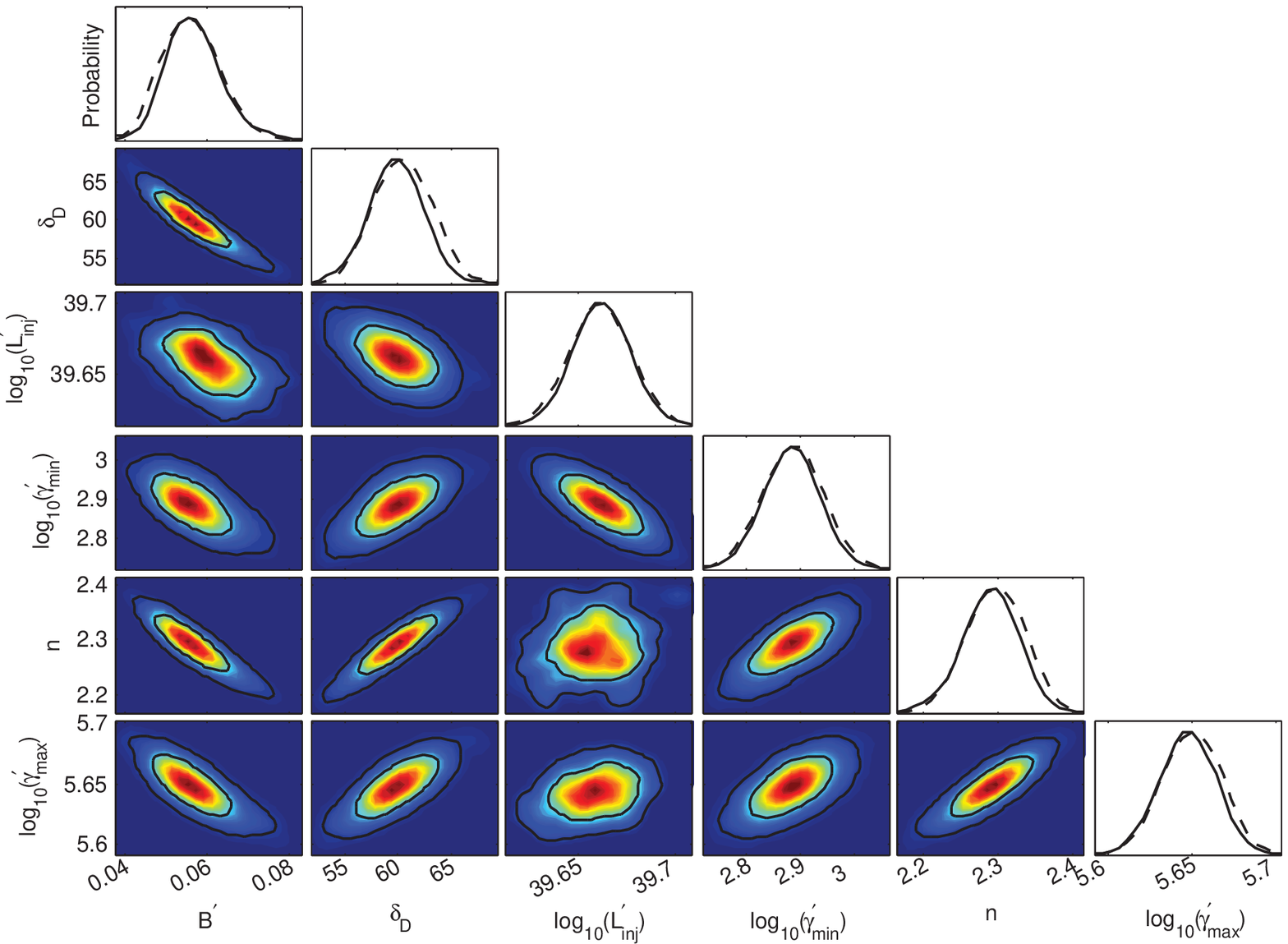} \includegraphics[width=0.43\textwidth]{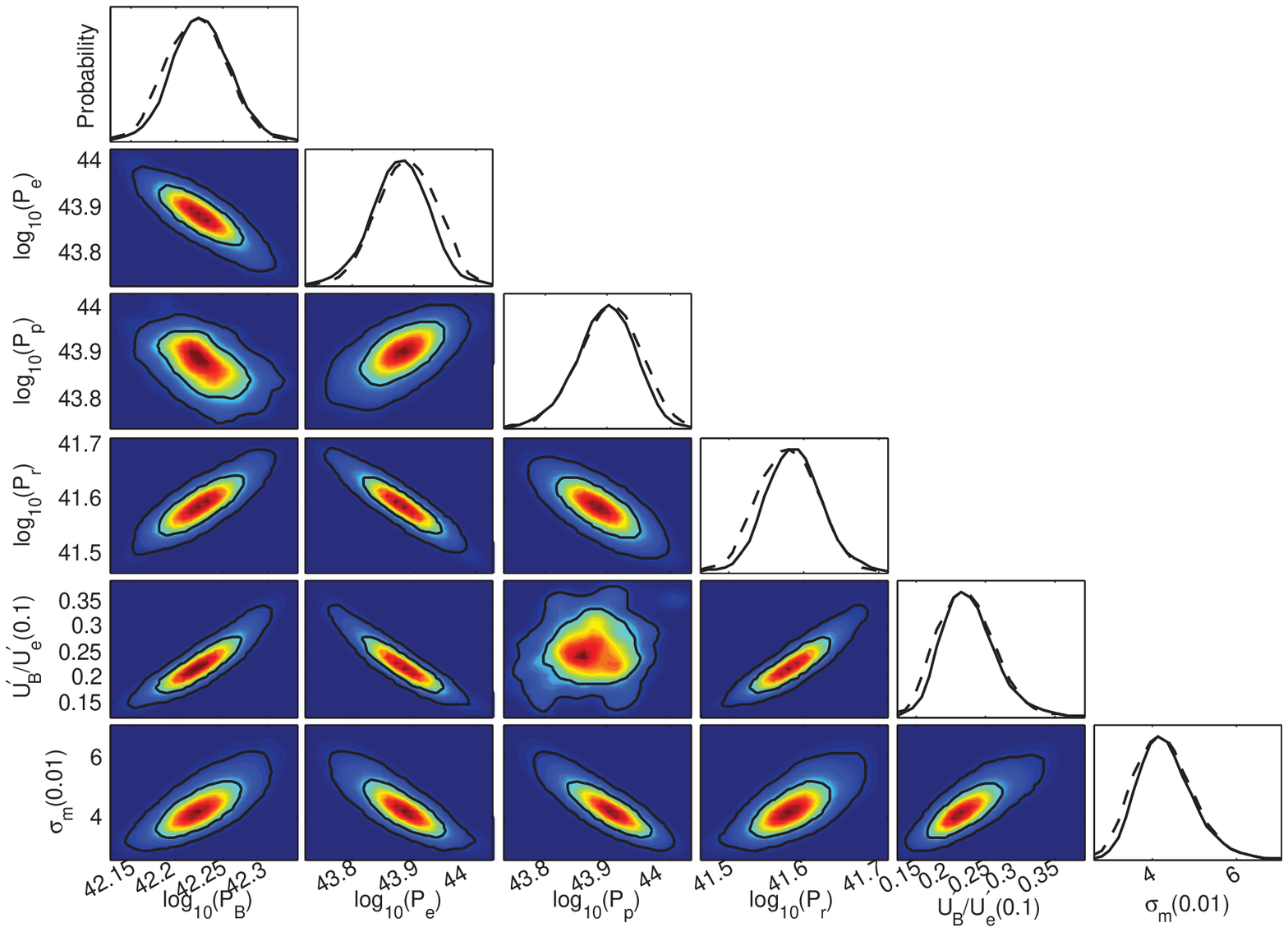}
 \includegraphics[width=0.43\textwidth]{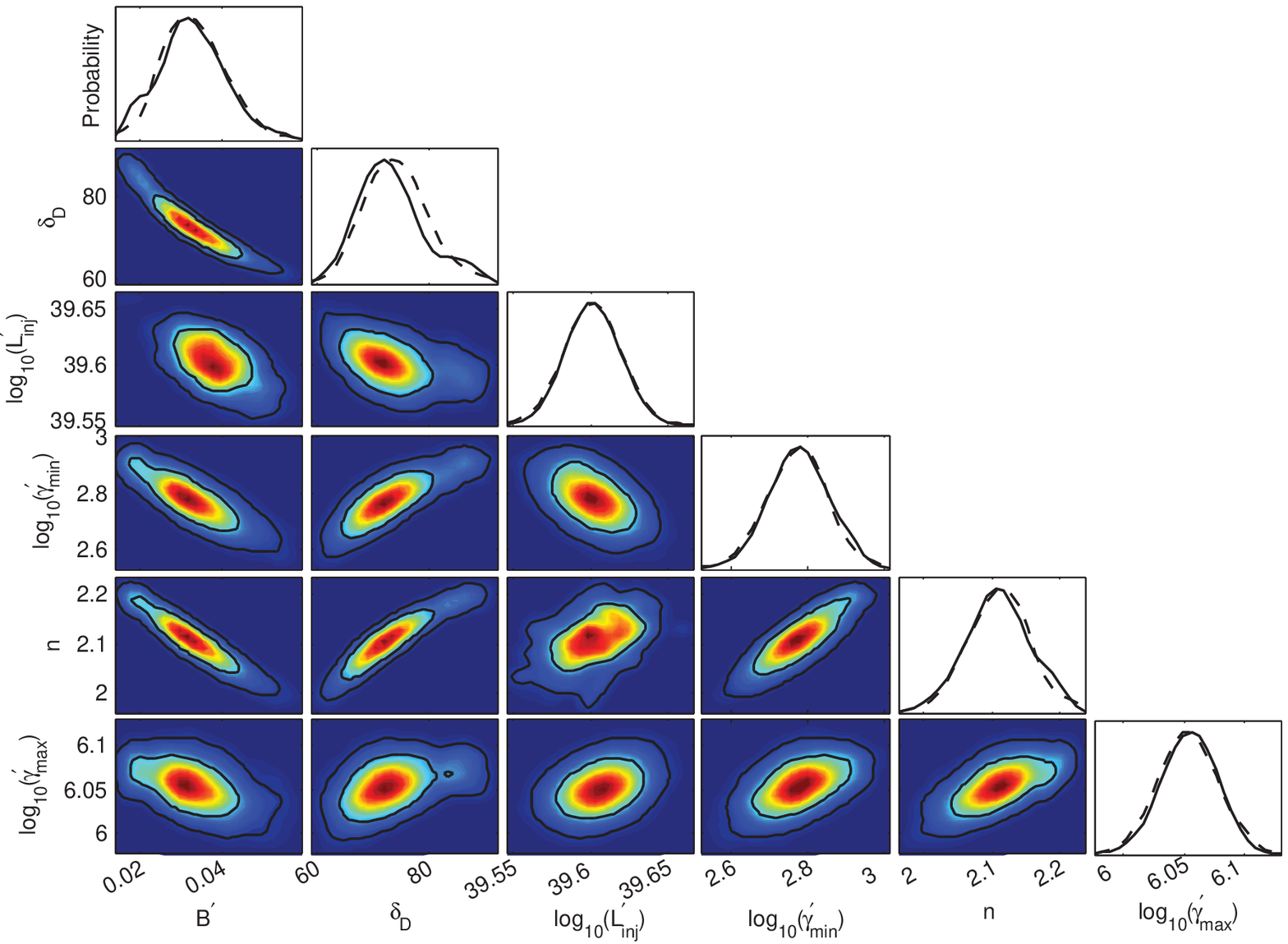} \includegraphics[width=0.43\textwidth]{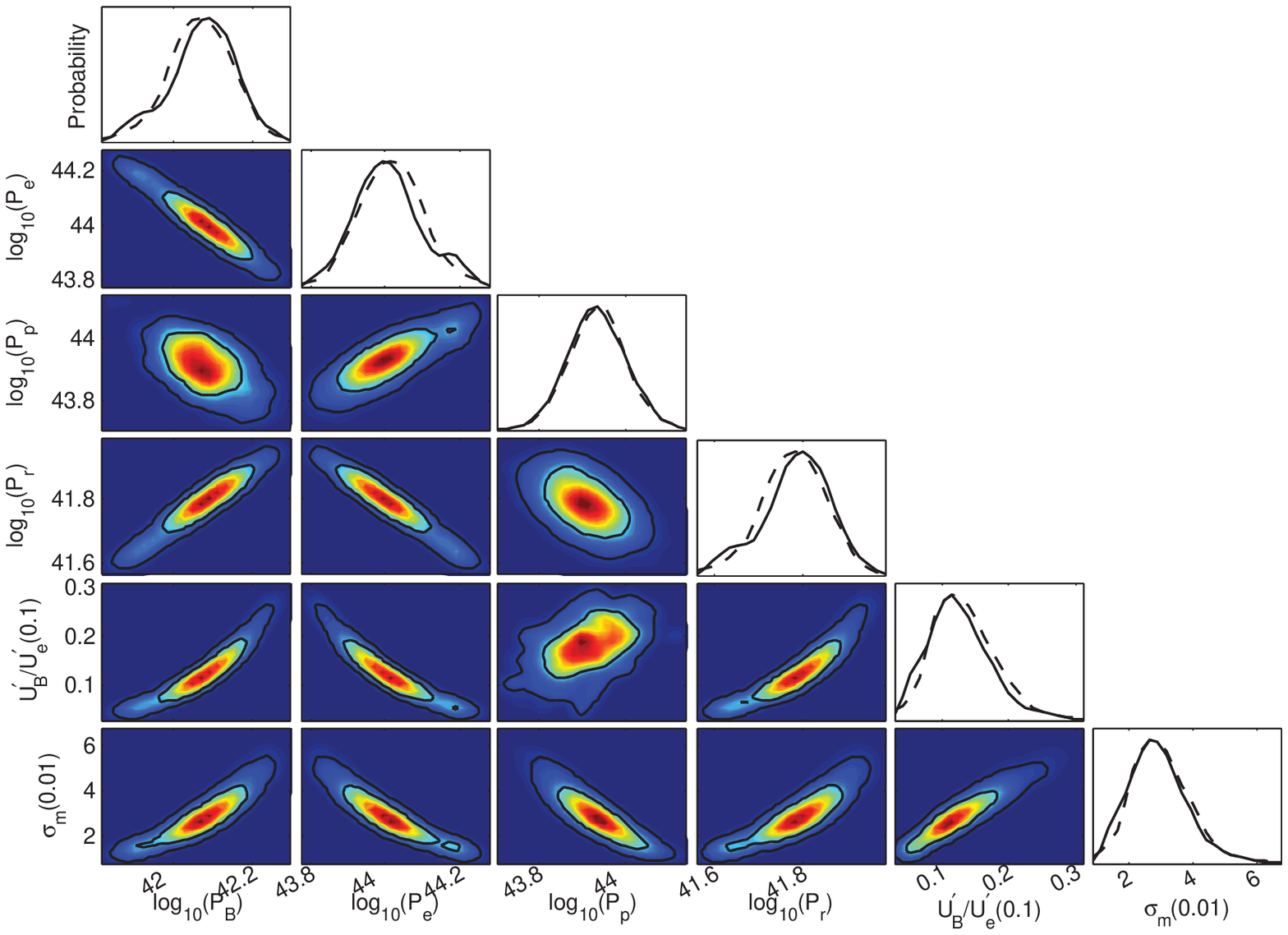}
 \includegraphics[width=0.43\textwidth]{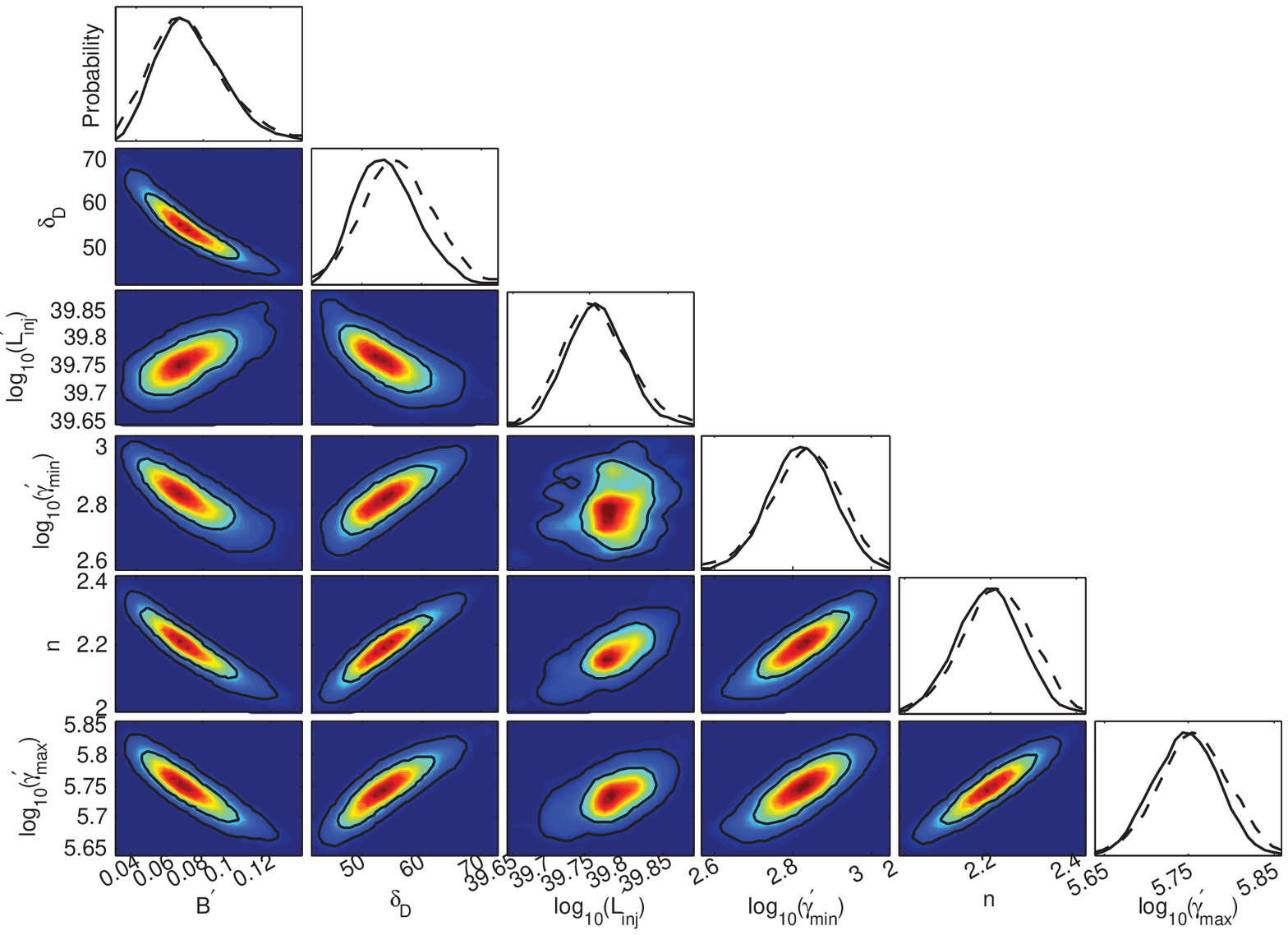} \includegraphics[width=0.43\textwidth]{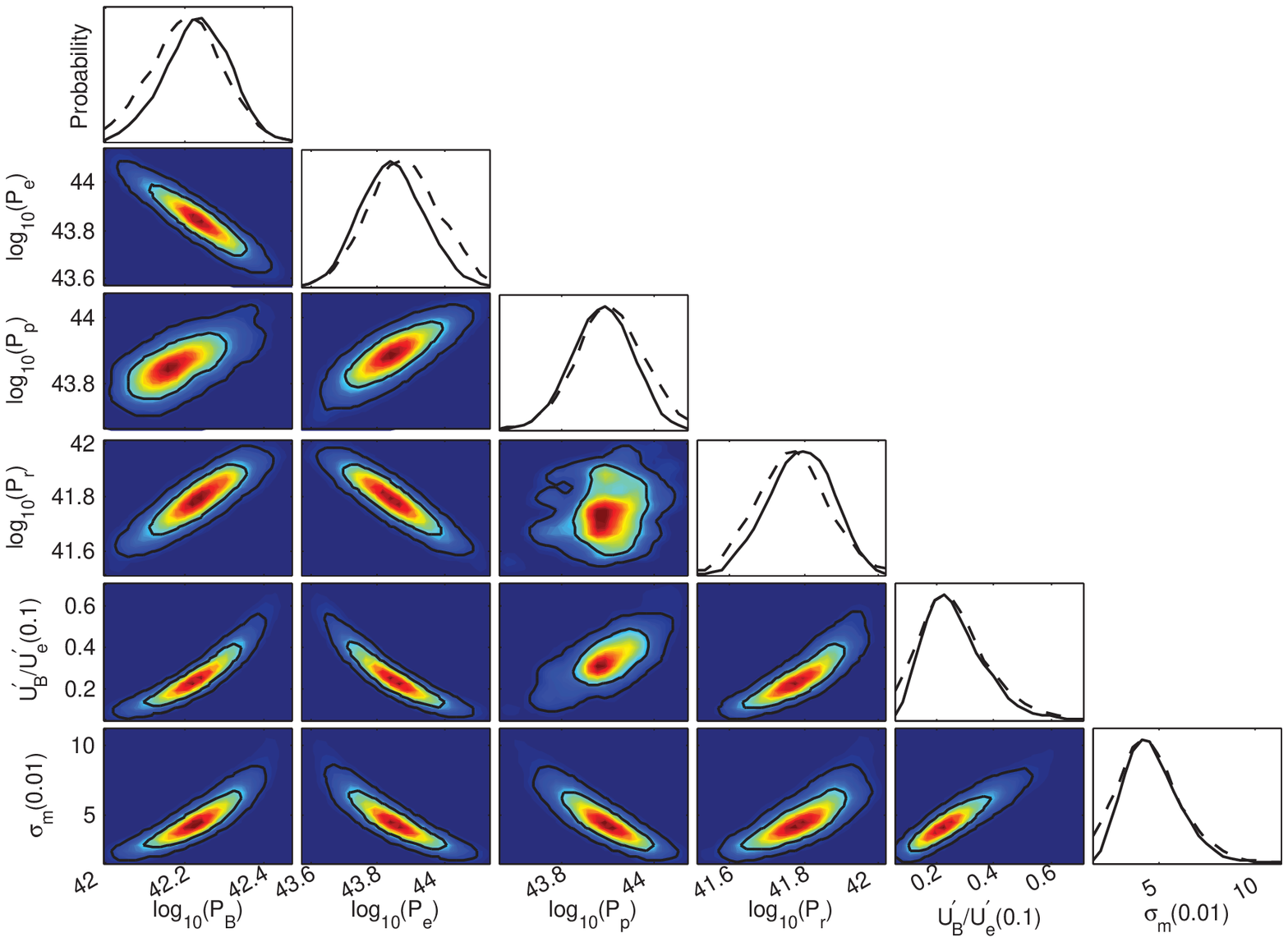}
 \includegraphics[width=0.43\textwidth]{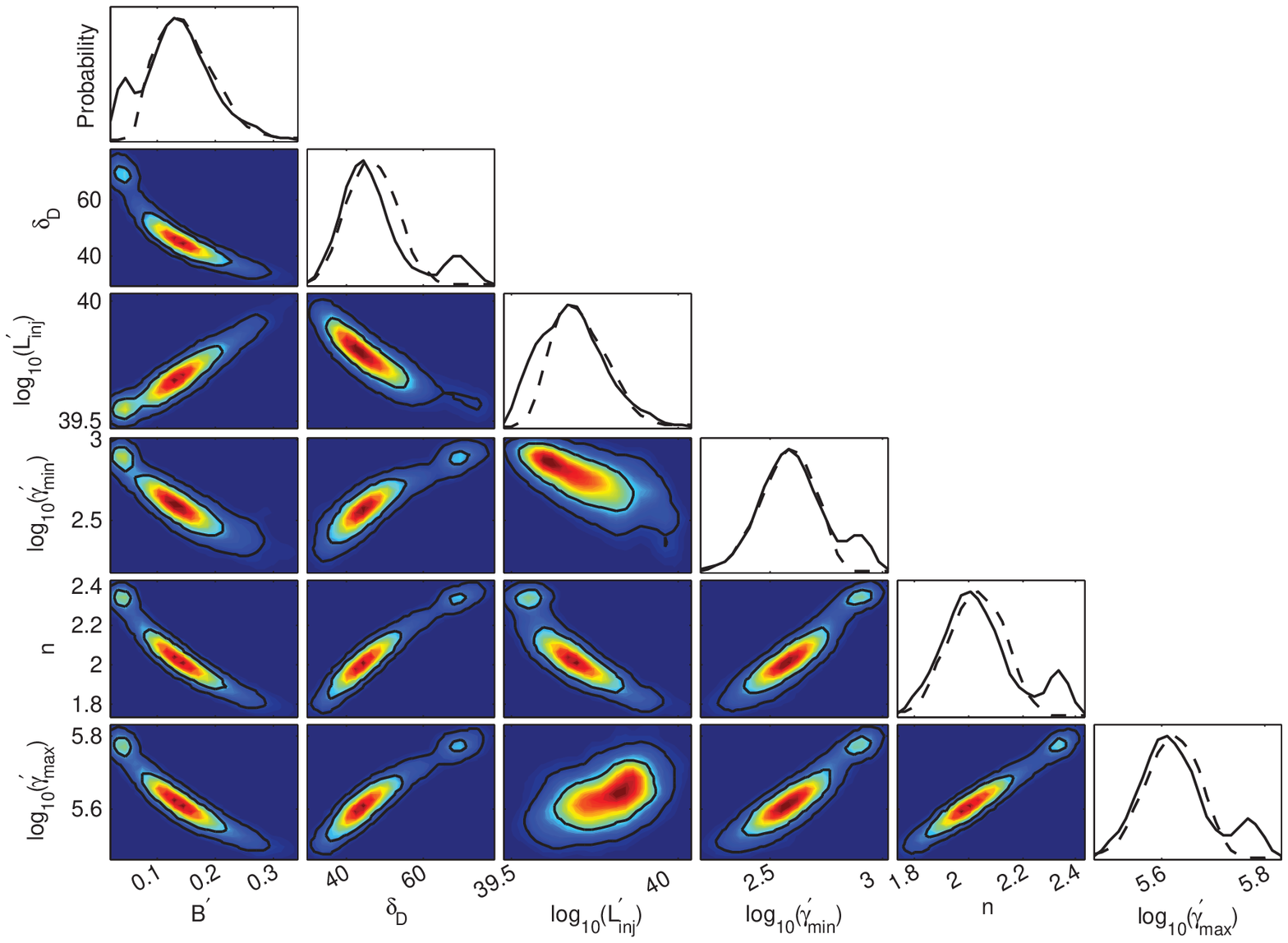} \includegraphics[width=0.43\textwidth]{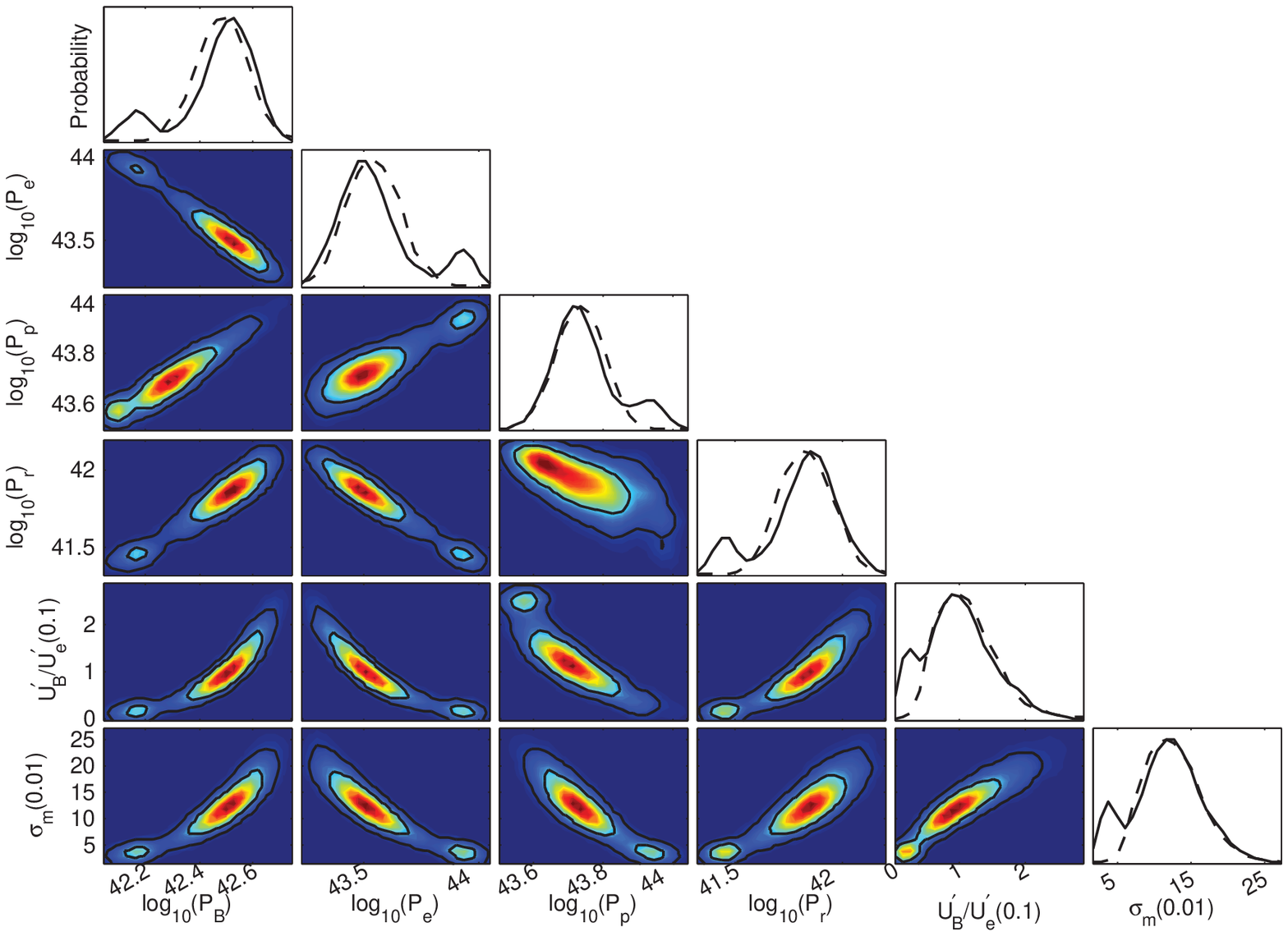}
\caption{Corner plots of the input (left) and output parameters (right) for HBL Mrk 421. 
From top to bottom, the plots are arranged in the following order quiescent state, 55266, 55270 and 55277.
 \label{distr3}}
\end{figure*}

\begin{figure*}
 \includegraphics[width=0.49\textwidth]{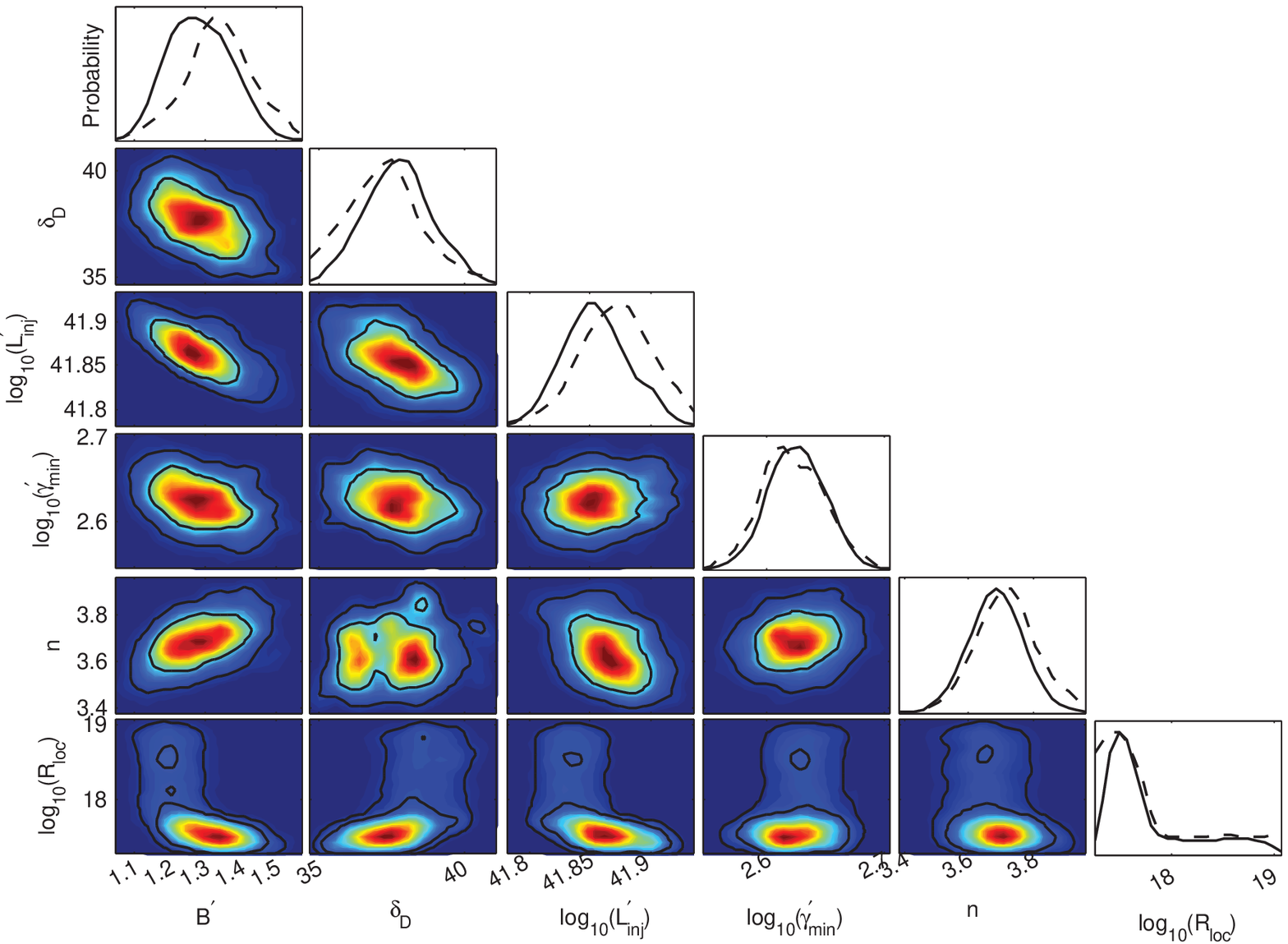} \includegraphics[width=0.49\textwidth]{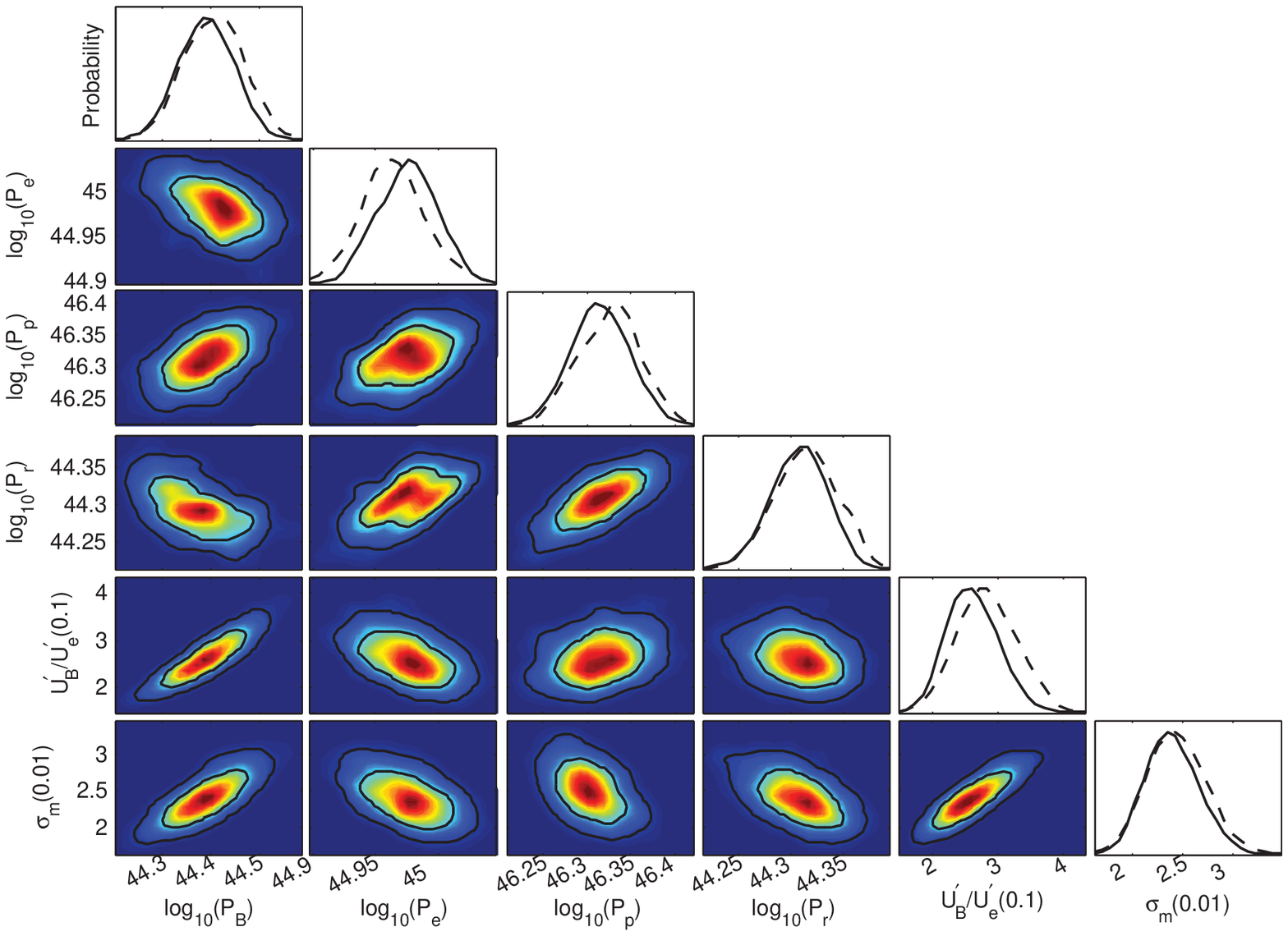}
 \includegraphics[width=0.49\textwidth]{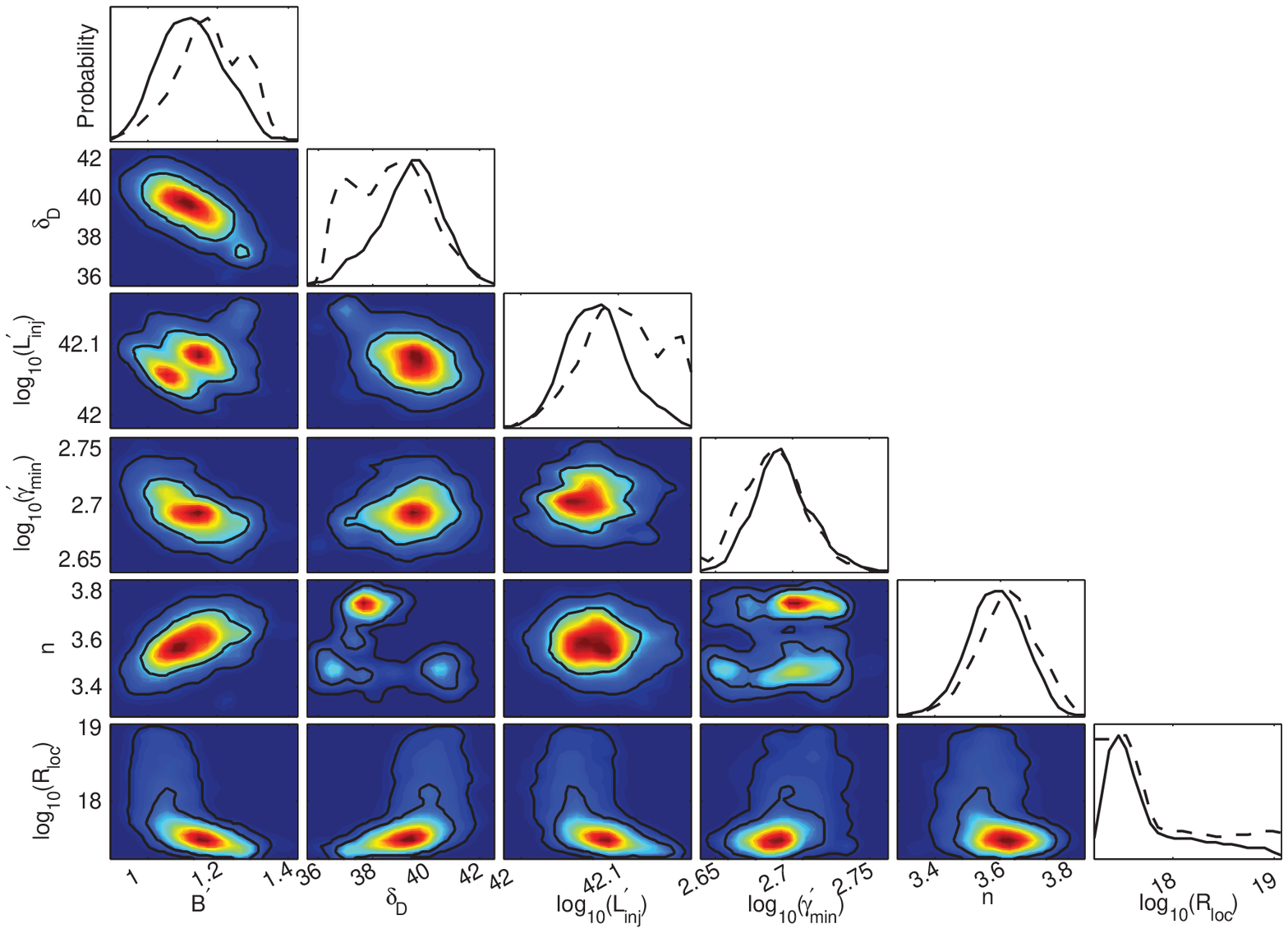} \includegraphics[width=0.49\textwidth]{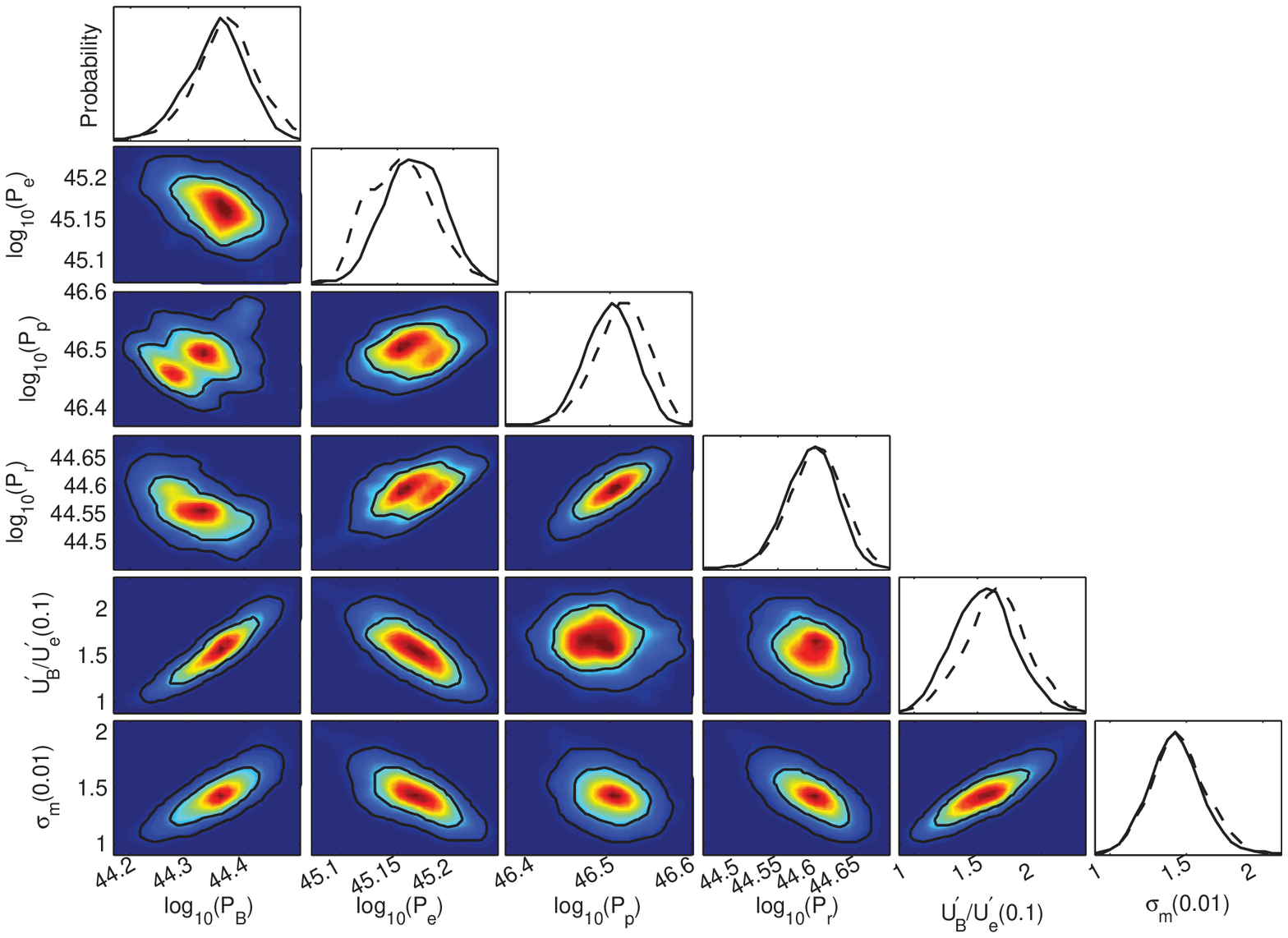}
 \includegraphics[width=0.49\textwidth]{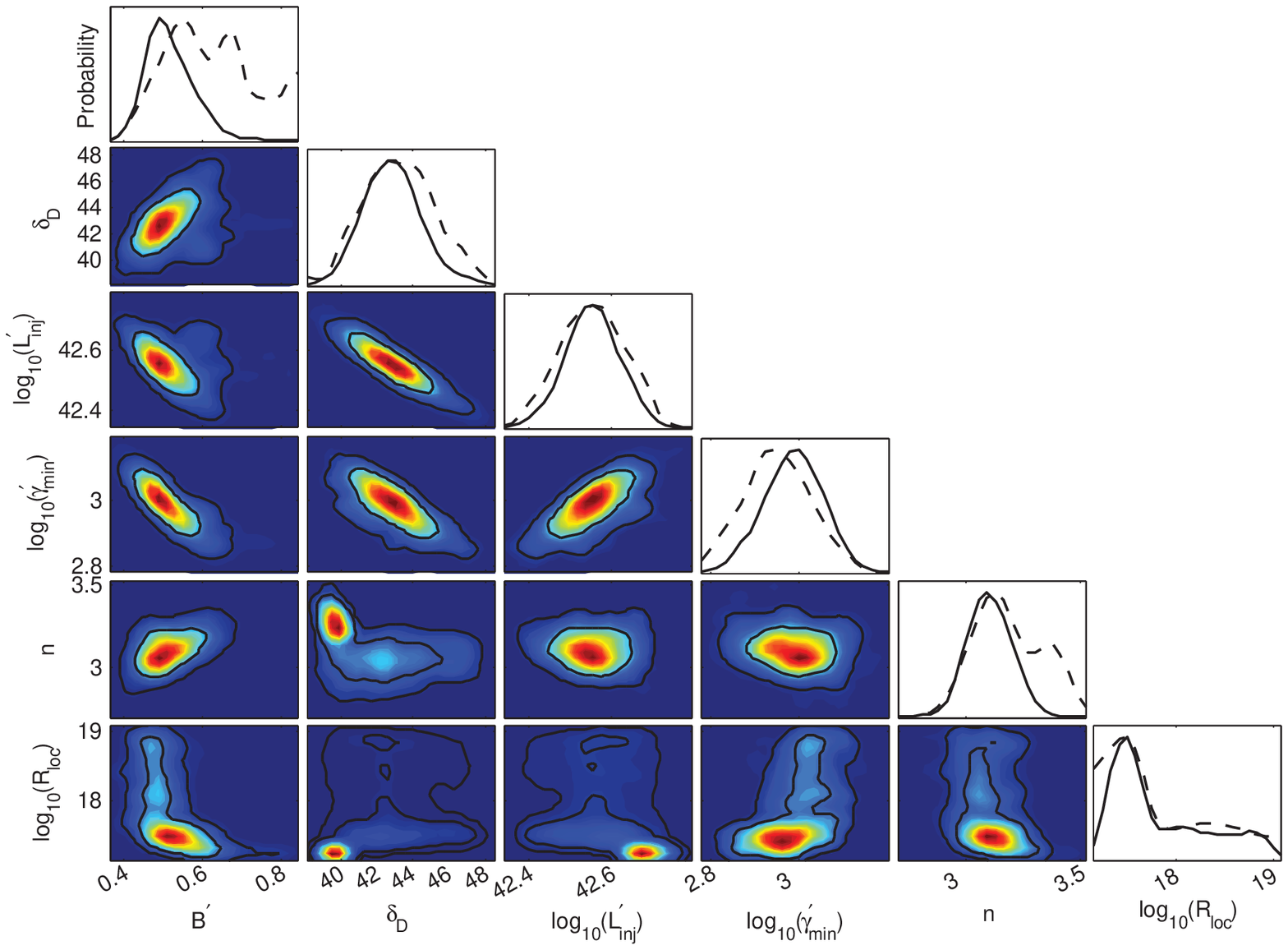} \includegraphics[width=0.49\textwidth]{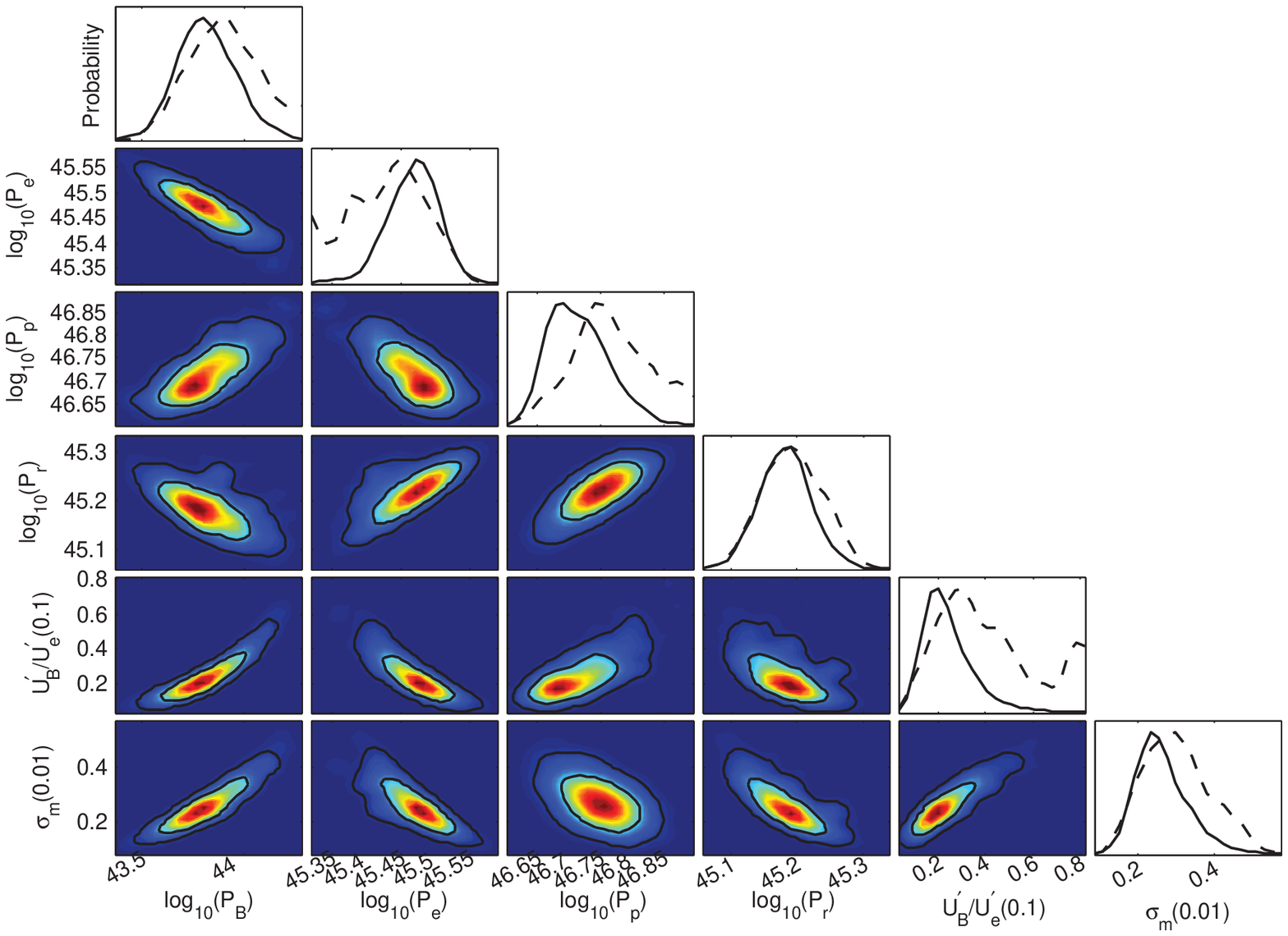}
\caption{
Corner plots of the free model parameters (left) and derived parameters (right) for FSRQs 3C 279.
From top to bottom, the plots are the results obtained from fitting SEDs in Periods A, C, D reported in Hayashida et al. (2015), respectively.
 \label{distr4}}
\end{figure*}

\section{SED fitting with the variability timescale of one day for quiescent states}
\begin{figure*}
  \centering
  \includegraphics[width=\textwidth]{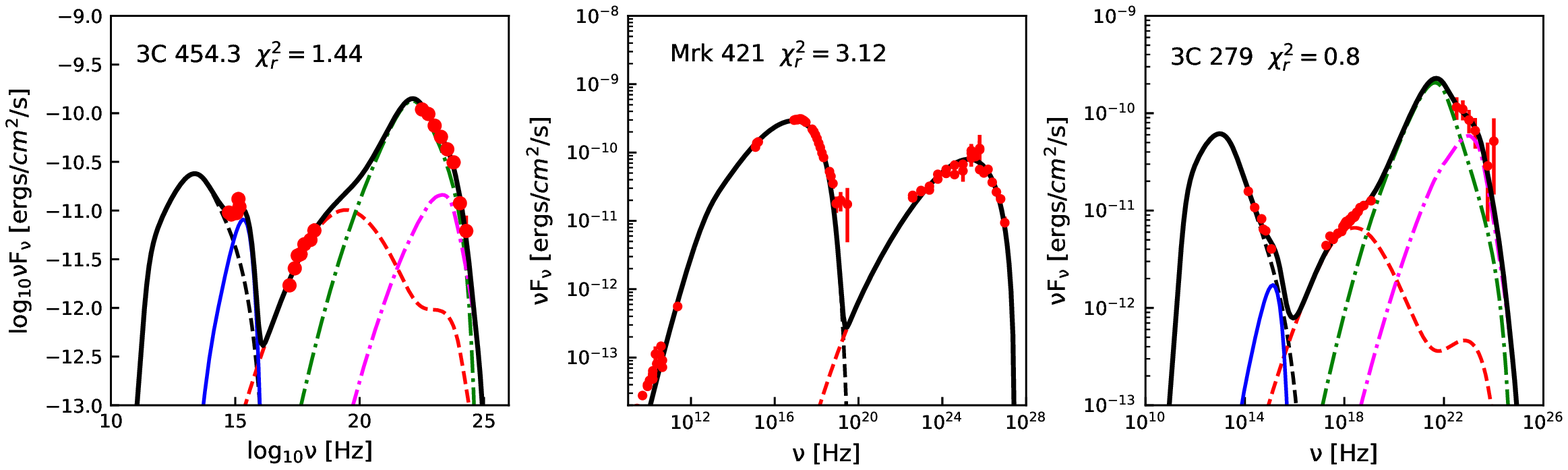}
  \caption{Comparisons of the best-fitting SEDs with observed data of 3C 454.3, Mrk 421 and 3C 279 at the quiescent states with $t_{\rm var}=1\ $day.
   }\label{figure6}
\end{figure*}

\begin{figure*}
 \includegraphics[width=0.49\textwidth]{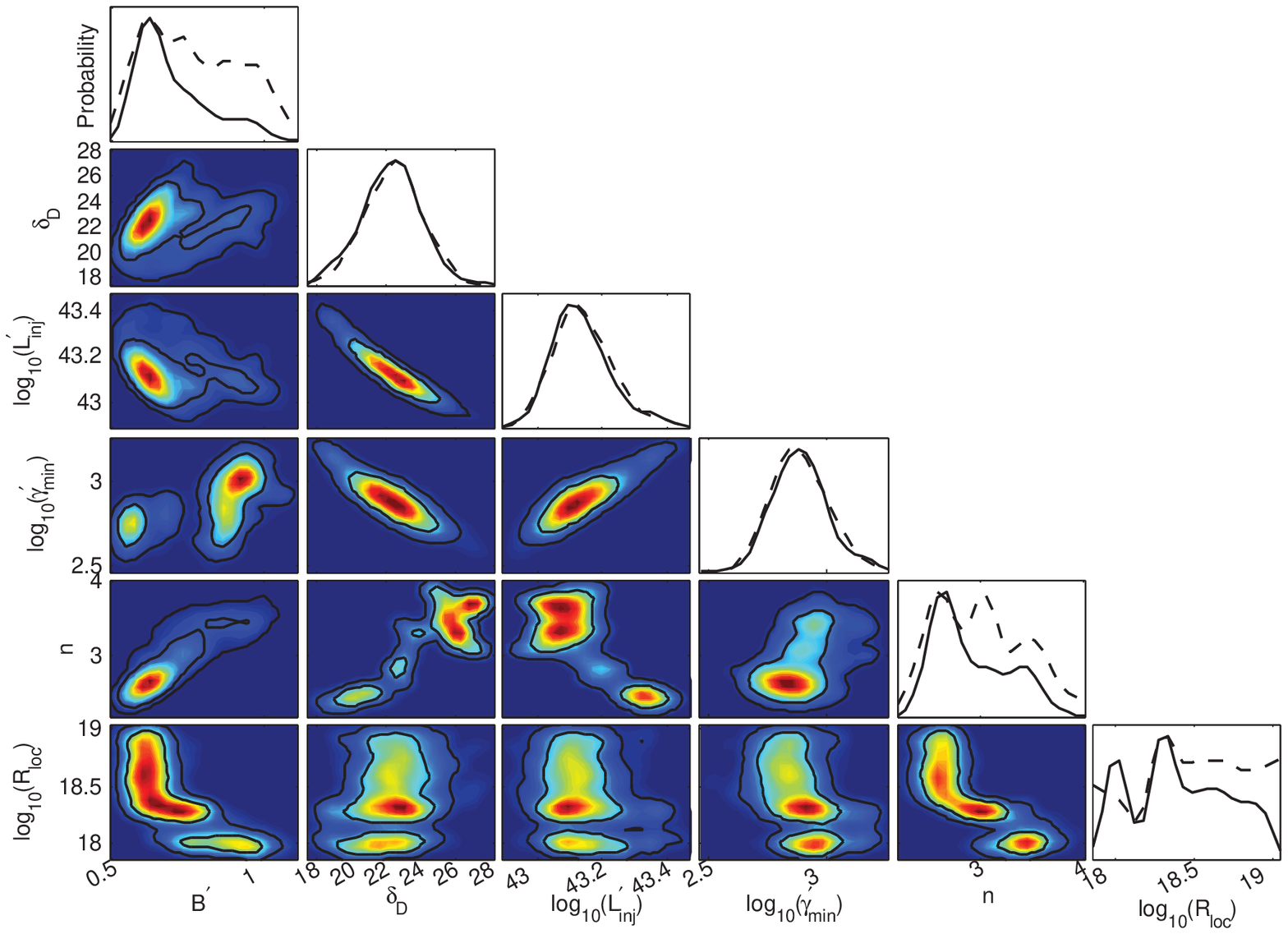} \includegraphics[width=0.49\textwidth]{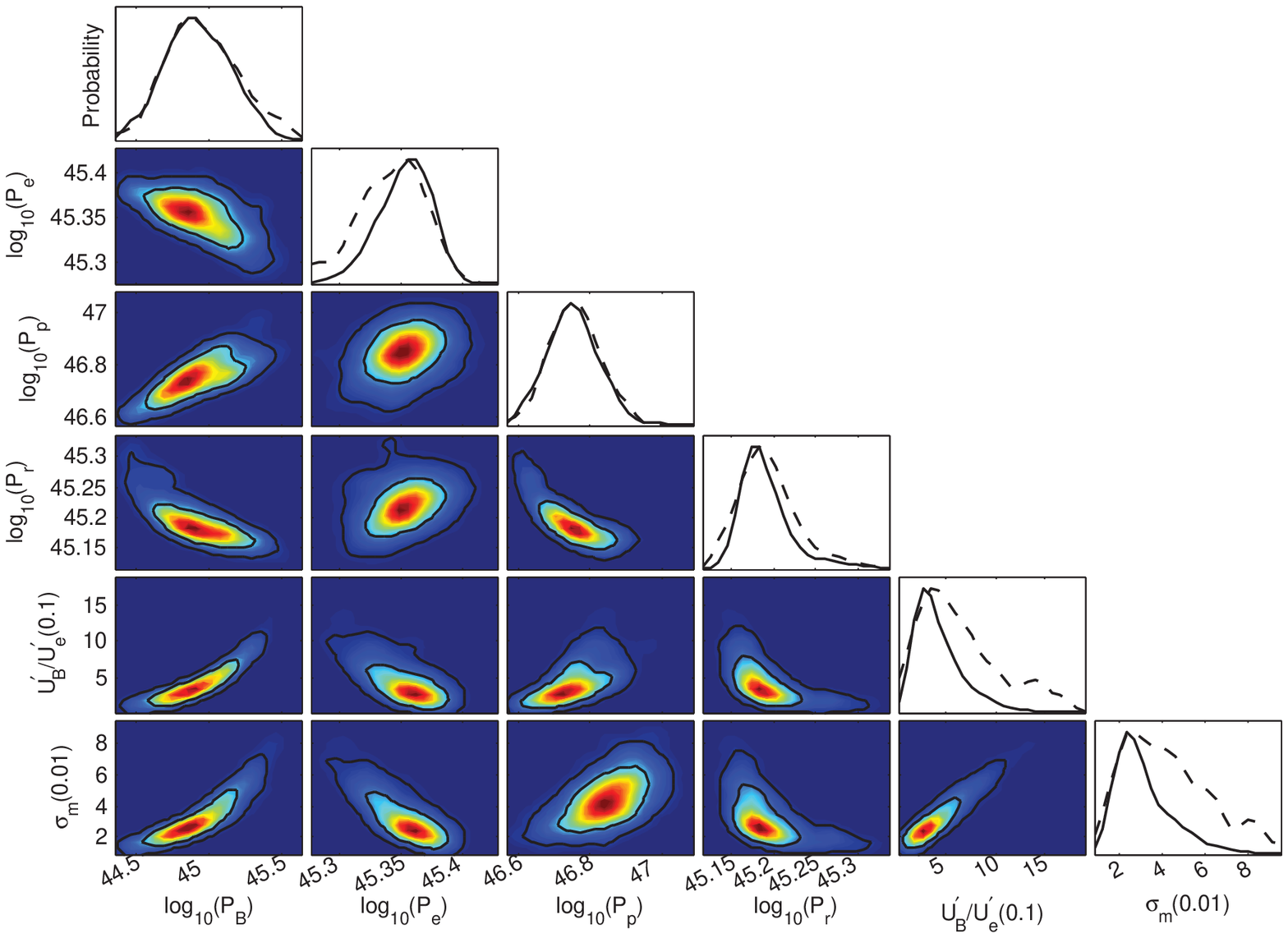}
 \includegraphics[width=0.49\textwidth]{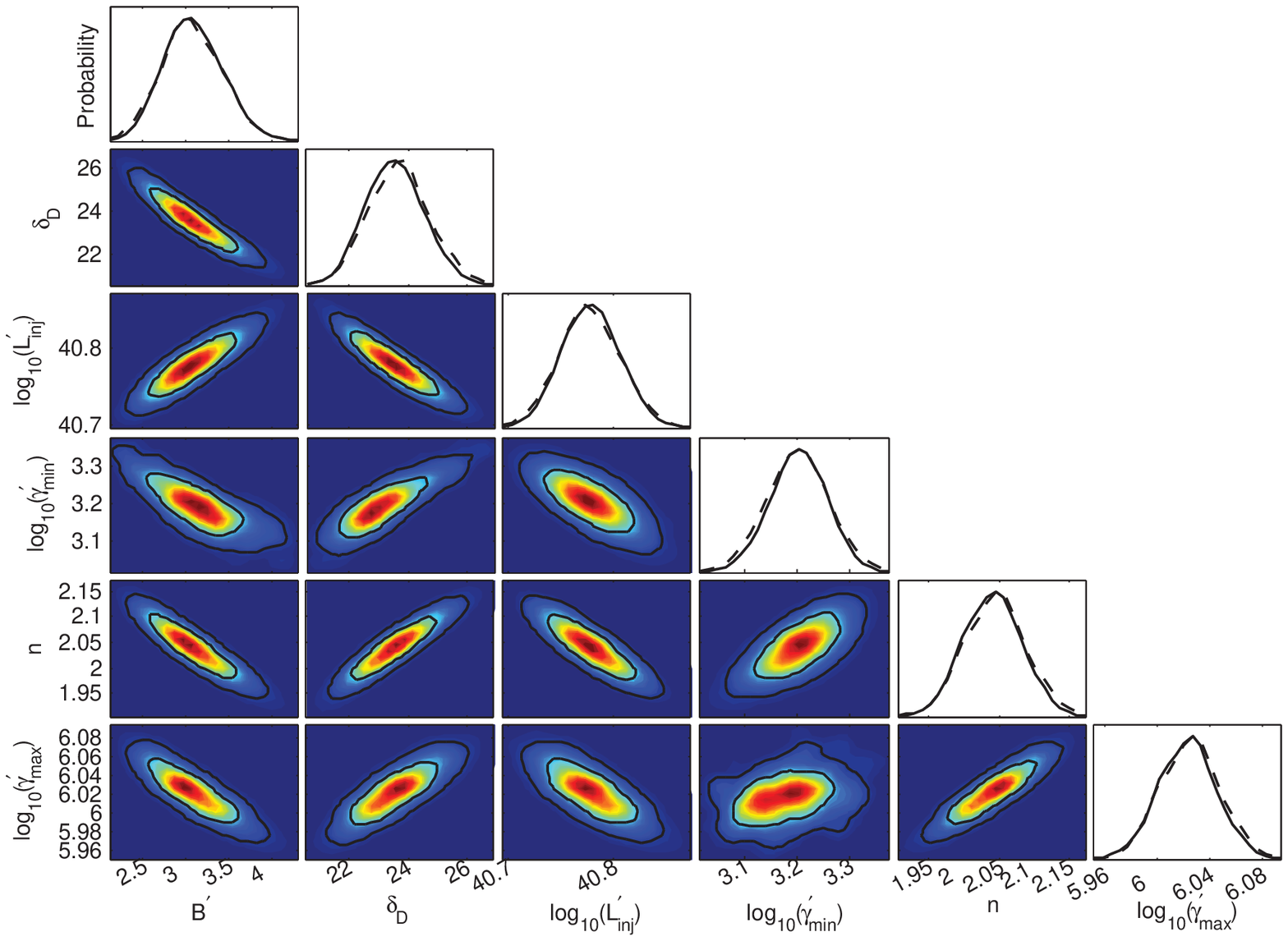} \includegraphics[width=0.49\textwidth]{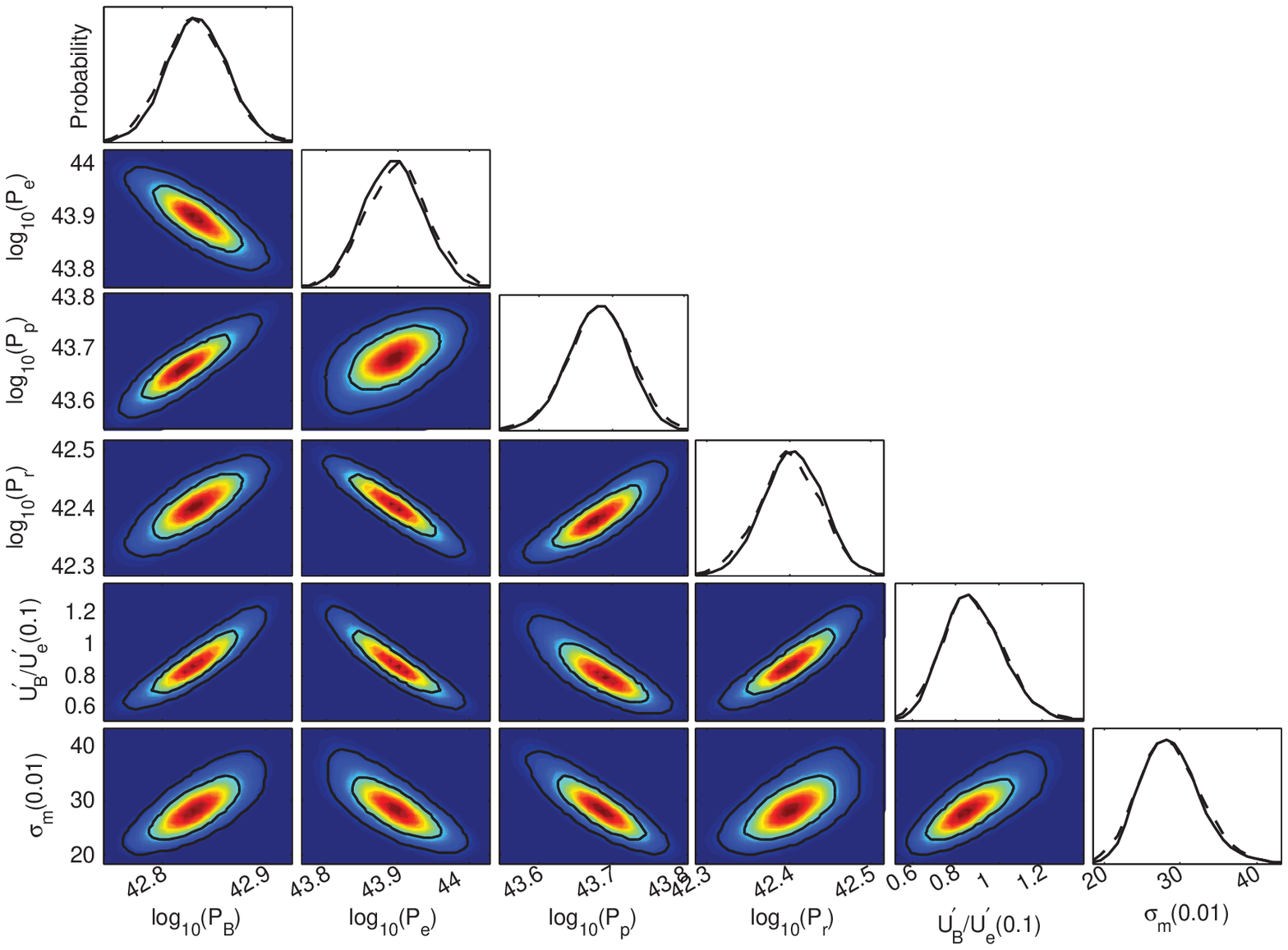}
 \includegraphics[width=0.49\textwidth]{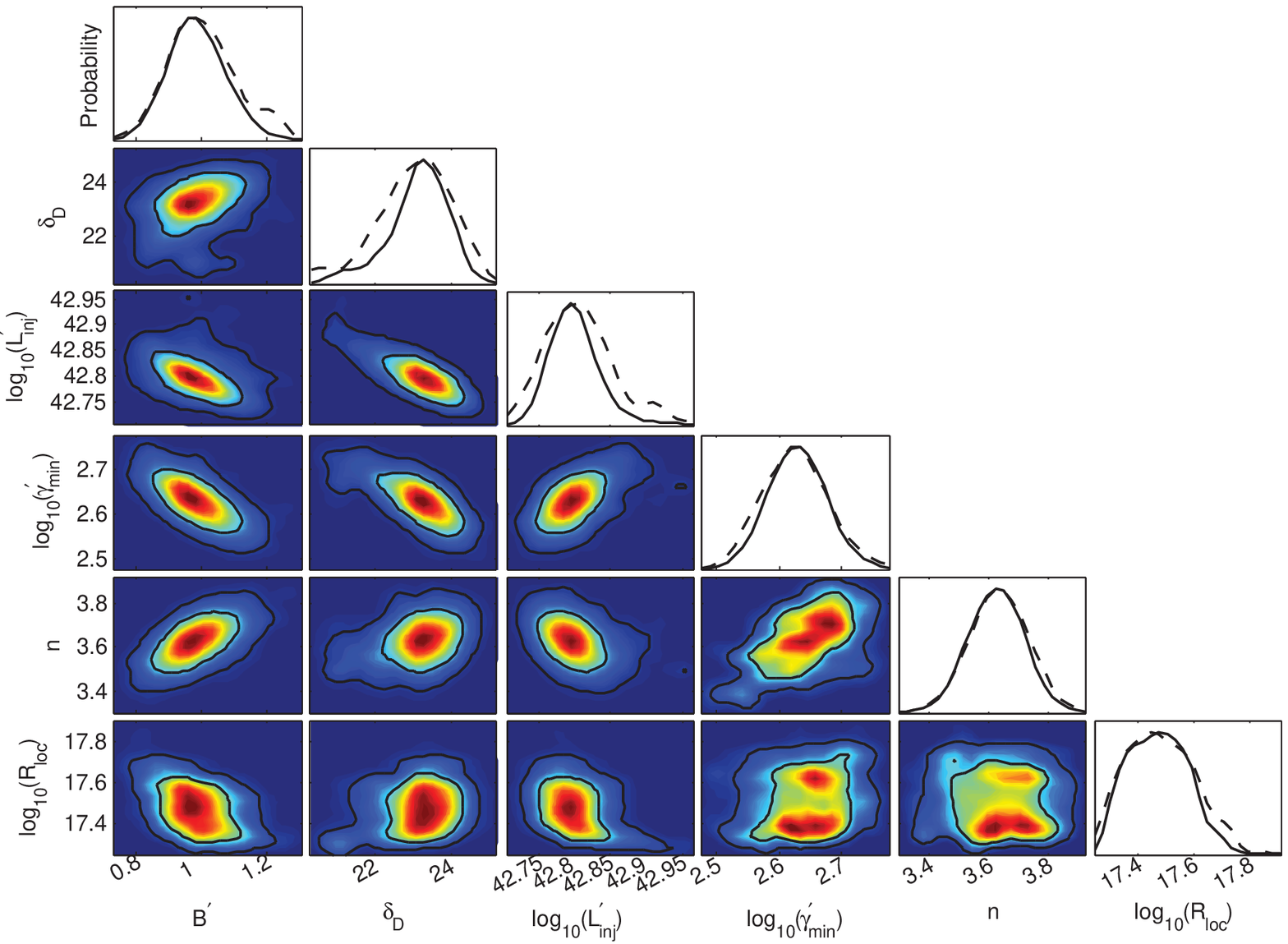} \includegraphics[width=0.49\textwidth]{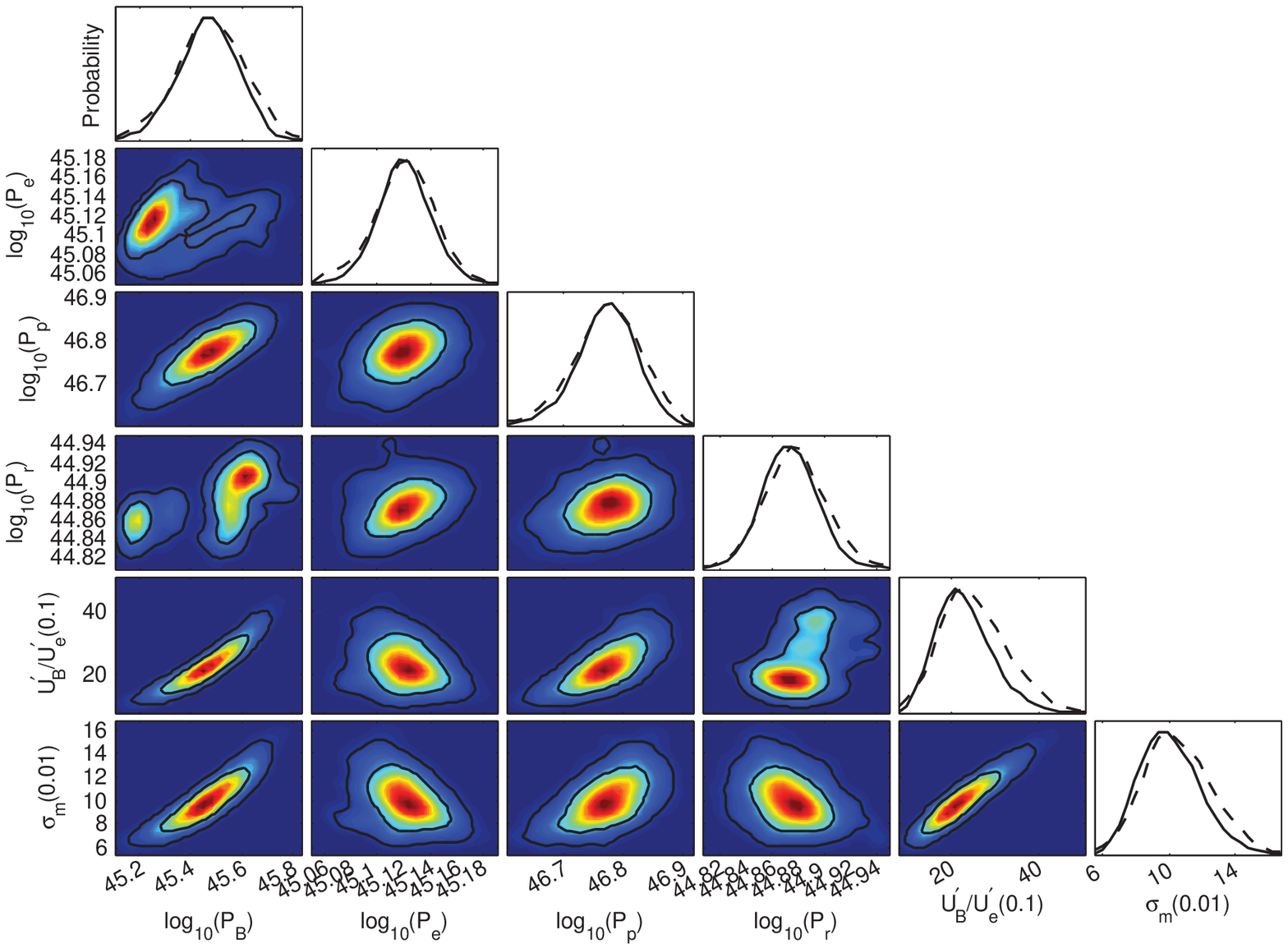}
\caption{
Corner plots of the free model parameters (left) and derived parameters (right) for the quiescent states with $t_{\rm var}=1\ $day.
From top to bottom, the plots are the results obtained from fitting SEDs for 3C 454.3, Mrk 421 and 3C 279, respectively.
 \label{distr5}}
\end{figure*}

Figure \ref{figure6} shows the best-fitting SEDs to the three sources at the quiescent states with $t_{\rm var}=1\ $day, 
and Figure \ref{distr5} shows the corner plots of the model parameters. 
The results of this analysis are discussed in detail in Section \ref{sec:summ}, and the values are shown in Tables \ref{tab1} and \ref{tab2}.

\end{document}